\documentclass[12pt,a4paper]{article}
\pdfoutput=1
 
\usepackage{jheppub-arxiv}
\usepackage[usenames,dvipsnames,table]{xcolor}
 
\usepackage{amssymb}
\usepackage{amsmath}
\usepackage{mathtools}
\usepackage{amsfonts}    
\usepackage{dsfont}
\usepackage{pdfpages}
\usepackage{verbatim}
\usepackage{stmaryrd}
\usepackage{graphicx}
\usepackage{fontawesome}
\usepackage{import}
\usepackage{etoc}
\usepackage{enumitem}
\usepackage{tensor}
\usepackage{mathrsfs}
\usepackage{textgreek} 
\usepackage[mathscr]{euscript}
\usepackage[smalltableaux]{ytableau}
\ytableausetup{centertableaux}
\usepackage{tikz}
\usetikzlibrary{decorations.pathreplacing, decorations.markings,calc,shapes.misc,decorations.pathmorphing,patterns.meta, math,positioning}
\usepackage{tabularx,ragged2e}
\usepackage{physics}

\usepackage{slashed}
\usepackage{datetime}
\usepackage{hyperref}
\hypersetup{
    pdfencoding=unicode,
	colorlinks=true,
	urlcolor=Maroon,
	linkcolor=RoyalBlue,
	citecolor=Maroon,
	pdftitle={Records from the S-Matrix Marathon: A Timeless History of Time},
	pdfauthor={},
	pdfdisplaydoctitle=true,
	pdfstartview=FitH,
	linktocpage=true
}

\usetikzlibrary{shapes}
\usepackage{enumitem}
\usepackage[export]{adjustbox}
\usepackage{nccmath}
\usepackage{bbm}
\usepackage{braket}
\usepackage{empheq}
\usepackage{cancel}
\usepackage{cleveref}
\usepackage[most]{tcolorbox}
\newcommand{\defQ}[1]{\textcolor{black}{\emph{#1}}}
\definecolor{charcoal}{HTML}{343837}
\newcommand{\sumint}{
  \mathop{%
    \mathchoice%
    {\ooalign{$\displaystyle\sum$\cr\hidewidth$\displaystyle\int$\hidewidth\cr}}%
    {\ooalign{$\textstyle\sum$\cr\hidewidth$\textstyle\int$\hidewidth\cr}}%
    {\ooalign{$\scriptstyle\sum$\cr\hidewidth$\scriptstyle\int$\hidewidth\cr}}%
    {\ooalign{$\scriptscriptstyle\sum$\cr\hidewidth$\scriptscriptstyle\int$\hidewidth\cr}}%
  }\displaylimits
}

\usepackage{bm}
\definecolor{yellowish}{rgb}{0.880722,0.611041,0.142051}

\newcommand{\ba}{\begin{align}}

\newcommand{\be}{\begin{equation}}
\newcommand{\ee}{\end{equation}}
\def\bd{\begin{tikzpicture}}
\def\ed{\end{tikzpicture}}

\renewcommand\Im{\mathop{\text{Im}}}

\renewcommand\Re{\mathop{\text{Re}}}

\renewcommand{\vec}[1]{\mathbf{#1}}
\allowdisplaybreaks

\def\XXint#1#2#3{{\setbox0=\hbox{$#1{#2#3}{\int}$}
     \vcenter{\hbox{$#2#3$}}\kern-.5\wd0}}

\definecolor{light-gray}{gray}{0.75}

\renewcommand\d{\text{d}}

\newcommand{\e}{\mathrm{e}}

\newcommand{\D}{\mathbb{D}}

\newcommand{\eps}{\varepsilon}

\renewcommand{\ge}{\geqslant}
\renewcommand{\leq}{\leqslant}
\renewcommand{\geq}{\geqslant}

\DeclareMathOperator\Disc{Disc}
\definecolor{dpurple}  {RGB} {189,  147,  249}

\usepackage{ntheorem}
\usepackage[framemethod=tikz,footnoteinside=false,hidealllines=true,topline=true,bottomline=true,linecolor=black!10!white,linewidth=1pt,innerleftmargin=0em,innerrightmargin=0em,frametitlerule=true,ntheorem]{mdframed}
\theorembodyfont{\upshape}

\newmdtheoremenv{mtheorem}{Theorem}[section]
\newmdtheoremenv[]{mdexample}{Example}[section]
\newmdtheoremenv{mdremark}{Remark}[section]
\newmdtheoremenv{mddefinition}{Definition}[section]
\newmdtheoremenv{mdcorollary}{Corollary}[section]
\newmdtheoremenv{mdproposition}{Proposition}[section]
\newmdtheoremenv{QA}{Audience question}[section]
\newcommand{\question}[1]{%
    #1\\[2ex] 
    \textit{Answer:} 
}

\title{{\Large\normalfont Records from the S-Matrix Marathon:}\\ A Timeless History of Time}

\author{{\normalfont Lecturers:}}
\author[1,2]{Mang~Hei~Gordon~Lee,}
\author[1]{Enrico~Pajer}

\author{\\ {\normalfont Notes written by:}}
\author[3]{Mathieu~Giroux,}
\author[4]{Holmfridur~S.~Hannesdottir,}
\author[4,5,6]{Sebastian~Mizera,}
\author[3]{Celina~Pasiecznik}

\affiliation[1]{Department of Applied Mathematics and Theoretical Physics,\\ University of Cambridge, Wilberforce
Road, Cambridge, CB3 0WA, UK}
\affiliation[2]{Leung Center for Cosmology and Particle Astrophysics, National Taiwan University,
Taipei 10617, Taiwan.}
\affiliation[3]{Department of Physics, McGill University, 3600 Rue University,\\ Montr\'eal, H3A 2T8, QC Canada}
\affiliation[4]{Institute for Advanced Study, Princeton, NJ 08540, USA}
\affiliation[5]{Department of Physics, Princeton University, Princeton, NJ 08544, USA}
\affiliation[6]{Princeton Center for Theoretical Science,\\ Princeton University, Princeton, NJ 08544, USA}

\abstract{By directly probing the initial conditions of our universe, cosmological surveys offer us a unique observational handle on quantum field theory in curved spacetime with dynamical gravity and might even allow us to glean information about a full theory of quantum gravity. Here we report on recent progress to study the natural observables in the problem, namely cosmological correlators. After setting the stage, we review results from three different approaches. First, we present the \textit{in-out formalism} as an interesting alternative to the well-known in-in formalism and stress some of its advantages, such as the derivation of recursion relations, correlators cutting rules and a proposal for a de Sitter scattering matrix. Second, we tackle the important open problem of constructing effective theories in curved spacetime, which generally requires an \textit{open quantum system} approach. Third, we provide an executive summary of general properties of the field-theoretic \textit{wavefunction} that follow from symmetries, unitarity, causality and locality. We describe how these properties can be leveraged to bootstrap all tree-level results and we discuss loop contributions.

These notes are based on a series of lectures held during the S-Matrix Marathon workshop at the Institute for Advanced Study on 11--22 March 2024.
}

\begin{document}

\maketitle
\setcounter{page}{1}

\setcounter{tocdepth}{4}
\setcounter{secnumdepth}{4}

\makeatletter
\g@addto@macro\bfseries{\boldmath}
\makeatother

\newpage
\section*{Preface}

This article is a chapter from the \emph{Records from the S-Matrix Marathon}, a series of lecture notes covering selected topics on scattering amplitudes~\cite{RecordsBook}. They are based on lectures delivered during a workshop on 11--22 March 2024 at the Institute for Advanced Study in Princeton, NJ. We hope that they can serve as a pedagogical introduction to the topics surrounding the S-matrix theory. 

These lecture notes were prepared by the above-mentioned note-writers in collaboration with the lecturers. For lecture notes on a closely related topic see also \cite{Benincasa:2022gtd,Benincasa:2024nwd} 

The results presented in Sec. \ref{sec2} were derived in collaboration with Yaniv Donath, those in Sec. \ref{sec3} in collaboration with Thomas Colas and Santiago Agui Salcedo and finally those in Sec. \ref{sec4} in collaboration with Sadra Jazayeri, Harry Goodhew, Scott Melville and David Stefanyszyn. 


\vfill
\section*{Acknowledgments}

M.G.’s and C.P.'s work is supported in parts by the National Science and Engineering Council of Canada (NSERC) and the Canada Research
Chair program, reference number CRC-2022-00421. Additionally, C.P. is supported by the Walter C. Sumner Memorial Fellowship.
H.S.H. gratefully acknowledges funding provided by the J. Robert Oppenheimer Endowed Fund of the Institute for Advanced Study.
S.M. gratefully acknowledges funding provided by the Sivian Fund and the Roger Dashen Member Fund at the Institute for Advanced Study. E.P.'s and M.H.G.L.'s work has been supported by the STFC consolidated grant ST/X001113/1, ST/T000694/1, ST/X000664/1 and EP/V048422/1. 
This material is based upon work supported by the U.S. Department of Energy, Office of Science, Office of High Energy Physics under Award Number DE-SC0009988.

The S-Matrix Marathon workshop was sponsored by the Institute for Advanced Study and the Carl P. Feinberg Program in Cross-Disciplinary Innovation.

\def\D{\mathrm{D}}
\makeatletter
\newenvironment{sqcases}{
  \matrix@check\sqcases\env@sqcases
}{
  \endarray\right.
}
\def\env@sqcases{
  \let\@ifnextchar\new@ifnextchar
  \left\lbrack
  \def\arraystretch{1.2}
  \array{@{}l@{\quad}l@{}}
}
\makeatother
\def\vac#1{\braket{0|#1|0}}

\tikzset{ma/.style={decoration={markings,mark=at position 0.5 with {\arrow[scale=0.7]{>}}},postaction={decorate}}}
\tikzset{mar/.style={decoration={markings,mark=at position 0.5 with {\arrowreversed[scale=0.7]{>}}},postaction={decorate}}}

\newpage


\section*{\label{ch:PajerLee}A Timeless History of Time\\
\normalfont{\textit{Mang Hei Gordon Lee, Enrico Pajer}}}

\setcounter{section}{0}

\noindent\rule{\textwidth}{0.25pt}
\vspace{-0.8em}
\etocsettocstyle{\noindent\textbf{Contents}\vskip0pt}{}
\localtableofcontents
\vspace{0.5em}
\noindent\rule{\textwidth}{0.25pt}
\vspace{1em}

\section[Five things all high-energy physicists should know about cosmology ]{\label{sec:five-things-cosmology}Five things all high-energy physicists should know about cosmology\\
\normalfont{\textit{Enrico Pajer}}}

In the following, we outline five facts about cosmology that all high-energy theorists should know about. These facts highlight the key role that early-universe cosmology plays in the quest to further our understanding of fundamental physics. We believe that these observations provide a strong motivation for high-energy theorists to care about cosmology.  

\subsubsection*{\emph{Fact 1} All cosmological perturbations are primordial}

One fact about our Universe is that perturbations in the distribution of everything that we observe on cosmological scale are primordial in origin. They can be causally generated only \textit{before} the hot big bang\footnote{By hot big bang we mean the phase during which the Universe is filled with a hot bath of relativistic standard model particles.}.

\begin{figure}
    \centering
    \includegraphics[scale=1.4]{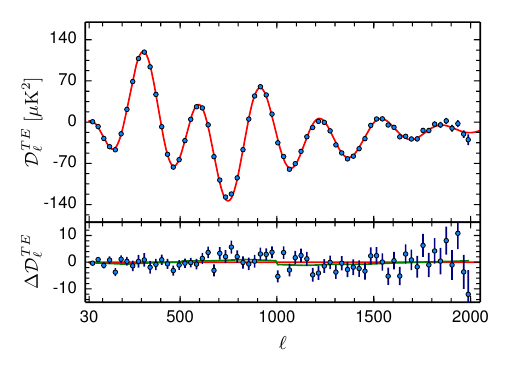}
    \caption{The plot shows the cross-correlation ($\mathcal{D}_{\ell}^{TE}$) between temperature and polarization anisotropies of the CMB as a function of $\ell$, the multiple moment, as measured by the Planck satellite \cite{Planck:2015mrs}. The multiple $l$ is related to the angular scale in the sky by $\theta \approx \frac{180^\circ}{\ell}$. High $\ell$ values correspond to smaller angular scales (fine details) and vice versa. The curve shows how the cross correlation varies with this scale. The red dots with blue error bars represent the actual measurements, whereas the solid line is the theoretical prediction based on the best-fit cosmological model. The anti-correlation around $l\lesssim 50$ is evidence that cosmological perturbations can be causally generated only before the hot big bang. The residuals (bottom plot) show that the differences between the observed data and the theoretical model are small at reasonably large distances.
}
    \label{fig:anasotropies}
\end{figure}

It is an experimental fact that the density of the Universe is slightly inhomogeneous at very early times. We see the consequences of this in the distribution of baryonic and dark matter and in the distribution of photons, the Cosmic Microwave Background (CMB). There are in fact several effects that link the inhomogeneities of the matter distribution to the CMB anisotropies. Fig.~\ref{fig:anasotropies} shows the correlation between temperature and polarization of the CMB, very roughly corresponding to density and velocity of the plasma respectively. On very large scales, at $l<50$, we notice a clear detection of an anti-correlation. This is direct evidence that cosmological perturbations are coherent on those scales \cite{Dodelson:2003ip}. If one assumes the standard hot big bang, these scales are found to be out of causal contact with each other at the time the CMB was released. This shows that a \textit{causal} explanation of initial conditions must be "primordial", i.e. pertaining to \textit{before} the universe became filled with a hot thermal bath of particles.  

\subsubsection*{\emph{Fact 2} Primordial perturbation are adiabatic} On large scales primordial perturbations are all proportional to a \emph{single} function of the spatial position $\vec{x}$:
\begin{equation}\label{eq:single}
    \delta_a(\Vec{x},t)\equiv\frac{\delta\rho_a(\Vec{x},t)}{\Bar{\rho}_a(t)+\Bar{p}_a(t)}\propto \zeta(\vec{x}) \qquad \forall~a\,,
\end{equation}
where $\rho_a$ denotes the density and $p_a$ the pressure (barred quantity denotes spatial average) of particles of kind $a$. Here, $a$ accounts for everything\footnote{Incidentally, also the velocities $\delta u_a$ of all these fluids are observed to be linearly proportional to $\zeta$, as demonstrated by the tight bounds on "velocity" isocurvature perturbations. The precise expression on superHubble scales during radiation domination is $\delta_a(\Vec{k},t)=k^2t^2\zeta(\Vec{k})/a^2=-9\delta u(\Vec{k},t)/(2t)$.} (baryon, dark matter, neutrinos, photons, etc.). For initial cosmological conditions, \eqref{eq:single} is experimentally true to the percent ($1\%$) accuracy level, as demonstrated by the tight bounds on isocurvature perturbations \cite{Planck:2018jri}. At face value, this suggests that the early Universe is not \emph{that} messy.

This observation gives rise to one of the biggest open questions in cosmology: Why is this so simple? There are essentially two paradigms: (i) single-field inflation, which guarantees everything follows this a single "clock", and (ii) multifield inflation followed by a thermalization epoch that makes the many degrees of freedom decay into a single one.

In a sense it is not just a distribution of all forms of matter to be the same, but also the distribution of space-time itself that is determined by the same function (in the comoving gauge)
\be
g_{ij} = a^2 \e^{2\zeta(\vec{x})} \delta_{ij} \quad \text{such that} \quad \delta g_{ii}=6a^2\zeta(\vec{x})\,,
\ee
where $a(t)$ is the FLRW scale factor, whose dynamics is dictated
by the Einstein's equations. Hence, to talk about cosmology we have to talk about quantizing the metric, at least perturbatively.

\subsubsection*{\emph{Fact 3} $\zeta$ is Gau\ss ian and scale invariant}

Based on the previous paragraph, we better understand $\zeta$ as best as we can. The perturbations $\zeta(\vec{x})$ are statistically Gau\ss ian to an accuracy better than $ 0.01\%$ (this is one of the most precise results in cosmology \cite{Planck:2019kim}). This means that the relevant physics is characterized by its two-point function
\be
\hspace{-0.5cm}
\langle \zeta(\vec{x}) \zeta(\vec{y}) \rangle \sim \substack{\text{constant}\\ \text{up to $3\%$}}
~~\text{or in Fourier space}~~ 
\langle \zeta(\vec{k}) \zeta(\vec{k}') \rangle {=} (2\pi)^3 \delta(\vec{k}{+}\vec{k}') \frac{10^{-9}}{k^{3+(1-n_s)}}\,,
\ee
where $n_s$ at the exponent is a the so-called \textit{spectral tilt}, for which the current best measurement gives $n_s =0.9649\pm 0.0042$ \cite{Planck:2018vyg}. The fact that the two-point function depends only very weakly on distance in position space, or equivalently that the power spectrum scales approximately as $\langle \zeta(\vec{k})\rangle \sim k^{-3}$ is known as the (approximate) \textit{scale invariance} of primordial perturbations.

Scale invariance gives us a hint about what the spacetime looked like during the generation of primordial perturbations. The idea is that scale invariance emerged from a spacetime isometry. The deal is that if we want a homogeneous and isotropic spacetime and on top we demand for this extra ``scale invariance" symmetry, we get a maximally symmetric spacetime called de Sitter spacetime with isometry group SO(4,1), corresponding to the conformal group in 3 Euclidean dimension. So in $1{+}3$ space-time dimensions, the geometry of the primordial universe was well approximated by the de Sitter spacetime (in flat slicing):
\be
\d s^2 = -\d t^2 + \e^{2Ht} \d x^2 = \frac{-\d\eta^2 + \d x^2}{\eta^2 H^2}\,,
\ee
where $\eta=-\e^{-Ht}/H$, $-\infty<\eta<0$, and where $H$ is the unknown value of the Hubble parameter during inflation. In fact, it is ``unknown'' by some 37 orders of magnitude! The upper bound comes from the non-observation of primordial gravitational waves and the lower bound from demanding that the universe reheats at a temperature above that of big bang nucleosynthesis, 
\be
10^{13}\text{GeV}>H\gg 10^{-24}\text{GeV}\,.
\ee
This is perhaps the second most unknown quantity in physics. 

\subsubsection*{\emph{Fact 4} Large scale evolution is linear}
Consider the distribution $\rho(\vec{x})$ of some substance and decompose it into a homogeneous background $\bar \rho(t)$ and some perturbations $\delta(t,\vec{x})=(\rho-\bar \rho)/\bar \rho$. Here $\delta$ could represent perturbations in the distribution of atoms, dark matter, photons, neutrinos, and so on. Then, on large scales we have
\be
\delta(\vec{k},t) = T^{(\delta)}(\vec{k},t) \zeta(\vec{k})\,,
\ee
where $T^{(\delta)}$ denotes the so-called \textit{transfer functions} for the perturbations $\delta$. In general, $T^{(\delta)}$ depends on the substance we are looking at and on the choice of cosmology. The transfer functions are well known for the $\Lambda$-Cold Dark Matter ($\Lambda$CDM) model and can be computed numerically in a fraction of a second using a Boltzmann code such as \href{https://camb.info/}{CAMB}.

\subsubsection*{\emph{Fact 5} Cosmological observations probe QFT in curved spacetime}

We are finally coming to the punchline. On large scales, cosmological surveys measure QFT correlators of metric fluctuations
\be\label{eq:important}
\underbracket[0.4pt]{\Big\langle \prod_{a=1}^n \delta(\vec{k}_{a},t)\Big\rangle}_{\text{from observations}}=\underbracket[0.4pt]{\Big[\prod_{a=1}^n T^{(\delta)}(\vec{k}_a,t)\Big]}_{\text{known}}\underbracket[0.4pt]{\Big\langle  \prod_{a=1}^n\zeta(\vec{k}_a)\Big\rangle}_{\text{QFT in de Sitter}}+~\underbracket[0.4pt]{\text{non-linearities}}_{\mathcal{O}(\zeta^{n+1})}\,.
\ee
The central goal of primordial cosmology is to understand QFT and quantum gravity (if there is such a thing) asymptotically in de Sitter. So far we only have access to experimental data for $n=2$, namely the primordial power spectrum (see e.g., Planck data \cite{Planck:2018vyg}). The hope is that within the next 500 years we will have access to $n=3,4$ and, why not, maybe even higher. The prospects for detecting $\langle \zeta^3 \rangle$ are not so bad because it has a known lower limit \cite{Cabass:2016cgp}, which is coming from the fact that gravity is a non-linear theory \cite{Einstein:1916vd}. A recent discussion of future observational prospects can be found in \cite{Achucarro:2022qrl}.

If there is one formula to remember, \eqref{eq:important} is the one\footnote{The non expert may simply neglect the non-linearities and assumed the transfer functions are known in the standard model, however it should be mentioned that a large research effort in the cosmology community is devoted to better understand and model non-linearities and to predict the transfer functions for more general models beyond $\Lambda$CDM.}. It tells us why we should care about cosmology as field theorists and why we should consider anything beyond QFT. This represents the path forward.

\section[The ins and outs of cosmological correlators]{The ins and outs of cosmological correlators\\
\normalfont{\textit{Enrico Pajer}}}\label{sec2}

The big picture is that we want to think about QFT in de Sitter because that allows us to compute the correlators in \eqref{eq:important}, and many different approaches to this central problem have been taken so far. In the last four or five years, many groups worldwide (recent work on the field theoretic wavefunction can be found, e.g., in \cite{Pimentel:2013gza,Maldacena:2002vr,Anninos:2014lwa,Ghosh:2014kba,Arkani-Hamed:2017fdk,Arkani-Hamed:2018kmz,Baumann:2019oyu,Hillman:2019wgh,Baumann:2020dch,Goodhew:2020hob,Jazayeri:2021fvk,Melville:2021lst,Goodhew:2021oqg,Bonifacio:2021azc,Cabass:2022jda,Stefanyszyn:2023qov,Cabass:2021fnw,Hillman:2021bnk,Cabass:2022rhr,Bonifacio:2022vwa,Pimentel:2022fsc,Wang:2022eop,Cespedes:2020xqq,Salcedo:2022aal,Lee:2023jby,DuasoPueyo:2023kyh,Ghosh:2024aqd,Stefanyszyn:2024msm,Goodhew:2024eup,Benincasa:2024ptf,Cespedes:2023aal,Jazayeri:2022kjy,Creminelli:2024cge}) have been exploring this problem from the perspective of the wavefunction of the Universe and its properties, and much of the progress in the field of the cosmological bootstrap has had the wavefunction itself as the main object of investigation. This is because the wavefunction is somehow a simpler and more primitive object: if we know the wavefunction of the Universe, we can compute whatever we want.

There is another approach, which is perhaps the oldest one: not computing the wavefunction but just computing the correlators. After all, these are the real observables. People have found that the computation of correlators is somewhat more involved for technical reasons, which we shall explain below.

\subsection{The in-in formalism}
\subsubsection*{Setting up the stage} 
\begin{figure}
    \centering
    \includegraphics[scale=0.7]{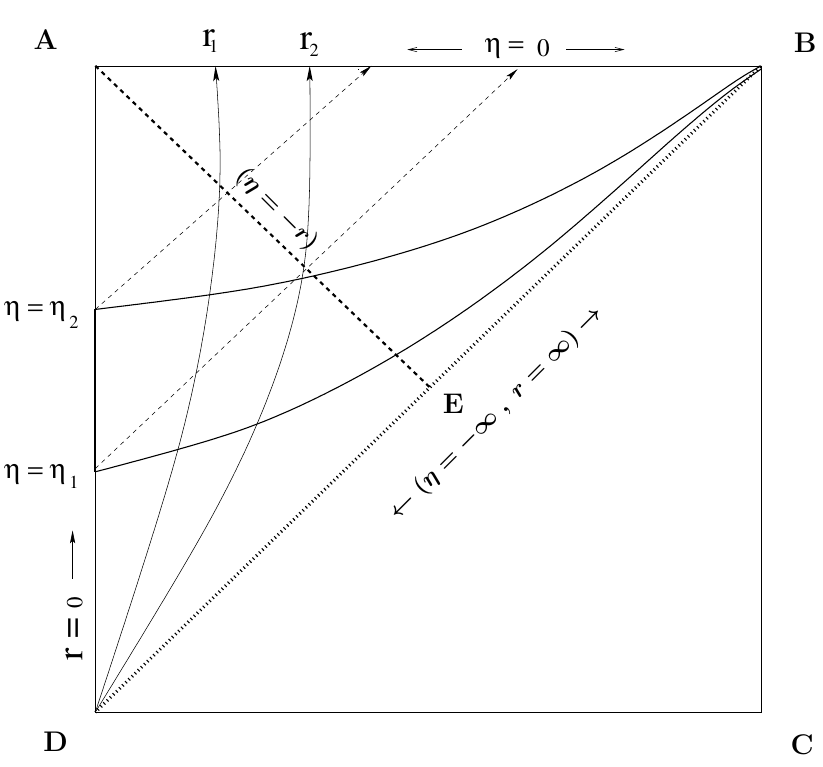}
    \caption{Displayed is the complete Penrose diagram for de Sitter space-time, where each point represents a 2-sphere. The Poincaré patch labeled with ABD is described by the conformal chart $(\eta, r, \theta, \phi)$. Line AB represents future infinity ($\mathscr{J}^+$), and lines BD (past null infinity where the Bunch--Davies vacuum is defined) and AE denote the cosmological horizon of a static observer at the south pole ($r=0$). The diagram illustrates two constant-$\eta$ spacelike hypersurfaces with $\eta_2 > \eta_1$, and two constant $r$, timelike hypersurfaces with $r_2 > r_1$. Dotted lines inclined at 45 degrees show paths of gravitational waves emitted at $\eta = \eta_1, \eta_2$ on the worldline at $r = 0$, emanating from the source. The source is active during the time interval $(\eta_1, \eta_2)$, emitting rapidly enough to fall within detectable frequencies. The AED region is static. Picture reproduced from \cite{Date:2015kma}.}
    \label{fig:patch}
\end{figure}

The system we are considering is that of a large semi-classical gravitational spacetime background that is very close to de Sitter space (and, in fact, for most of this section, we will work in exactly de Sitter space, although small perturbations can be described in perturbation theory). Later, we will allow quantum perturbations on top of it, so that the framework is that of quantum field theory on curved spacetime, which has a conformal diagram depicted as a square in Fig.~\ref{fig:patch}.
The part of the square that is relevant for cosmology is the upper triangle and it is called the Poincaré patch. It is charted by the Poincar\'e coordinates 
\begin{equation}
    \d s^2=\underbracket[0.4pt]{-\d t^2+a^2 \d x^2}_{\text{FLRW coordinates}}=\underbracket[0.4pt]{a^2(-\d \eta+\d x^2)}_{\text{conformal coordinates}}\,,
\end{equation}
where $a=\e^{H t}=-1/(\eta H)$ with constant Hubble parameter $H$. 

In principle, if we knew everything about the Universe, we could find a full wavefunction that describes all degrees of freedom, including the metric. Provided we knew the initial conditions at the beginning of inflation in the infinite past (BD line in Fig.~\ref{fig:patch}), and understood the theory of the Universe, we would be able to evolve it in time to find predictions on the AB line of the same figure. The AB line represents what, for de Sitter, is the future conformal\footnote{The word "conformal" here is just a reminder that this "boundary" line can only be reached from a bulk point in an infinite proper time.} boundary and, in the context of inflation, represents the reheating surface where inflation transitions into the hot Big Bang. Let us emphasize once more that if we know the wavefunction on the AB line, then we essentially know everything about the Universe and can directly compute correlators on that future conformal boundary. 

In this section, we will assume that the initial state is the Bunch--Davies vacuum. This state satisfies two important properties: (i) it is de Sitter invariant, and (ii) it asymptotes the Minkowski vacuum on very short scales. In principle, we could have started with a different initial condition, since at the level of QFT in curved spacetime, other states in the Hilbert space are also allowed, albeit less motivated initial states\footnote{An example would be the Bogliubov initial state, a set of states related to the Bunch--Davies vacuum by Bogliubov transformations \cite{deBoer:2004nd, Einhorn:2003xb}. See \cite{Ghosh:2022cny, Chopping:2024oiu,Ghosh:2024aqd} for recent development.}.

\subsubsection*{Correlators} While the wavefunction describes the entire system, what we actually measure, as mentioned earlier, are correlators. By correlators, we mean the expectation value of the product of local operators $\mathscr{O}$ on the state of the Universe $\Omega$ (presumably a vector in some Hilbert space of the QFT of everything), which, in momentum space, reads as
\begin{equation}\label{eq:corr0}
    \lim_{\eta\to 0}\braket{\Omega|\prod_{a=1}^n\mathscr{O}(\vec{k}_a,\eta)|\Omega}\equiv \braket{\Omega|\prod_{a=1}^n\mathscr{O}(\vec{k}_a)|\Omega}\equiv \braket{\mathscr{O}^n}\,.
\end{equation}
Here, $\eta$ denotes a time variable. Most of the time, cosmologists are interested in equal-time products, and usually the time is taken to be the asymptotic future, $\eta \to 0$, because it is assumed that what we measure is the end result of inflation, since none of us was present during inflation to measure unequal-time correlators at finite $\eta$.

In practice, we know how to compute \eqref{eq:corr0} in perturbation theory. The framework that makes this possible is known as the in-in formalism, in which we imagine evolving from the infinite past by turning on interactions adiabatically. More concretely, we start from the vacuum of the free theory and turn on the interaction slowly enough with an $i\varepsilon$ rotation of the time integral contour, and then evolve in time until some point (this generates the ket) where we insert an operator $\mathscr{O}_I(\eta)$, and then evolve back to generate the bra. This is summarized in Fig.~\ref{fig:schem} or in the following expression:
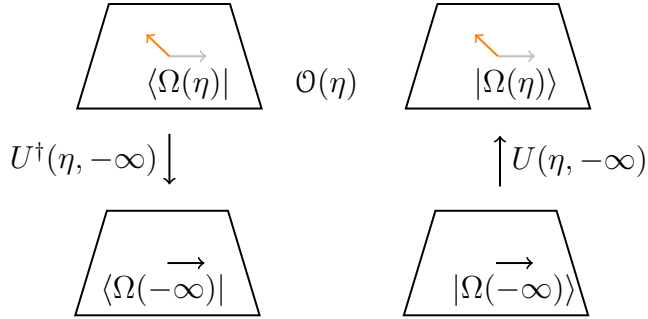
\begin{figure}
    \centering
    \tikzset{every picture/.style={line width=0.75pt}} 
\begin{tikzpicture}[x=0.75pt,y=0.75pt,yscale=-1,xscale=1]
\draw   (105,97.29) -- (120.77,44.71) -- (181.23,44.71) -- (197,97.29) -- cycle ;
\draw   (104,201.29) -- (119.77,148.71) -- (180.23,148.71) -- (196,201.29) -- cycle ;
\draw   (268,201.29) -- (283.77,148.71) -- (344.23,148.71) -- (360,201.29) -- cycle ;
\draw   (269,97.29) -- (284.77,44.71) -- (345.23,44.71) -- (361,97.29) -- cycle ;
\draw[->]    (151,110) -- (151,135.5) node[midway,left]{$U^\dagger(\eta,-\infty)$};
\draw[<-]    (316,111) -- (316,136.5) node[midway,right]{$U(\eta,-\infty)$};
\draw[->,color=gray!50]    (151,71) -- (169.5,71) node[midway,below,color=black]{$\bra{\Omega(\eta)}$};
\draw[<-,color=orange]    (139.5,60) -- (151,71) ;
\draw[->,color=gray!50]    (315,71) -- (333.5,71) node[midway,below,color=black]{$\ket{\Omega(\eta)}$} node[midway,below,color=black,xshift=-2.5cm]{$\mathscr{O}(\eta)$};
\draw[<-,color=orange]     (303.5,60) -- (315,71) ;
\draw[->]    (150,175) -- (168.5,175) node[midway,below,color=black,xshift=-0.3cm]{$\bra{\Omega(-\infty)}$};
\draw[->]    (314,175) -- (332.5,175) node[midway,below,color=black]{$\ket{\Omega(-\infty)}$};
\end{tikzpicture}
    \caption{On the bottom, we start with both a bra and a ket, which we evolve from minus infinity, where they are initially the Bunch--Davies states. After some time, we insert the operator that we want to use for a measurement. This is called an in-in observable because both the bra and the ket are prepared from the past. The arrows represent some vector in the Hilbert space and $U$ denotes unitary evolution, which is geometrically just a rotation of the vectors.}
    \label{fig:schem}
\end{figure}
\begin{equation}\label{eq:corr1}
    \braket{\mathscr{O}(\eta)}=\braket{0|\Big[\overline{\mathcal{T}}\e^{i\int_{-\infty(1+i\varepsilon)}^{\eta}\d \eta' H_{\text{int}}(\eta')}\Big]\mathscr{O}_I(\eta)\Big[\mathcal{T}\e^{-i\int_{-\infty(1-i\varepsilon)}^{\eta}\d \eta' H_{\text{int}}(\eta')}\Big]|0}\,,
\end{equation}
in the interaction picture, which is computed order-by-order via Feynman diagrams (note that \eqref{eq:corr1} is not an amplitude). This strategy was introduced about twenty years ago by Maldacena and then Weinberg, and has since been used by a variety of people to make calculations in a variety of models and to study formal properties of de Sitter. There are straightforward diagrammatic rules for calculations, which are reviewed in many references, for example \cite{Chen:2017ryl,Donath:2024utn}. Here is an explicit example to illustrate what a typical calculation would require.
\begin{mdexample}\label{ex:Paj1}
We consider a contact diagram with one vertex $H_{\text{int}}\sim \dot{\phi}^3$. To leading order in the coupling constant of this interaction, there are two diagrams to compute; one where the perturbations interact once on the bra and the other where the perturbation interacts on the ket. In this simple case, the two diagrams are just the complex conjugate of each other and so the sum can be written as a real part:
\begin{equation}
\begin{split}
    \adjustbox{valign=c}{
        \tikzset{every picture/.style={line width=0.75pt}}
\begin{tikzpicture}[x=0.75pt,y=0.75pt,yscale=-1,xscale=1]
\draw    (180,61) -- (203.25,105.75) ;
\draw    (203.25,105.75) -- (195,61.5) ;
\draw    (237,60.5) -- (203.25,105.75) ;
\draw    (50,62) -- (137,61.5) ;
\draw    (161,61) -- (248,60.5) ;
\draw    (67,62) -- (90.25,106.75) ;
\draw  [fill={rgb, 255:red, 0; green, 0; blue, 0 }  ,fill opacity=1 ] (93,106.75) .. controls (93,105.23) and (91.77,104) .. (90.25,104) .. controls (88.73,104) and (87.5,105.23) .. (87.5,106.75) .. controls (87.5,108.27) and (88.73,109.5) .. (90.25,109.5) .. controls (91.77,109.5) and (93,108.27) .. (93,106.75) -- cycle ;
\draw  [fill={rgb, 255:red, 255; green, 255; blue, 255 }  ,fill opacity=1 ] (206,105.75) .. controls (206,104.23) and (204.77,103) .. (203.25,103) .. controls (201.73,103) and (200.5,104.23) .. (200.5,105.75) .. controls (200.5,107.27) and (201.73,108.5) .. (203.25,108.5) .. controls (204.77,108.5) and (206,107.27) .. (206,105.75) -- cycle ;
\draw    (90.25,106.75) -- (82,62.5) ;
\draw    (124,61.5) -- (90.25,106.75) ;
\draw  [dash pattern={on 0.84pt off 2.51pt}]  (137,61.5) -- (161,61) ;
\draw[->]    (67,62) -- (78.63,84.38) node[pos=0,above]{$\vec{k}_1$};
\draw[->]    (82,62.5) -- (86.13,84.63) node[pos=0,above]{$\vec{k}_1$};
\draw[->]    (124,61.5) -- (107.13,84.13) node[midway,left]{$...$} node[pos=0,above]{$\vec{k}_n$};
\draw[->]    (180,61) -- (191.63,83.38) node[pos=0,above]{$\vec{k}_1$};
\draw[->]    (195,61.5) -- (199.13,83.63) node[pos=0,above]{$\vec{k}_2$};
\draw[->]    (237,60.5) -- (220.13,83.13) node[midway,left]{$...$} node[pos=0,above]{$\vec{k}_n$};
\end{tikzpicture}
    }&\stackrel{n=3}{ =}2\text{Re}\Big[-i\lambda_1\int_{-\infty}^{0}\frac{\d \eta }{(H \eta)^4}(-H \eta)^3\prod_{a=1}^3G_r'(\eta,k_a)\Big]
    \\&
    =2\text{Re}\Big[i\lambda_1\int_{-\infty}^{0}\frac{\d \eta }{H \eta}\prod_{a=1}^3\frac{H^2}{2c_s k_a^3}c_s^2k_a^2 \eta\, \e^{i c_s k\eta}\Big]
    \\&=
    2\text{Re}\Big[i\frac{\lambda_1 H^5 c_s^3}{8k_1k_2k_3}\int_{-\infty}^{0}\d \eta\,\eta^2\,\e^{i E_T\eta}\Big]\\&=\frac{\lambda_1 H^5 c_s^3}{2k_1k_2k_3 E_T^3}\,.
\end{split}
\end{equation}
$G_r'$ denotes the time derivative of the mode functions for a generic mass, which are Hankel functions. To go from the first to the second line, we assumed that $m=0$ and used the massless mode functions for which 
\be 
G(\eta,k)=\frac{H^2}{2c_s k^3} (1-ik\eta) e^{ik\eta}\,.
\ee
It is crucial to add both of these diagrams (according to \eqref{eq:corr1}, there are two Hamiltonians to keep track of, one inside an anti-time order and one inside a time ordering) to obtain a real correlator, which is necessary to ensure that the expectation value of a Hermitian operators is real. The three-point correlator computed above is typically what people search for in the sky in the form of non-Gau{\ss}ianities in galaxy correlations, for example the so-called equilateral and orthogonal bispectrum templates (see \cite{Planck:2019kim} for recent bounds). 
\end{mdexample}

Although we know how to perform the above calculation given a Lagrangian, things get more complicated. (In this chapter, the notation of energies $E$ and $\omega$ will be used interchangeably. Unless specified otherwise, both symbols mean the same thing.)
\begin{mdexample}\label{ex:Paj2}
    For example, the next-to-the-simplest diagram involves a particle exchange. There are four diagrams, corresponding to picking two interactions $H_\text{int}$ from left-left ($B^{(ll)}$), left-right ($B^{(lr)}$), right-left ($B^{(rl)}$) and right-right ($B^{(rr)}$) in \eqref{eq:corr1}. Since diagrams that differ from each other by exchanging all left and right vertices are complex conjugates of each other, we just need to compute two contributions
    \begin{equation}
        \adjustbox{valign=c}{
        \tikzset{every picture/.style={line width=0.75pt}} 
\begin{tikzpicture}[x=0.75pt,y=0.75pt,yscale=-1,xscale=1]
\draw    (281,105) -- (229.25,105.75) ;
\draw    (281,105) -- (286,61) ;
\draw    (281,105) -- (302,61) ;
\draw    (54,62) -- (175,61.5) ;
\draw    (67,62) -- (90.25,106.75) ;
\draw  [fill={rgb, 255:red, 0; green, 0; blue, 0 }  ,fill opacity=1 ] (93,106.75) .. controls (93,105.23) and (91.77,104) .. (90.25,104) .. controls (88.73,104) and (87.5,105.23) .. (87.5,106.75) .. controls (87.5,108.27) and (88.73,109.5) .. (90.25,109.5) .. controls (91.77,109.5) and (93,108.27) .. (93,106.75) -- cycle ;
\draw  [fill={rgb, 255:red, 255; green, 255; blue, 255 }  ,fill opacity=1 ] (283.75,105) .. controls (283.75,103.48) and (282.52,102.25) .. (281,102.25) .. controls (279.48,102.25) and (278.25,103.48) .. (278.25,105) .. controls (278.25,106.52) and (279.48,107.75) .. (281,107.75) .. controls (282.52,107.75) and (283.75,106.52) .. (283.75,105) -- cycle ;
\draw    (90.25,106.75) -- (82,62.5) ;
\draw    (142,106) -- (90.25,106.75) ;
\draw    (142,106) -- (147,62) ;
\draw    (142,106) -- (163,62) ;
\draw  [fill={rgb, 255:red, 0; green, 0; blue, 0 }  ,fill opacity=1 ] (144.75,106) .. controls (144.75,104.48) and (143.52,103.25) .. (142,103.25) .. controls (140.48,103.25) and (139.25,104.48) .. (139.25,106) .. controls (139.25,107.52) and (140.48,108.75) .. (142,108.75) .. controls (143.52,108.75) and (144.75,107.52) .. (144.75,106) -- cycle ;
\draw    (193,61) -- (314,60.5) ;
\draw    (206,61) -- (229.25,105.75) ;
\draw  [fill={rgb, 255:red, 0; green, 0; blue, 0 }  ,fill opacity=1 ] (232,105.75) .. controls (232,104.23) and (230.77,103) .. (229.25,103) .. controls (227.73,103) and (226.5,104.23) .. (226.5,105.75) .. controls (226.5,107.27) and (227.73,108.5) .. (229.25,108.5) .. controls (230.77,108.5) and (232,107.27) .. (232,105.75) -- cycle ;
\draw   (229.25,105.75) -- (221,61.5) ;
\draw[->]    (67,62) -- (78.63,84.38) node[left,midway]{$G_r$};
\draw[->]    (82,62.5) -- (86.13,84.63) node[right,midway]{$G_r$};
\draw[->]    (90.25,106.75) -- (116.13,106.38) node[above]{$G_{rr}$};
\draw[->]    (147,62) -- (144.5,84) node[left,midway]{$G_r$};
\draw[->]    (163,62) -- (152.5,84) node[right,midway]{$G_r$};
\draw[->]    (206,61) -- (217.63,83.38) node[left,midway]{$G_r$};
\draw[->]    (221,61.5) -- (225.13,83.63) node[right,midway]{$G_r$};
\draw[->]    (229.25,105.75) -- (255.13,105.38) node[above]{$G_{rl}$};
\draw[->]    (286,61) -- (283.5,83) node[left,midway]{$G_l$};
\draw[->]    (302,61) -- (291.5,83) node[right,midway]{$G_l$};
\end{tikzpicture}
    }
    \end{equation}
    This example already gets algebraically involved in de Sitter so we discuss it in Minkowski, where the mode functions are simply
\be
G(t,k)=\frac{e^{iEt}}{2E}\,,
\ee
with $E=\sqrt{k^2+m^2}$. The correlator is then found to be
    \begin{equation}
    \begin{split}
        B_{4,s}^{(rr)}&=(-i\lambda)^2\int \d t_1\d t_2 G_{r}(t_1,k_1)G_{r}(t_1,k_2)G_{rr}(t_1,t_2,E_s)G_{r}(t_2,k_3)G_{r}(t_2,k_4)\,,\\
        B_{4,s}^{(r\ell)}&=\lambda^2\int \d t_1\d t_2 G_{r}(t_1,k_1)G_{r}(t_1,k_2)G_{r\ell}(t_1,t_2,E_s)G_{\ell}(t_2,k_3)G_{\ell}(t_2,k_4)\, ,\\
        B_{4,s}&{=}B_{4,s}^{(rr)}{+}B_{4,s}^{(r\ell)}{+}B_{4,s}^{(\ell r)}{+}B_{4,s}^{(\ell \ell)}{=}2\text{Re}\Big[B_{4,s}^{(rr)}{+}B_{4,s}^{(r\ell)}\Big]{=}\frac{2(E_T+E_s)\lambda^2}{E_LE_RE_TE_s \prod_{a}^4 (2E_a)}\,,
    \end{split}
    \end{equation}
    where the partial energies are $E_R=E_3+E_4+E_s$ and $E_L=E_1+E_2+E_s$. We emphasize that \textit{the number of integrals to perform for a single tree-level diagram increases exponentially in the number of vertices}: for a process involving $V$ vertices, there are $2^V$ labeling of the diagram to consider, and each of them correspond to a different nested time integral over Hankel functions already at tree-level. Each diagram describes one way in which the interactions change the bra and/or the ket through time evolution. The increase in combinatorial complexity emerges from the need to keep track of both the right and left vertices in \eqref{eq:corr1}. Thus, the calculation quickly becomes intractable for a large enough $V$ even though the integrated result is manifestly simple. In particular, the singularity structure of $ B_{4,s}$ is physical while the individual $B_{4,s}^{(\bullet \star)}$ contributions have spurious poles that cancel in the sum. This indicates that doing the calculation in this way is probably not optimal.
\end{mdexample}
The formalism outlined above was not originally developed for cosmology, but was instead developed for entirely different reasons by Schwinger, and later by Keldysh, then by Feynman and Vernon, to calculate out-of-equilibrium quantum field theories or quantum field theories of open systems (e.g., Brownian motion of particles) rather than closed ones. This is known as the Schwinger--Keldysh formalism, where schematically for equal time correlators one encounters a path integral over a \textit{closed time contour} where each field in the theory has been doubled,
\begin{equation}\label{SK}
    \braket{\phi^n}=\int \d \phi \, \phi^n\, \int_{\rho_0}^{\phi} D\Phi_+\int_{\rho_0}^{\phi} D\Phi_-\,\e^{i(S(\Phi_+)-S(\Phi_-)+F(\Phi_+,\Phi_-))}\,.
\end{equation}
In cosmology, we use this formalism for closed systems (i.e., there are no fluctuations, no dissipation, no Feynman--Vernon influence $F(\Phi_+,\Phi_-)=0$ and one assumes a pure state of a closed system)\footnote{This last assumption is an essential one for in-in $=$ in-out.} in which we are trying to describe the whole Universe (all the degrees of freedom), so we never really allow this description to be applied to an open system. In the presence of an open quantum system, we further need to couple the two branches with $F(\Phi_+,\Phi_-)\neq 0$ of the path integral, and the situation becomes much more interesting. For the rest of this section we assume a close system and no interaction between the branches of the path integral \eqref{SK}, i.e. $F=0$. The open system case will be discussed in Sec.~\ref{sec:open-effective-field-theories}.

Many different advances in the study of correlators have been made on various fronts, and we will describe some exciting progress below, starting with \cite{Donath:2024utn}. Before providing the full details, let us summarize the three main takeaway points:

\begin{itemize}[label=$\diamond$]
    \item Cosmological correlators in de Sitter (and Minkowski) spaces can be computed using the in-in formalism (see Ex.~\ref{ex:Paj1} and \ref{ex:Paj2}). We propose that there is a simpler method for such computations using the \textit{in-out formalism} -- namely, using only the familiar time-ordered (Feynman) propagators:
        \begin{equation}\label{eq:ininEqinout}
        \adjustbox{valign=c}{
        \includegraphics[scale=0.15]{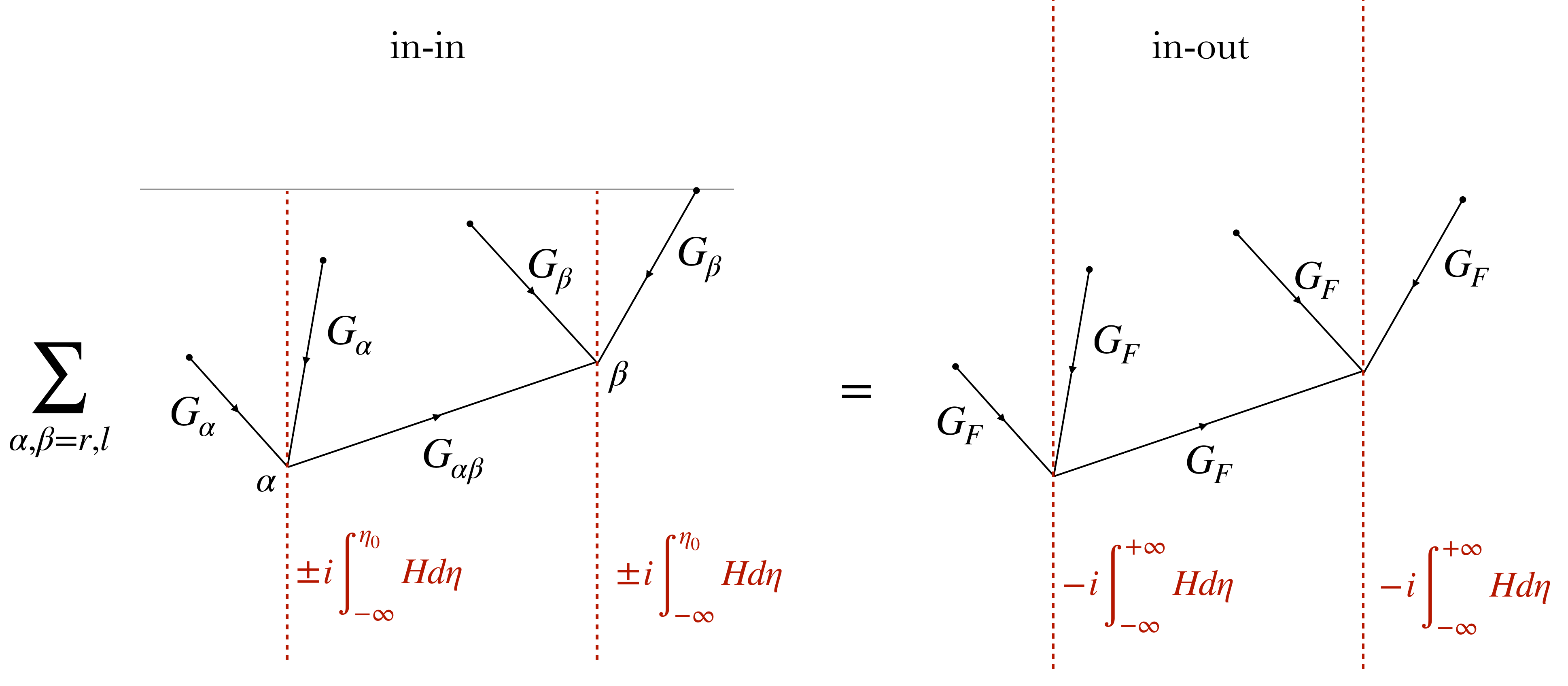}
    }
    \end{equation}
     This method of computation is convenient because it eliminates the need to sum over vertices from the right and left time-evolution operators, as it involves only one time-evolution operator. From the perspective of the integral, the only change is that the range of integration in $\eta$ now extends from $-\infty$ to $\infty$. Because this approach leads to significant technical simplifications, we will use it in the context of cosmology to derive formal results that were previously far-fetched. One of our main goals, for example, will be to make the general consequences of unitarity, locality, and causality more manifest.
    \item For the equality in \eqref{eq:ininEqinout} to hold, some technical assumptions need to be made. One is that the interactions are IR finite. This is the case, for example, for single-field inflation, where interactions always have a sufficient number of derivatives\footnote{Non-derivative interactions such as $\phi^3$ do appear for the inflaton, but not for the relevant quantity $\zeta$.} so that they become very soft in the future and thus go to zero as $\eta \to 0$, which is the future. We also assume that the evolution is unitary (a closed system in the Bunch--Davies vacuum). Moreover, we notice that the equality holds for any number of fields, regardless of their spin and mass.
    \item Finally, the in-out formalism leads to significant simplifications practically (many applications: new recursion relations, cutting rules, pole bagging) and conceptually (S-matrix technology, de Sitter S-matrix, non-perturbative optical theorem).
\end{itemize}
At this stage, let us note that there are different ways to interpret the above messages: 

\begin{figure}
    \centering
    \includegraphics[scale=0.2]{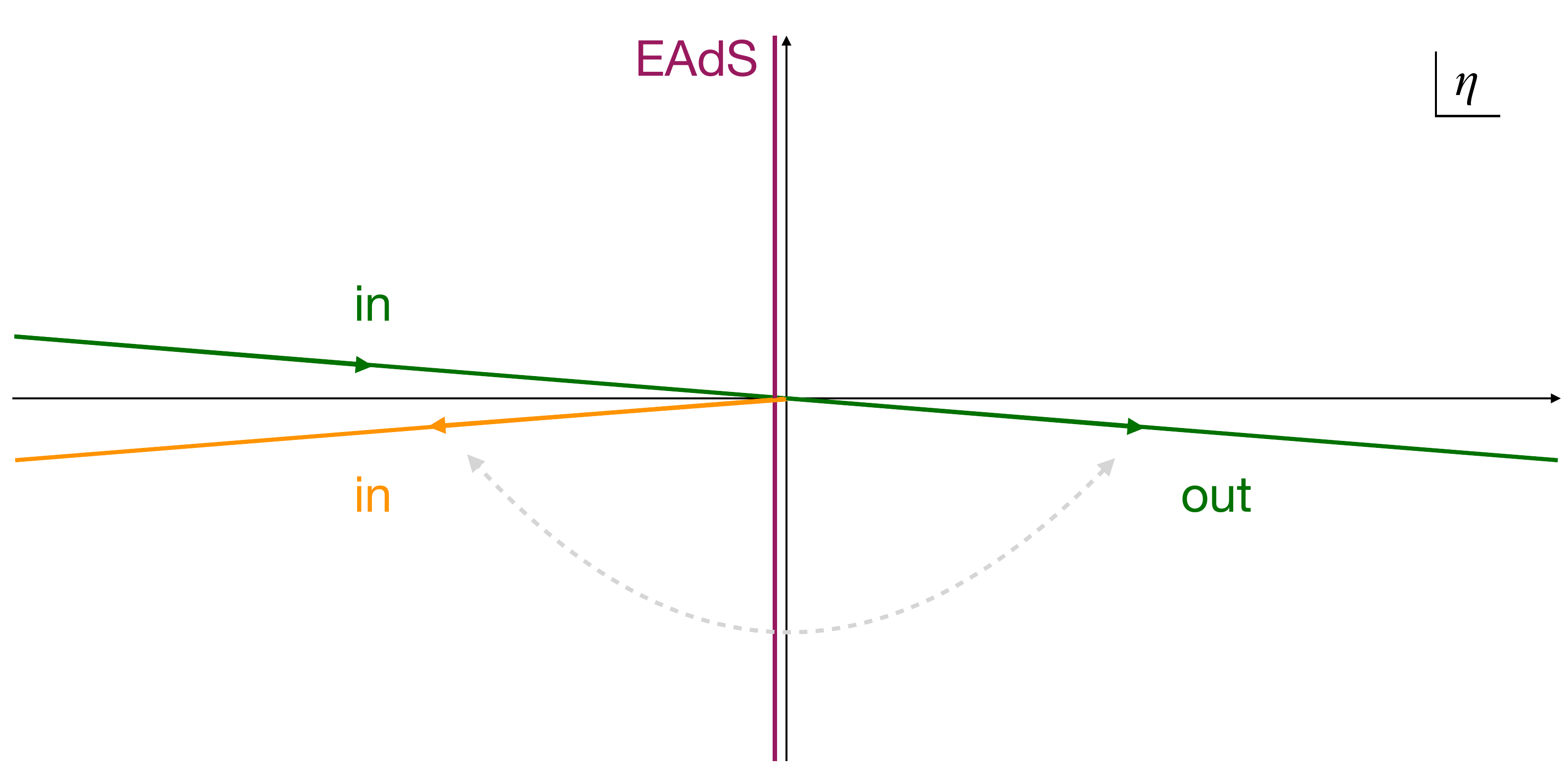}
    \caption{The complex $\eta$ (conformal time) plane with $-\infty<\eta<0$ and the traditional path integral describe the theory as being performed over the yellow and green lines; these are the in-in contours of the path integral. These contours prepare the two indices of the density matrix of the system, allowing for the computation of any correlator.}
    \label{fig:contours}
\end{figure}

\begin{itemize}[label=$\diamond$]
    \item The in-in time-evolution contour (green and yellow in Fig.~\ref{fig:contours}), which lives exclusively at $\eta <0$, can be deformed into an in-out contour (green) by adding a second spacetime (contracting Poincaré patch, where $\eta >0$) that prepares the bra from the future and the ket from the past. Thus, in this picture, we have two different Poincar\'e patched of de Sitter spaces ($-\infty<\eta<0$ and $0<\eta<\infty$) that are merged together; one is an expanding de Sitter patch (on the left) and the other is a contracting de Sitter patch associated with positive conformal time instead of negative.
    \item The end result is a straight contour (green), just like in Euclidean anti de Sitter (vertical line) and in Minkowski amplitudes. 
\end{itemize}
There is an conceptual advantage in the in-out formalism, as depicted in Fig.~\ref{fig:Minkowski-deSitter}:
\begin{figure}
    \centering
    \includegraphics[scale=0.2]{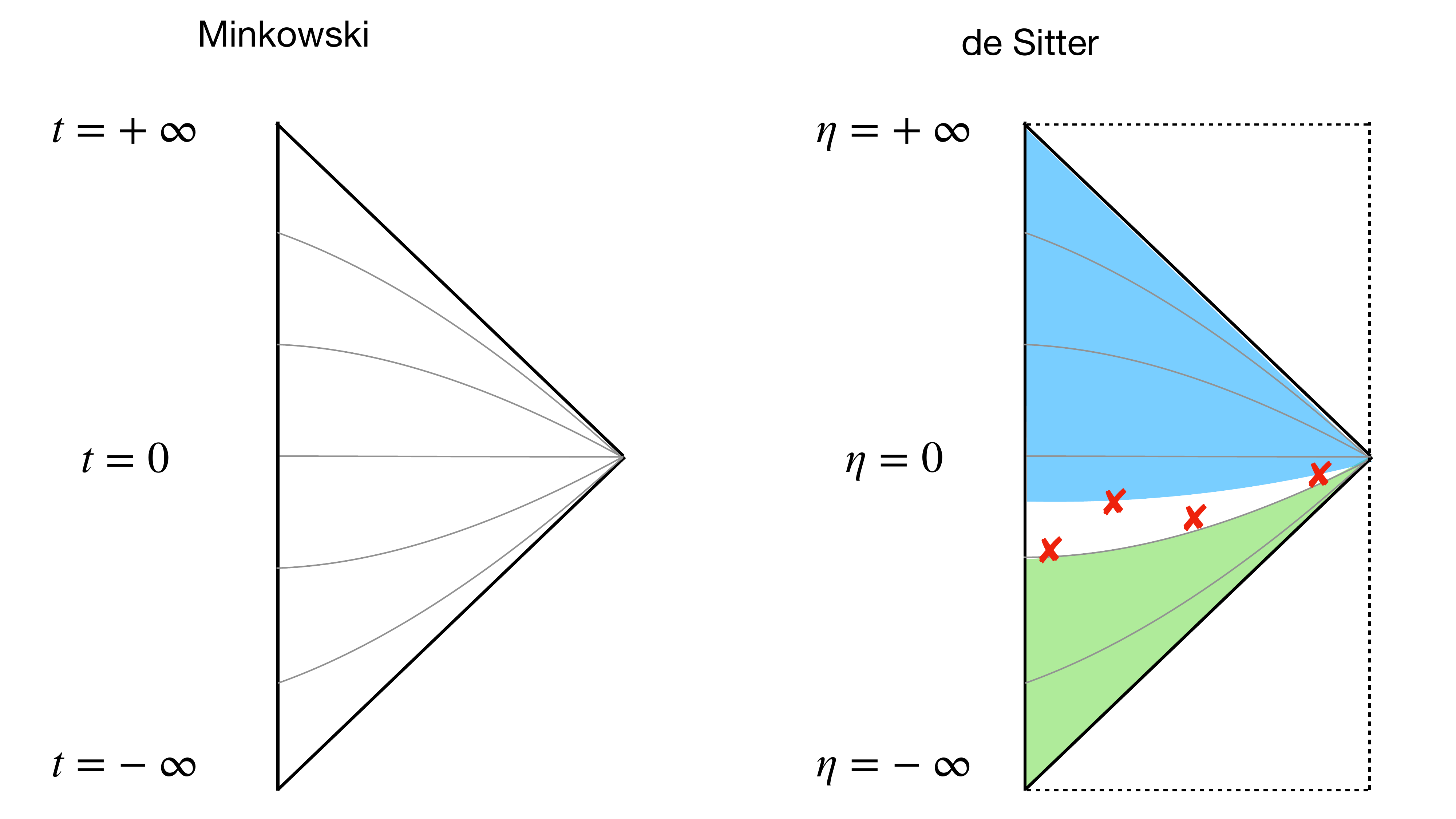}
    \caption{\label{fig:Minkowski-deSitter}Comparison between Minkowski (left) and de Sitter (right). On the right, the lower square is where the operators (red crosses) are inserted, and we aim to compute their expectation value. Typically, we prepare both the bra and the ket from the infinite past. Here, we highlight that the bra can be prepared from the infinite future of another copy of de Sitter (bra = blue region and ket = green region). As long as the bra is prepared to be the correct Bunch--Davies evolved, the in-in or in-out calculation yields the same result. The advantage of preparing the bra from the future is that it allows a single time ordering from $-\infty$ to $\infty$, resulting in a single (Feynman) propagator, which simplifies the technical process of obtaining the result.}
\end{figure}
\begin{itemize}[label=$\diamond$]
    \item Conceptually, the in-out formalism leads to a natural proposal for a de Sitter S-matrix, from a null past (null boundary of the green region) to a null future (null boundary of the blue region).  This is also the case for massive particles.
\item Unitarity-time evolution is imprinted in the de Sitter S-matrix via the
standard optical theorem (both perturbative and non-perturbatively).
\end{itemize}
We now get to some more technical details.


\subsection{In-in \texorpdfstring{$=$}{=} in-out}

Let us slightly generalize the definition of an in-in correlator allowing for unequal time operators inside a time ordering\footnote{When discussing general results we use $\eta$ and $t$ interchangeably. Then, when we consider explicit calculations and check we will use $t$ in Minkowski and $\eta$ in de Sitter.}
\begin{align}\label{eq:corr2}
\hspace{-0.55cm}B_{\text{in-in}}^{(n)}  =\langle 0| & \overline{\mathcal{T}}\Big[\e^{i\int_{-\infty(1+i\varepsilon)}^{t_0}\d t\, H_{\text{int}}(t)}\Big]\,
\\ & \mathcal{T}\Big[\mathscr{O}(t_1,\vec{x}_1)\ldots \mathscr{O}(t_n,\vec{x}_n) \e^{-i\int_{-\infty(1-i\varepsilon)}^{t_0}\d t\, H_{\text{int}}(t)}\Big]|0 \rangle'\,,
\nonumber
\end{align}
where the prime removes the Dirac delta of momentum conservation. If we set $t_1 = \ldots = t_n = t$, then we recover \eqref{eq:corr1}. We will refer to \eqref{eq:corr2} as an in-in correlator. In this case, the $i\varepsilon$ rotation of the contour selects the Fock vacuum as the initial state (Bunch--Davies state) in the infinite past by adiabatically turning off interactions. Furthermore, $t_0$ is any time after all operator insertions.

Now, let us define an a priory different object, which we call an in-out correlator in de Sitter or Minkowski space:
\begin{equation}\label{eq:corr3}
    B_{\text{in-out}}^{(n)}=\frac{\braket{0|\mathcal{T}\Big[\mathscr{O}(t_1,\vec{x}_1)\cdots \mathscr{O}(t_n,\vec{x}_n) \e^{-i\int_{-\infty(1-i\varepsilon)}^{\infty(1-i\varepsilon)}\d t\, H_{\text{int}}(t)}\Big]|0}'}{\braket{0|\mathcal{T}\Big[ \e^{-i\int_{-\infty(1-i\varepsilon)}^{\infty(1-i\varepsilon)}\d t\, H_{\text{int}}(t)}\Big]|0}'}\,,
\end{equation}
where we assume that the operators are inserted at negative conformal time (the standard expanding de Sitter patch). The denominator removes the vacuum-to-vacuum bubbles such that $\braket{\mathbbm{1}}_{\text{in-out}}=1$.

Now, there is a single time ordering that goes from $-\infty<\eta<\infty$, which includes the standard expanding Poincaré patch ($-\infty<\eta<0$) and an extra contracting Poincaré patch ($0<\eta<\infty$). Furthermore, we note that the $i\varepsilon$ rotation of the contour turns off the interactions adiabatically at the
past and future null boundaries (notice the different signs of the $i\varepsilon$'s). 

The claim now is that, for all IR-finite interactions, for which the time
integral converges around $\eta=0$, we have 
\begin{align}
\eqref{eq:corr2}=B_{\text{in-in}}^{(n)}=B_{\text{in-out}}^{(n)}=\eqref{eq:corr3}    
\end{align}
This is a known fact in Minkowski (see e.g. \cite{Kamenev:2011aa}). Here we claim that it applies to de Sitter too (and probably to any accelerated FLRW but we have not checked yet). We next provide a formal argument and some explicit checks.

\subsubsection*{A formal argument} 
A formal argument relies on the observation that infinite time evolution changes the ground (Bunch--Davies) state only by a phase
\begin{equation}
    U(+\infty,-\infty)\ket{0}=\ket{0}\braket{0|U(+\infty,-\infty)|0}\,.
\end{equation}
Consequently, an infinite amount of time evolution from minus infinity to plus infinity does not result in a state excited with many particles, but rather returns us to the Bunch--Davies vacuum. This can be explicitly verified by projecting the left-hand side onto any
$n$-point excited state and observing that the projection is zero for physical momenta.

Thus, our claim can be verified at all orders in perturbation theory by projecting onto any excited state. The result is a derivative of a delta function that enforces energy conservation, and which has zero support on physical perturbations. Then
\begin{equation}
    \begin{split}
B_{\text{in-out}}&=\frac{\vac{ T \left[ \prod^{n}_{a=1}\phi(t_{a})\,\e^{-i\int_{-\infty_-}^{+\infty_-} H_{\text{int}} \d t } \right]}'}{\vac{ T\left[ \e^{-i\int_{-\infty_-}^{+\infty_-} H_{\text{int}} \d t } \right]}' } \\
&= \vac{ U^{\dagger}(+\infty_+,-\infty_+) T\left[ \prod^{n}_{a=1}\phi(t_{a})\, \e^{-i\int_{-\infty_-}^{+\infty_-} H_{\text{int}} \d t } \right]}'\\
&=\vac{ \overline{\mathcal{T}} \left[ \e^{ +i\int_{-\infty_+}^{+\infty_+} H_{\text{int}} \d t } \right] \mathcal{T}\left[ \e^{-i\int^{+\infty_-}_{t_{0}} H_{\text{int}} \d t } \right] \mathcal{T}\left[\prod^{n}_{a=1}\phi(t_{a})\, \e^{-i\int_{-\infty_-}^{t_{0}} H_{\text{int}} \d t } \right]}'\\
&= \vac{ \overline{\mathcal{T}} \left[ \e^{+i\int^{t_{0}}_{-\infty_+} H_{\text{int}} \d t } \right] U^{\dagger}(+\infty,t_{0})U(+\infty,t_{0}) \mathcal{T}\left[\prod^{n}_{a=1}\phi(t_{a}) \,\e^{-i\int^{t_{0}}_{-\infty_-} H_{\text{int}} \d t } \right]}'\\
&=B_{\text{in-in}}\,,
    \end{split}
\end{equation}
where $\infty_\pm=\infty\pm i\varepsilon$. In terms of path integral contours, this is simply illustrated as follows:
    \begin{equation}
        \adjustbox{valign=c}{
        \tikzset{every picture/.style={line width=0.75pt}} 
\begin{tikzpicture}[x=0.75pt,y=0.75pt,yscale=-1,xscale=1]
\begin{scope}[xshift=-0.9cm,yshift=-0.14cm]
\draw  [draw opacity=0] (78.25,42) .. controls (79.91,42) and (81.25,43.34) .. (81.25,45) .. controls (81.25,46.66) and (79.91,48) .. (78.25,48) -- (78.25,45) -- cycle ; \draw   (78.25,42) .. controls (79.91,42) and (81.25,43.34) .. (81.25,45) .. controls (81.25,46.66) and (79.91,48) .. (78.25,48) ;  
\draw[mar]    (15.5,42) -- (78.25,42) node[left,pos=0]{in} node[right,pos=1]{out};
\draw[ma]    (15.5,48) -- (78.25,48) ;
\end{scope}
\begin{scope}[xshift=0cm,yshift=-0.14cm]
\draw  [draw opacity=0] (172.25,46) .. controls (172.25,46) and (172.25,46) .. (172.25,46) .. controls (172.25,46) and (172.25,46) .. (172.25,46) .. controls (173.91,46) and (175.25,47.34) .. (175.25,49) .. controls (175.25,50.66) and (173.91,52) .. (172.25,52) -- (172.25,49) -- cycle ; \draw   (172.25,46) .. controls (172.25,46) and (172.25,46) .. (172.25,46) .. controls (172.25,46) and (172.25,46) .. (172.25,46) .. controls (173.91,46) and (175.25,47.34) .. (175.25,49) .. controls (175.25,50.66) and (173.91,52) .. (172.25,52) ;  
\draw[mar]    (109.5,46) -- (172.25,46) ;
\draw[ma]   (109.5,52) -- (172.25,52) ;
\draw[ma]    (110,41) -- (237,41) node[pos=0.25,above,color=Maroon]{cancels};
\draw  [draw opacity=0][fill=Maroon  ,fill opacity=0.19 ] (107.5,40.4) .. controls (107.5,39.35)  and (108.35,38.5) .. (109.4,38.5) -- (175.6,38.5) .. controls (176.65,38.5) and (177.5,39.35) .. (177.5,40.4) -- (177.5,46.1) .. controls (177.5,47.15) and (176.65,48) .. (175.6,48) -- (109.4,48) .. controls (108.35,48) and (107.5,47.15) .. (107.5,46.1) -- cycle;
\end{scope}
\draw[ma]    (168.5,71) -- (178,64);
\draw[ma]    (110,71) -- (168.5,71) ;
\draw[ma]    (178,64) -- (237.5,64);
\draw (81,36.4) node [anchor=north west][inner sep=0.75pt]    {$=$};
\draw (81,60.4) node [anchor=north west][inner sep=0.75pt]    {$=$};
\end{tikzpicture}
    }
    \end{equation}
Here, starting with the in-in contour, we consider adding an infinitely long contour for free, because acting on the vacuum amounts to just a phase. Then, the red-shaded parts of the contour cancel each other out, and we are left with the in-out contour, which gives the same result as the in-in one, up to a phase.

So, clearly, the formal argument holds, but now we would like to see how it works in practice.

\subsubsection*{Explicit checks} 
There are many checks that we can perform. Here, we only go over a few of them (for more, see \cite{Donath:2024utn}) just to see how it works in perturbation theory. 

We examine a basic theory involving a scalar field with a cubic interaction
\begin{equation}
     H_{\text{int}}(\eta) =\int_{\vec{x}} \frac{\lambda}{(n!)} \,F(\eta \partial_i,\eta \partial_\eta) \phi^{n}(\eta)\,,
\end{equation}
where $F$ represents a generic set of space and time derivatives acting on any of the fields, and $\phi$ can be any set of fields of any mass and spin. To simplify our discussion, we will focus on a single massive scalar. We will compute the $n$-point function to $\mathcal{O}(\lambda)$, with fields inserted at times $\eta_a \leq 0$ for $a=1,\dots,n$. In the context of the in-in formalism, and for simplicity setting $H=1$, we find
\begin{equation}
\begin{split}
    B_{\text{in-in}}&=B_{\text{in-in}}^{r}+B_{\text{in-in}}^{\ell}\\
&=-i\lambda \int_{-\infty(1-i\varepsilon)}^{0} \frac{\d\eta}{\eta^{4}}\, F \,\prod_{a=1}^{n} G_{F}(\eta,\eta_{a};k_{a})+i\lambda \int_{-\infty(1+i\varepsilon)}^{0} \frac{\d\eta}{\eta^{4}} \,F \,\prod_{a=1}^{n} G^{+}(\eta,\eta_{a};k_{a})^*\,.
\end{split}
\end{equation}

As long as the total number of $\eta$ factors from $F$ and the propagators exceeds four, the result converges at $\eta=0$. This condition is met in scenarios like a local interaction with over three conformally coupled scalars, or interactions among massless scalars where $2n_{\partial_\eta} + n_{\partial_i} \geq 4$. In the in-out formalism, we find
\begin{align}\label{dsio}
B_{\text{in-out}}&=-i\lambda \int_{-\infty(1-i\varepsilon)}^{+\infty(1-i\varepsilon)} \frac{\d\eta}{\eta^{4}}\, F\, \prod_{a=1}^{n} G_{F}(\eta,\eta_{a};k_{a})\,.
\end{align}
where $G_F$ is the time-ordered propagator (the curved spacetime analog of the Feynman propagator). To verify that this is equivalent to the in-in expression, we compute the difference. The portion of the in-out time integral from $-\infty$ to $0$ precisely cancels out the right contribution of $B_{\text{in-in}}$, leaving us with
\begin{equation}
    \begin{split}
       B_{\text{in-out}}-B_{\text{in-in}}&=-i\lambda \int_{0}^{+\infty_-} \!\frac{\d\eta}{\eta^4}  F\left[ \prod_{a=1}^{n} f_a(\eta) f_a^{*}(\eta_{a}) \! \right]-i\lambda \int_{-\infty_+}^{0} \!\frac{\d\eta}{\eta^4}  F \left[ \prod_{a=1}^{n} f_a(\eta) f_a^{*}(\eta_{a}) \! \right] \\
&=-i\lambda \prod_{a=1}^{n} f_a^{*}(\eta_{a}) \int_{-\infty (1+i\varepsilon)}^{+\infty(1-i\varepsilon)} \frac{\d\eta}{\eta^4} \, F f_a(\eta)\,.\label{argument}
    \end{split}
\end{equation}
Here, the label $a$ on the mode function corresponds to the various momenta $\vec{k}_a$; it can also denote different fields, each with distinct mass values. 

Because of the $i\epsilon$ deformations, the difference in \eqref{argument} thus becomes an integral that can be closed in the lower half of the complex plane, where the integrand is analytic. According to Cauchy's theorem, this integral is zero, and we reach the desired conclusion. Higher-order diagrams are more complicated, but follow similar logic.

\subsection{Applications}
Next we discuss why this mathematical observation between in-in and in-out is \emph{useful} in cosmology by considering a few concrete applications. These results are based on \cite{Donath:2024utn}.

\subsubsection{Pole bagging}
As previously mentioned, one reason why the correspondence between in-in and in-out is useful is because the in-out formalism involves the computation of fewer diagrams. Another example in which it is manifestly useful is the idea of ``pole bagging.''  Indeed, a major simplification of the in-out formalism is that there is a single, time-ordered propagator, just like for amplitudes. 

In Minkowski this is especially familiar in energy-momentum domain
$(E,\vec{p})=p^\mu$, where $G_F=1/(p^2+m^2-i\varepsilon)$ (recall that we use the mostly plus signature) in which the time ordering is conveniently built in by the $i\varepsilon$ prescription. Thus, it is clear that $B_{\text{in-out}}$ are just the time-ordered correlators appearing in the LSZ reduction formulae for scattering amplitudes, i.e., $B_{\text{in-out}}$ are simply unamputated Feynman diagrams. 

In cosmology, we tend to work in time-momentum domain $(t,\vec{k})$. The reason we do not Fourier transform time is that the background's expansion breaks time translation invariance and so different frequencies couple to each other already at linear order and energy/frequency is not conserved. Therefore, in
Minkowski, this is simply a sum of poles, i.e., in-out correlators in Minkowski are computed by repeatedly using
\begin{equation}\label{eq:residue0}
    \int_{-\infty}^\infty \frac{\d \omega_i}{2\pi} \frac{1}{(\omega_i^2-\Omega_i^2)((\omega_i+\omega_X)^2-\Omega_j^2)}=-i\frac{\Omega_{ij}}{2\Omega_i \Omega_j(\omega_X^2-\Omega_{ij}^2)}\,,
\end{equation}
where $\Omega_{ij}=\Omega_i + \Omega_j$ and $\Omega_{a}^{2}\equiv |\vec{k}_{a}|^{2}+M_{a}^{2}$.
\begin{mdexample} \textbf{(Pole bagging in Minkowski)}
    For example, we consider the simplest case of a cubic polynomial interaction $\lambda \phi^3/3!$, where $F=\lambda$ and $n=3$. The Fourier transform of \textit{equal-time} correlators then yields
\begin{equation}
    \begin{split}
        B_3^{\text{flat}}&=\int_{-\infty}^{\infty} \frac{\d\omega_{1}\d\omega_{2}\d\omega_{3} }{(2\pi)^{3}} (2\pi)\delta\left(\sum_{a=1}^3 \omega_{a} \right) \e^{it \sum_{a=1}^{3} \omega_{a} } \lambda \prod_{b=1}^3 \frac{1}{p_{b}^{2}+i\varepsilon}\\
&= \int_{-\infty}^{\infty} \frac{\d\omega_{1}\d\omega_{2}}{(2\pi)^2} \frac{\lambda}{(\omega_{1}^{2}-\Omega_{1}^{2}+i\varepsilon)(\omega_{2}^{2}-\Omega_{2}^{2}+i\varepsilon)((\omega_{1}+\omega_{2})^{2}-\Omega_{3}^{2}+i\varepsilon)}\,.
    \end{split}
\end{equation}
Here, the first line is just the product of three Feynman propagators in the frequency domain, and we recall that the prime on the correlator \eqref{eq:corr3} indicates that we have omitted $(2\pi)^3\delta^{(3)}\left( \sum \vec{k}_a \right)$. 

The integrals can be evaluated using the residue theorem (see \eqref{eq:residue0}), closing the contour in the upper half-plane. The first integral has poles at $\omega_{1}=-\Omega_{1}$ and $\omega_{1}=-\omega_{2}-\Omega_{3}$. This results in
\begin{equation}
     B_3^{\text{flat}}=-i \lambda\int \frac{\d\omega_{2}}{2\pi} \frac{\Omega_{13}}{2\Omega_{1}\Omega_{3}(\omega_{2}-\Omega_{2} )(\omega_{2}+\Omega_{2} ) (\omega_{2}+\Omega_{13})(\omega_{2}-\Omega_{13})}\,,
\end{equation}
where we have left the $i\varepsilon$'s implicit. It should be noted that there is a cancellation between two residues, which eliminates two of the zeros in the denominator (namely, $\omega_{2} = \pm(\Omega_{1} - \Omega_{3})$), retaining only the zeros at $\omega_{2} = \pm \Omega_{2}$ and $\omega_{2} = \pm(\Omega_{1} + \Omega_{3})$. The $i\varepsilon$ prescription directs us to consider only the two poles on the negative real axis, $-\Omega_{2}$ and $-\Omega_{13}$ (a strategy to maintain clarity in this calculation is to omit the $i\varepsilon$s in the integrand and introduce a slight counterclockwise rotation of the integration contours, ensuring that only the residues from poles on the negative real axis are picked up), which yields
\begin{equation}
    B_3^{\text{flat}} = - \frac{\lambda}{4\Omega_{1}\Omega_{2}\Omega_{3}(\Omega_{1}+\Omega_{2}+\Omega_{3})}\,.
\end{equation}
This yields the anticipated results from the bulk time integral, characterized by the simple $E_{T}$ pole and the appropriate normalization factor for each $1/\Omega_{a}$. This can of course be done for any $n$-point function in Minkowski.
\end{mdexample}
Below we illustrate how something very similar also holds in a de Sitter cosmology.
\begin{mdexample} \textbf{(Pole bagging in de Sitter)}
A similar approach works also in de Sitter for massless and conformally-coupled (c.c.) scalars using the above ``dispersive'' representation of the propagator. For example, we note that the Feynman (time-ordered) propagator of a massless or conformally coupled scalar in de Sitter can be written as 
\begin{equation}
    G_F^{\text{de Sitter}}(\eta,\eta', k) = \left(\frac{1}{2 \pi i}\right)\int_{-\infty}^{\infty} \d\omega \frac{2 \omega f_{\omega}(\eta)f^*_{\omega}(\eta')}{\omega^2 -k^2+i \varepsilon}\,,
\end{equation}
 Hence, a similar computation as in Minkowski yields
\begin{equation}
    \begin{split}
        B_5^{\text{c.c.}}&= -i H^6\left(\frac{ \eta_0}{2 \pi i}\right)^5 \int_{-\infty}^{\infty}\left(\prod_{i=1}^5 \frac{\d\omega_i}{\omega_i^2 -k^2+i \varepsilon}\right)\e^{i \omega_T\eta_0}\int_{-\infty}^{\infty}\d\eta\, \eta\, \e^{-i \omega_T\eta} 
        \\&
        = 2\pi H^6 \left(\frac{ \eta_0}{2 \pi i}\right)^5 \int_{-\infty}^{\infty}\left(\prod_{i=1}^5 \frac{\d\omega_i}{\omega_i^2 -k_i^2+i \varepsilon}\right)\e^{i \omega_T\eta_0}\delta'(\omega_T) 
        \\&
        = -2\pi i H^6 \left(\frac{ \eta_0}{2 \pi i}\right)^5 \int_{-\infty}^{\infty} \frac{\d\omega_1\d\omega_2\d\omega_3\d\omega_4(\eta_0((\omega_T-\omega_5)^2 -k_5^2))-2i(\omega_T-\omega_5))}{\left(\prod_{i=1}^4\left(\omega_i^2 -k_i^2+i \varepsilon\right)\right)\left((\omega_T-\omega_5)^2 -k_5^2+i \varepsilon\right)^2}
        \\&
        = \frac{ (H \eta_0)^6}{(k_1+k_2+k_3+k_4+k_5)(16 k_1 k_2 k_3 k_4 k_5)}\,.
    \end{split}
\end{equation}
(Note that now, there is no delta function in the $\omega$'s because energy is not conserved in de Sitter.) One should be able to extend this to all masses in de Sitter, perhaps using embedding space coordinates, for which the propagators are simpler.
\end{mdexample}
Let us now move on to a second application.

\subsubsection{Recursion relations}
So far, the recursion relations we discuss below are found to work only in Minkowski. The key observation is that pole bagging has many similarities with integrations by part, and we
can exploit this as follows. We first introduce \textit{chains} to simplify correlators with many external lines, that is, we (schematically) write
\begin{equation}\label{eq:chains0}
    \text{correlator}=\frac{\prod_{i=1}^m 2x_i}{\prod_{i=1}^n 2E_i}\times \text{chains}\,.
\end{equation}
\begin{mdexample}
What we mean by \eqref{eq:chains0} is, for example,
\begin{equation}
    \prod_{i=1}^5 2E_i~\adjustbox{valign=c}{\begin{tikzpicture}[baseline={([yshift=2ex]current bounding box.center)},scale=1.0001]
    \coordinate (a) at (0,0);
    \coordinate (b) at (0.5,0);
    \coordinate (bt) at (1.5,0);
    \coordinate (c) at (1.5,0);
    \coordinate (d) at (2.5,0);
    \coordinate (e) at (3,0);
    \coordinate (la) at (0.5,-1);
    \coordinate (lat) at (1.5,-1);
    \coordinate (lb) at (1.5,-1);
    \coordinate (lc) at (2.5,-1);
    \draw[thick] (a) -- (e) (a) --++ (180:0.2) (e) --++ (0:0.2);
    \draw[thick, color=RoyalBlue] (a) -- (la) node[left,midway]{$1$}; 
    \draw[thick, color=RoyalBlue] (b) -- (la) node[right,midway]{$2$};
    \draw[thick, color=RoyalBlue] (bt) -- (lat) node[right,midway]{$3$};
    \draw[thick, color=RoyalBlue] (d) -- (lc) node[left,midway]{$4$};
    \draw[thick, color=RoyalBlue] (e) -- (lc) node[right,midway]{$5$}; 
    \draw[thick, color=Maroon] (la) -- (lc) node[below,midway,xshift=-0.5cm]{$12$} node[below,midway,xshift=0.5cm]{$45$};
    \draw[thick,fill=black] (a) circle (1pt)  (b) circle (1pt)  (d) circle (1pt)  (e) circle (1pt);
    \draw[thick,fill=black] (la) circle (1pt) (lc) circle (1pt) (lat) circle (1pt) (bt) circle (1pt);
    \end{tikzpicture}}=\prod_{i=1}^32 x_i~\adjustbox{valign=c}{\begin{tikzpicture}[baseline={([yshift=2ex]current bounding box.center)},scale=1.0001]
    \coordinate (a) at (0,0);
    \coordinate (b) at (0.5,0);
    \coordinate (bt) at (1.5,0);
    \coordinate (c) at (1.5,0);
    \coordinate (d) at (2.5,0);
    \coordinate (e) at (3,0);
    \coordinate (la) at (0.5,-1);
    \coordinate (lat) at (1.5,-1);
    \coordinate (lb) at (1.5,-1);
    \coordinate (lc) at (2.5,-1); 
    \draw[thick, color=Maroon] (la) -- (lc) node[below,midway,xshift=-0.5cm]{$12$} node[below,midway,xshift=0.5cm]{$45$};
    \draw[thick,fill=black] (la) circle (1pt) node[above]{$x_1$} (lc) circle (1pt) node[above]{$x_3$} (lat) circle (1pt) node[above]{$x_2$};
    \end{tikzpicture}}
\end{equation}
\end{mdexample}
Here, the chains are computed by nested time integrals
\begin{equation}
    \begin{split}
        \label{eq:TreeInt}
 \adjustbox{valign=c}{
 \begin{tikzpicture}[ball/.style = {circle, draw, align=center, anchor=north, inner sep=0}]
 \node[ball,text width=.18cm,fill,color=black, label=left:$\mbox{\scriptsize $x_1$}$] at (0,0) (x1) {};
 \node[ball,text width=1cm,right=2.1cm of x1.east,fill=gray!15] (S1) {$\mathcal{B}$};
 \node[ball,text width=.18cm,fill,color=black,right=2cm of x1.east, label=left:$\mbox{\scriptsize $\hspace{-.2cm}x_2$}$] (x2) {};
\draw[thick] (x1) to[bend right=75] node[below] {\scriptsize $y_1$} (x2);
 \draw[thick] (x1) to[bend right=30] node[below,yshift=0.05cm] {\scriptsize $y_2$} (x2);
 \draw[thick] (x1) to[bend left=75] node[above] {\scriptsize $y_{n-1}$} (x2);
 \draw[thick] (x1) to[bend left=30] node[above,yshift=-0.1cm] {\scriptsize $y_n$} (x2);
 \node[rotate=90] at (1.1,-0.1) {\scriptsize ...};
 \end{tikzpicture}}
&\equiv\int_{-\infty}^\infty \hspace{-0.25em}\prod_{v\in\mathcal{B}\setminus\{2\}} \hspace{-0.6em} i\, \d t_v \,G_F(t_v,t_0,x_t) \\&  \times \int_{-\infty}^\infty \hspace{-0.7em}i\, \d t_2 G_{F}(t_2,t_0,x_2)
 \prod_{e\in\mathcal{B}}G_{F}(t_{v_e},t_{v'_e},y_{v_e})I_n(\{y_i\},t_2),
    \end{split}
\end{equation}
where $\mathcal{B}$ could be anything and with
\begin{equation}
    I_n(\{y_i\},t_2)\equiv\int_{-\infty}^{\infty}i\, \d t_1\, G_F(t_1,t_0,x_1)\prod_{i=1}^nG_F (t_1,t_2,y_{i})=\frac{2 y_T}{\prod_{i=1}^n 2 y_i}I_1(y_T,t_2)\,.
\end{equation}
The $t_1$ integral can be done and re-written by explicitly performing one time integral we obtain the recursion relation as
\begin{equation}
     \adjustbox{valign=c}{
 \begin{tikzpicture}[ball/.style = {circle, draw, align=center, anchor=north, inner sep=0}]
 \node[ball,text width=.18cm,fill,color=black, label=left:$\mbox{\scriptsize $x_1$}$] at (0,0) (x1) {};
 \node[ball,text width=1cm,right=2.1cm of x1.east,fill=gray!15] (S1) {$\mathcal{B}$};
 \node[ball,text width=.18cm,fill,color=black,right=2cm of x1.east, label=left:$\mbox{\scriptsize $\hspace{-.2cm}x_2$}$] (x2) {};
\draw[thick] (x1) to[bend right=75] node[below] {\scriptsize $y_1$} (x2);
 \draw[thick] (x1) to[bend right=30] node[below,yshift=0.05cm] {\scriptsize $y_2$} (x2);
 \draw[thick] (x1) to[bend left=75] node[above] {\scriptsize $y_{n-1}$} (x2);
 \draw[thick] (x1) to[bend left=30] node[above,yshift=-0.1cm] {\scriptsize $y_n$} (x2);
 \node[rotate=90] at (1.1,-0.1) {\scriptsize ...};
 \end{tikzpicture}}=\frac{2 y_T}{\prod_{i=1}^n 2 y_i}\frac{1}{x_1^2-y_{T}^2}\left[\frac{x_1+x_2}{ 2 x_1 x_2} 
 \adjustbox{valign=c}{
 \begin{tikzpicture}[ball/.style = {circle, draw, align=center, anchor=north, inner sep=0}]
 \node[ball,text width=1cm,right=2.1cm of x1.east,fill=gray!15] (S1) {$\mathcal{B}$};
 \node[ball,text width=.18cm,fill,color=black,right=2cm of x1.east, label={below,yshift=-0.3cm,xshift=0.6cm}:$\mbox{\scriptsize $\hspace{-.2cm}(x_1+x_2)$}$] (x2) {};
 \end{tikzpicture}}
 -\frac{y_{T}+x_2}{ 2 y_{T} x_2}
  \adjustbox{valign=c}{
 \begin{tikzpicture}[ball/.style = {circle, draw, align=center, anchor=north, inner sep=0}]
 \node[ball,text width=1cm,right=2.1cm of x1.east,fill=gray!15] (S1) {$\mathcal{B}$};
 \node[ball,text width=.18cm,fill,color=black,right=2cm of x1.east, label={below,yshift=-0.3cm,xshift=0.6cm}:$\mbox{\scriptsize $\hspace{-.2cm}(y_T+x_2)$}$] (x2) {};
 \end{tikzpicture}}
 \right]
\end{equation}
The last equation tells us how diagrams with $V$ vertices are related to those with $V-1$ vertices in a simple (rational) way. In practice, this tells us that what we need to compute at the end of the day are one-vertex diagrams. For a straightforward application, consider a one-loop exchange diagram.
\begin{mdexample}
    We have
    \begin{equation}\label{eq:2s1l}
    \begin{split}
         \adjustbox{valign=c,scale=0.75}{\begin{tikzpicture}[ball/.style = {circle, draw, align=center, anchor=north, inner sep=0}]
 \node[ball,text width=.18cm,fill,color=black] at (0,0) (x1) {};
 \node[left=.0cm of x1.west,label=below:$x_1$] (x1l) {};
 \node[ball,text width=.18cm,fill,color=black,right=1.5cm of x1.east] (x2) {};
 \node[right=.0cm of x2.east,label=below:$x_2$] (x2l) {};
 \draw[thick] (.85,0) circle (.85);
 \node at (.85,0.5) {$\displaystyle y_a$};
 \node at (.85,-0.5) {$\displaystyle y_b$}; 
 \end{tikzpicture}} &=\frac{y_a{+}y_b}{ 2 y_a y_b}\frac{1}{x_1^2{-}(y_a{+}y_b)^2}\Bigg[\frac{x_1{+}x_2}{ 2 x_1 x_2} \frac{-1}{(x_1{+}x_2)^2}{-}\frac{y_a{+}y_b{+}x_2}{ 2(y_a{+}y_b) x_2}\frac{-1}{(y_a{+}y_b{+}x_2)^2} \Bigg]\\&
 =\frac{x_1+x_2+y_a+y_b}{4 x_1 x_2 y_a y_b (x_1+x_2)(x_1+y_a+y_b)(x_2+y_a+y_b)}\,.
    \end{split}
\end{equation}
\end{mdexample}
Let us mention that the recursions relations for correlators given above are
very similar to the wavefunction recursion relations of \cite{Arkani-Hamed:2017fdk}. For the wavefunction this relation becomes quite complicated at loop level, while it remains simple at the correlator level.

We now move on to the last application that we want to discuss.


\subsubsection{Correlator cutting rules}
We understand that unitarity for amplitudes is embodied in the optical theorem, which provides an explicit relationship between the imaginary part of the amplitude and the total cross-section in the forward limit. For correlators, a similar argument was not known until recently \cite{Donath:2024utn, Ema:2024hkj}. Here, we aim to demonstrate how representing the correlator as an in-out object enables us to apply the same derivation of the Cutkosky cutting rules to correlators, just as with scattering amplitudes. Along the way, we will emphasize all the small technical modifications required.

Our starting point is the field identity
\begin{align}\label{CFTopt}
\sum_{r=0}^n (-1)^r \sum_{\sigma\in\Pi(r,n-r)} \overline{\mathcal{T}} \left[ \mathscr{O}_{\sigma(1)}...\mathscr{O}_{\sigma(r)} \right] \mathcal{T}\left[ (\mathscr{O}_{\sigma(r+1)}...\mathscr{O}_{\sigma(n)}\right]=0\,.
\end{align}
which may be recognized as Veltman's largest time equation \cite{Veltman:1994wz} and it has already been discussed many times so far (see the contributions~\cite{chapterHaehlRangamani} and \cite{chapterCaronHuotGiroux}). With $\Pi(r,n-r)$, we refer to the set of partitions of $\{1, \ldots, n\}$ into two subsets of sizes $r$ and $n-r$. Thus, the sum involves $2^n$ terms. The fields $\mathscr{O}_i\equiv\mathscr{O}_{i}(t_{i})$ represent arbitrary products of operators at the same time point. Our primary focus will be on scenarios where these operators are monomials in the fields of the theory and their derivatives. 

As we shall see, \eqref{CFTopt} leads to infinitely many propagator identities, which in turn
become correlator cutting rules. (These appear to be equivalent to the wavefunction cutting rules of \cite{Arkani-Hamed:2017fdk}, the advantage is that one works directly with observables, i.e.
correlators, rather than with the more primitive wavefunction; perhaps a more explicit relation can be established along the lines of \cite{Stefanyszyn:2024msm}) These identities are more easily derived when it is additionally assumed that all the operators in~\eqref{CFTopt} are Hermitian, and the resulting equation is real, which may not be obvious from its original form. However, we can combine terms pairwise to rewrite it as
\begin{align}\label{even}
 \sum_{r=0}^{n/2-1}& \sum_{\sigma\in\Pi(r,n-r)} (-1)^r 2\Re \vac{ \overline{\mathcal{T}} \left[ \prod_{a=1}^r\mathscr{O}_{\sigma(a)} \right] T\left[ \prod_{b=r+1}^n \mathscr{O}_{\sigma(b)} \right] }\\ &+\sum_{\sigma\in\Pi(n/2,n/2)} (-1)^{n/2} \Re \vac{ \overline{\mathcal{T}} \left[ \prod_{a=1}^{n/2} \mathscr{O}_{\sigma(a)} \right] T\left[ \prod_{b=n/2+1}^n \mathscr{O}_{\sigma(b)} \right]} =0 \,, \nonumber
\end{align}
for $n$ even and as
\begin{align}\label{odd}
 \sum_{r=0}^{(n-1)/2} \sum_{\sigma\in\Pi(r,n-r)} (-1)^r \Im \vac{ \overline{\mathcal{T}}\left[ \prod_{a=1}^r \mathscr{O}_{\sigma(a)} \right] T\left[ \prod_{b=r+1}^n \mathscr{O}_{\sigma(b)} \right]}=0\,,
\end{align}
for $n$ odd.

Next, we use~\eqref{CFTopt} to derive propagator identities. The initial step involves expanding the time-evolution operator within an in-out correlator up to a certain order in perturbation theory. We then consider the various powers of $\phi$ and $H_{\text{int}}$ as the distinct operators featured in~\eqref{CFTopt}. This approach yields a set of identities. To illustrate this process, we first consider diagrams containing a single vertex and then those with two vertices.
\begin{mdexample}
    \textbf{(Propagator identities: contact diagrams)} A diagram with one vertex contains a single power of $H_{\text{int}}$ (i.e., there is only one interaction) and $n$ instances of the field $\phi$. As a simple example, consider the following choice
    \begin{equation}\label{eq:expOp1}
        \mathscr{O}_1=\phi(\vec{x}_1,t_0)^m\,,\quad \mathscr{O}_2=\phi(\vec{x}_2,t_0)^{n-m}\,, \quad \mathscr{O}_3=H_{\text{int}}(t)\,.
    \end{equation}
    Here, $t_0$ represents an arbitrary time. Since our analysis is confined to equal-time correlators, we will disregard the time dependence from now on. Assuming the operators in \eqref{eq:expOp1} are Hermitian, we can insert them into~\eqref{odd} to obtain the following position space expression
    \begin{equation}\label{eq:Im0}
         \Im \left\{ \textcolor{Maroon}{\langle \mathcal{T} [\phi^n H_{\text{int}}]\rangle}-\textcolor{RoyalBlue}{\langle \phi^m \mathcal{T}[\phi^{n-m} H_{\text{int}}]\rangle} -\textcolor{ForestGreen}{\langle \phi^{n-m} \mathcal{T}[\phi^{m} H_{\text{int}}]\rangle} \right\} \simeq 0\,.
    \end{equation}
    Here, we have omitted the time-ordering or anti-time-ordering symbols when only a single time is involved. We used $\simeq 0$ to indicate that this identity is true only after performing time integration, due to the omission of the term $H_{\text{int}}\mathcal{T}[\phi^n]$. This simplification is allowed because any term where a Hamiltonian interaction does not share the same time ordering as a field will integrate to zero. It would be nice to understand this in simple physical terms. 

    Now, each objects in \eqref{eq:Im0} can be rewritten as a correlator and the result can be manipulated as follows after changing integration variables and using the properties of the Feynman propagator
    \begin{equation}
    \begin{split}
           &\textcolor{Maroon}{B^{\text{c}}_n(\{E_i\}_{i=1}^n)}-\textcolor{RoyalBlue}{\frac{1}{2}\left[B^{\text{c}}_n(\{E_i\}_{i=1}^n)+(-1)^mB^{\text{c}}_n(\{-E_i\}_{i=1}^m,\{E_i\}_{i=m+1}^n))\right]}\\&\qquad-\textcolor{ForestGreen}{\frac{1}{2}\left[B^{\text{c}}_n(\{E_i\}_{i=1}^n)+(-1)^{n-m}B^{\text{c}}_n(\{E_i\}_{i=1}^m,\{-E_i\}_{i=m+1}^n))\right]}=0\,.
    \end{split}
    \end{equation}
    Terms involving two sets have been analytically continued from respective positive to negative energies. Moreover, we observe that $B^{\text{c}}_n({E_i}_{i=1}^n)$ cancels out and if we subsequently invert the first $m$ energies, we obtain
    \begin{equation}\label{notuni}
        B^{\text{c}}_n(\{E_i\}_{i=1}^n)+(-1)^n B^{\text{c}}_n(\{-E_i\}_{i=1}^n)=0\,.
    \end{equation}
    This is a necessary condition for this correlator to come from a unitary theory. It provides a useful constraint from unitarity, which operates directly at the level of observables. This is valuable from the standpoint of the cosmological bootstrap, which attempts to define good correlators based on their general properties rather than computing them from an explicit model.
    
    There is a handy graphical notation for the cut identities correlators have to obey if the theory is unitary, e.g., 
    \begin{equation}
        \label{rules fin c}
\adjustbox{valign=c,scale=0.7}{\begin{tikzpicture}[ball/.style = {circle, draw, align=center, anchor=north, inner sep=0}]
 \draw (-1.2,0) -- (3.4,0);
 \foreach \x/\label in {-0.6/\hspace{1em}E_1\,\,\ldots,-0.3/ , 0.6/\hspace{-1em}E_m, 1.6/\hspace{1.5em}E_{m+1}\,\ldots, 2.5/,2.8/\hspace{1em}E_n} {
 \draw (\x,0) -- (1.1,-1.7);
 \filldraw (\x,0) circle (2pt) node[above] {\footnotesize $\label$};}
 \filldraw (1.1,-1.7) circle (2pt) ;
 \foreach \x in {0.5, 0.62,0.8} {
 \draw (1.1,-1.7-\x) circle (\x);}
 \draw[Orange, dashed, line width = 1.3] (1.2,0.7) -- (-0.1,-1.7);
 \end{tikzpicture}}= \frac{1}{2}\left[B^{\text{c}}_n(\{E_i\}_{i=1}^n)+(-1)^mB^{\text{c}}_n(\{-E_i\}_{i=1}^m,\{E_i\}_{i=m+1}^n))\right].
    \end{equation}
The orange dashed line represents the energies that need to be analytically continued. The contact identity \eqref{eq:Im0} thus becomes 
    \begin{equation}
        \adjustbox{valign=c,scale=0.85}{\begin{tikzpicture}[scale=0.8,ball/.style = {circle, draw, align=center, anchor=north, inner sep=0}]
 \draw (-0.2,0) -- (2.4,0);
 \foreach \x/\label in {0.1/E_1,0.4/ , 0.7/\hspace{1.7em}\ldots,1.3/,
 1.7/,2.1/\hspace{1em}E_n} {
 \draw (\x,0) -- (1.1,-1.7);
 \filldraw (\x,0) circle (2pt) node[above] {\footnotesize $\label$};
 }
 \filldraw (1.1,-1.7) circle (2pt) ;
 
 \foreach \x in {0.5, 0.62,0.8} {
 \draw (1.1,-1.7-\x) circle (\x);
 }
 \end{tikzpicture}} -
\adjustbox{valign=c,scale=0.85}{\begin{tikzpicture}[scale=0.8,ball/.style = {circle, draw, align=center, anchor=north, inner sep=0}]
 \draw (-1.2,0) -- (3.4,0);
 \foreach \x/\label in {-0.6/\hspace{1em}E_1\ldots,-0.3/ , 0.6/\hspace{-0.5em}E_m, 1.6/\hspace{1.5em}E_{m+1}\,\ldots, 2.5/,2.8/\hspace{1em}E_n} {
 \draw (\x,0) -- (1.1,-1.7);
 \filldraw (\x,0) circle (2pt) node[above] {\footnotesize $\label$};
 }
 \filldraw (1.1,-1.7) circle (2pt) ;
 
 \foreach \x in {0.5, 0.62,0.8} {
 \draw (1.1,-1.7-\x) circle (\x);
 }
 \draw[Orange, dashed, line width = 1.3] (1.2,0.7) -- (-0.1,-1.7);
 
 \end{tikzpicture}}-
 \adjustbox{valign=c,scale=0.85}{\begin{tikzpicture}[scale=0.8,ball/.style = {circle, draw, align=center, anchor=north, inner sep=0}]
 \draw (-1.2,0) -- (3.4,0);
 \foreach \x/\label in {-0.6/\hspace{1em}E_1\ldots,-0.3/ , 0.6/\hspace{-0.5em}E_m, 1.6/\hspace{1.5em}E_{m+1}\,\ldots, 2.5/,2.8/\hspace{1em}E_n} {
 \draw (\x,0) -- (1.1,-1.7);
 \filldraw (\x,0) circle (2pt) node[above] {\footnotesize $\label$};
 }
 \filldraw (1.1,-1.7) circle (2pt) ;
 
 \foreach \x in {0.5, 0.62,0.8} {
 \draw (1.1,-1.7-\x) circle (\x);
 }
 \draw[Orange, dashed, line width = 1.3] (0.8,0.7) -- (2.3,-1.7);
 
 \end{tikzpicture}}= 0\,.
    \end{equation}
    where the diagram without a cut should be interpreted as having a cut extending all the way to the right, or equivalently, to the left. It is important to note that this relation holds for the loop integrand at all loop orders, but only when considering a single vertex. In other words, we can contract any number of pairs of fields in $H_{\text{int}}$ because they remain associated with the same time point in the largest time equation.
\end{mdexample}
This works for more complicated diagrams.
\begin{mdexample}
    \textbf{(Propagator identities: exchange diagrams)} For example, we can consider the exchange graph with melonic exchange
    \begin{equation}
        \adjustbox{valign=c,scale=0.82}{
\begin{tikzpicture}[scale=0.7, ball/.style = {circle, draw, align=center, anchor=north, inner sep=0}]
 \draw (-0.1,0) -- (4.3,0);
 \foreach \x/\label in {0.1/\hspace{1em}E_1\ldots,0.4/ , 1/\hspace{1em}E_m} {
 \draw (\x,0) -- (1.1,-1.5);
 \filldraw (\x,0) circle (2pt) node[above] {\fontsize{8}{10}\selectfont $\label$};
 }
 \foreach \x/\label in {3.2/E_{m+1}\ldots, 3.8/,4.1/\hspace{1em}E_n} {
 \draw (\x,0) -- (3.1,-1.5);
 \filldraw (\x,0) circle (2pt) node[above] {\fontsize{8}{10}\selectfont $\label$};
 }
 \filldraw (1.1,-1.5) circle (2pt);
 \filldraw (3.1,-1.5) circle (2pt);
 \draw (1.1,-1.5) to[out=80, in=100] (3.1,-1.5);
 \draw (1.1,-1.5) to[out=-80, in=-100] (3.1,-1.5);
 \draw (1.1,-1.5) to[out=30, in=150] (3.1,-1.5);
 \draw (1.1,-1.5) to[out=-30, in=-150] (3.1,-1.5);
 \node at (2.1,-1.5) {\fontsize{8}{10}\selectfont $...$};
 \node[] at (2.1,-2.5) {\color{white}.};
\end{tikzpicture}} {-}
 \adjustbox{valign=c,scale=0.82}{%
\begin{tikzpicture}[scale=0.7, ball/.style = {circle, draw, align=center, anchor=north, inner sep=0}]
 \draw (-0.1,0) -- (4.3,0);
 \foreach \x/\label in {0.1/\hspace{1em}E_1\ldots,0.4/ , 1/\hspace{1em}E_m} {
 \draw (\x,0) -- (1.1,-1.5);
 \filldraw (\x,0) circle (2pt) node[above] {\fontsize{8}{10}\selectfont $\label$};
 }
 \foreach \x/\label in {3.2/E_{m+1}\ldots, 3.8/,4.1/\hspace{1em}E_n} {
 \draw (\x,0) -- (3.1,-1.5);
 \filldraw (\x,0) circle (2pt) node[above] {\fontsize{8}{10}\selectfont $\label$};
 }
 \filldraw (1.1,-1.5) circle (2pt);
 \filldraw (3.1,-1.5) circle (2pt);
 \draw (1.1,-1.5) to[out=80, in=100] (3.1,-1.5);
 \draw (1.1,-1.5) to[out=-80, in=-100] (3.1,-1.5);
 \draw (1.1,-1.5) to[out=30, in=150] (3.1,-1.5);
 \draw (1.1,-1.5) to[out=-30, in=-150] (3.1,-1.5);
 \node at (2.1,-1.5) {\fontsize{8}{10}\selectfont $...$};
 \draw[Orange, dashed, line width = 1.3] (2.0,0.2) -- (0.2,-1.6);
 \node[] at (2.1,-2.5) {\color{white}.};
\end{tikzpicture}} {-}
 \adjustbox{valign=c,scale=0.82}{%
\begin{tikzpicture}[scale=0.7, ball/.style = {circle, draw, align=center, anchor=north, inner sep=0}]
 \draw (-0.1,0) -- (4.3,0);
 \foreach \x/\label in {0.1/\hspace{1em}E_1\ldots,0.4/ , 1/\hspace{1em}E_m} {
 \draw (\x,0) -- (1.1,-1.5);
 \filldraw (\x,0) circle (2pt) node[above] {\fontsize{8}{10}\selectfont $\label$};
 }
 \foreach \x/\label in {3.2/E_{m+1}\ldots, 3.8/,4.1/\hspace{1em}E_n} {
 \draw (\x,0) -- (3.1,-1.5);
 \filldraw (\x,0) circle (2pt) node[above] {\fontsize{8}{10}\selectfont $\label$};
 }
 \filldraw (1.1,-1.5) circle (2pt);
 \filldraw (3.1,-1.5) circle (2pt);
 \draw (1.1,-1.5) to[out=80, in=100] (3.1,-1.5);
 \draw (1.1,-1.5) to[out=-80, in=-100] (3.1,-1.5);
 \draw (1.1,-1.5) to[out=30, in=150] (3.1,-1.5);
 \draw (1.1,-1.5) to[out=-30, in=-150] (3.1,-1.5);
\node at (2.1,-1.5) {\fontsize{8}{10}\selectfont $...$};
\draw[Orange, dashed, line width = 1.3] (2.2,0.2) -- (4,-1.6);
\node[] at (2.1,-2.5) {\color{white}.};
\end{tikzpicture}}
{+}
\adjustbox{valign=c,scale=0.82}{%
\begin{tikzpicture}[scale=0.7,ball/.style = {circle, draw, align=center, anchor=north, inner sep=0}]
 \draw (-0.1,0) -- (4.3,0);
 \foreach \x/\label in {0.1/\hspace{1em}E_1\ldots,0.4/ , 1/\hspace{1em}E_m} {
 \draw (\x,0) -- (1.1,-1.5);
 \filldraw (\x,0) circle (2pt) node[above] {\fontsize{8}{10}\selectfont $\label$};
 }
 \foreach \x/\label in {3.2/E_{m+1}\ldots, 3.8/,4.1/\hspace{1em}E_n} {
 \draw (\x,0) -- (3.1,-1.5);
 \filldraw (\x,0) circle (2pt) node[above] {\fontsize{8}{10}\selectfont $\label$};
 }
 \filldraw (1.1,-1.5) circle (2pt);
 \filldraw (3.1,-1.5) circle (2pt);
 \draw (1.1,-1.5) to[out=80, in=100] (3.1,-1.5);
 \draw (1.1,-1.5) to[out=-80, in=-100] (3.1,-1.5);
 \draw (1.1,-1.5) to[out=30, in=150] (3.1,-1.5);
 \draw (1.1,-1.5) to[out=-30, in=-150] (3.1,-1.5);
 \node at (1.7,-1.5) {\fontsize{8}{10}\selectfont $...$};
 \draw[Orange, dashed, line width = 1.3] (2.1,0.2) -- (2.1,-2.4);
 \node[] at (2.1,-2.5) {\color{white}.};
\end{tikzpicture}}=0\,,
    \end{equation}
    which reads, in formulae, as 
    \begin{equation}
        \begin{split}
            &B^{\text{ex},s}_n(\{E_i\}_{i=1}^n)+(-1)^{n}B^{\text{ex},s}_n(\{-E_i\}_{i=1}^n)\\ &\qquad \quad=2\int_{\vec{p}_1...\vec{p}_{L+1}}\hspace{-1em}\frac{B^{\text{c,cut}}_{m,L+1}(\{E_i\}_{i=1}^m,\{y_i\}_{i=1}^{L+1}) B^{\text{c,cut}}_{n-m,L+1}(\{E_i\}_{i=m+1}^n,\{y_i\}_{i=1}^{L+1})}{\prod_{i=1}^{L+1}P(y_i)}\,.
        \end{split}
    \end{equation}
\end{mdexample}
We emphasize that we have not yet found a suitable combinatorial structure to formulate the most general correlator cutting rule. 

\subsection{Scattering in de Sitter}

Scattering in de Sitter space is very exciting because it provides a non-perturbative notion of unitarity in de Sitter space. We believe that this could be used to derive positivity bounds for cosmology within the next few years, indicating which effective field theories used in cosmology admit a consistent UV completion. The discussion below is based on \cite{Donath:2024utn}, but see also \cite{Marolf:2012kh,Melville:2023kgd,Melville:2024ove} for other discussions of a de Sitter S-matrix. See also \cite{Baumann:2015nta,Grall:2020tqc,Grall:2021xxm,Pajer:2020wnj,Hui:2023pxc,Creminelli:2022onn,Creminelli:2023kze,Creminelli:2024lhd} for discussions of constraints on boost breaking effective field theories.

The in-out formalism we have developed offers us a clear strategy to build a scattering amplitude: take the in-out correlator and amputate its external legs with an LSZ-like projection. The result of this procedure automatically makes us think of an object similar to an S-matrix. Alternative, we can try to define a scattering matrix independently of correlators, while still using the in-out formalism. More formally, if we assume that we start with a state $\ket{n}$ in the infinite past ($\eta=-\infty$) that has $n$-particles
\begin{equation}
     \underbracket[0.4pt]{\ket{n}=\bigotimes_a^n \ket{\Delta_a,\vec{k}_a,s_a,\sigma_a}}_{\substack{\text{In and out states are tensor}\\\text{ products of unitary de Sitter irreps}}}\,,
\end{equation}
(classified by the first Casimir/scaling dimension, momentum and spin) then we evolve for an infinite amount of time (from $\eta=-\infty$ to $\eta=+\infty$) and subsequently project onto a state with $n'$ particles:
\begin{equation}
    S_{n,n'}=\bra{n'}U(+\infty,-\infty)\ket{n}=\bra{n'}\mathcal{T}\e^{-i\int_{-\infty}^{+\infty}H_{\text{int}}(\eta)\d\eta}\ket{n}\,,
\end{equation}
This is exactly what we would do in Minkowski, but in our formalism we can also do it in de Sitter.

Of course, people have discussed amplitudes in de Sitter space for around 20 years (see, e.g., \cite{Marolf:2012kh}), and there is a variety of criticism regarding the existence of scattering amplitudes in de Sitter space. Let us address some of them from our setup:
\begin{itemize}[label=$\diamond$]
    \item \emph{IR divergences prevent a de Sitter S-matrix}. Possibly this is an issue, but we assume with a sufficient number of derivatives and/or in the presence of massive field the IR divergences do not arise. 
\item \emph{Particles are unstable, so there are no asymptotic states}. Our $i\varepsilon$ prescription turns
on/off interaction adiabatically at $\eta=\pm \infty$. Therefore, at least in perturbation theory we don't observe any dramatic instability.
\item \emph{Blue-shifted particles near null infinity (the ``Big Bang'') lead to a large
backreaction}. This is a coordinate artifact. In global coordinates, a
particle can cross the ``Big Bang'' null hypersurface that bounds the Poincar\'e patch from the past.
\item \emph{Particle creation prevents an out state at $\eta=0$}. Possibly, but we work at $\eta=\pm \infty$.
\end{itemize}

Now that we have defined an S-matrix, let us compute it and see if the results are interesting. We define a (scalar) one-particle state in the relativistic normalization using
\begin{align}\label{contactA}
 \ket{\Delta,\vec{k}}=\sqrt{2|\vec{k}|} a^\dagger_\vec{k}\ket{0}\,.
\end{align}
We define the connected part of the amplitude to be
\begin{align}\label{Adef}
 \bra{f}U(+\infty,-\infty)-1\ket{i}=i(2\pi)^4\delta^{(3)}(\vec{k}_i-\vec{k}_f)A_{if}\,,
\end{align}
where we do not factor out the ``energy conserving'' delta function because energy is not conserved in de Sitter. Indeed as we will see $A_ij$ will turn out to be proportional to derivatives of a delta function. An interesting property that the answers will exhibit is a display of unitarity at the non-perturbative level, a fact which we check below in perturbative examples. 
\begin{mdexample}
    Let’s compute the simplest process: contact scattering of $n$ conformally coupled scalars $(m^2=2H^2)$ because they are very simple mode functions. Then to linear order in $\lambda$ the result is
\begin{align}\label{restreeamp}
  \adjustbox{valign=c,scale=1}{\tikzset{every picture/.style={line width=0.75pt}}
\begin{tikzpicture}[x=0.75pt,y=0.75pt,yscale=-1,xscale=1]
\draw    (38.25,50.25) -- (79.75,72.25) ;
\draw    (38.5,72) -- (79.5,50.5) node[left,xshift=0.7cm,yshift=-0.2cm]{$\vdots~n'$} node[right,xshift=-1.7cm,yshift=-0.2cm]{$n~\vdots$};
\end{tikzpicture}}
=A_{nn'}=- \lambda \, (-iH \partial_{E_T})^{n+n'-4}\delta(E_T)\,,
\end{align}
where $E_T$ represents the total energy, accounting for the opposite signs of incoming and outgoing particles
\begin{align}
 E_T=- \sum_a^n |\vec{k}_a| + \sum_b^{n'} |\vec{k}_b|\,.
\end{align}
We found that the $(n{+}n'{-}4)$-th derivative of the energy Dirac delta ($n+n'=4$ would therefore correspond to a Minkowski amplitude, which is consistent with the fact that $\phi^4$ is a conformal interaction). It is interesting to note that, at least in perturbation theory, the S-matrix exhibits crossing symmetry. This allows us to consider all particles as outgoing (as also noted in \cite{Melville:2023kgd}), so that $E_T$ is the usual positive sum of norms.

As a more complex case, take the exchange diagram that enables the elastic scattering of $3+r$ particles through the interaction $\lambda \phi^{4+r}/(4+r)!$:
\begin{equation}
    A_{3+r,3+r}=
    \adjustbox{valign=c,scale=1}{
    \tikzset{every picture/.style={line width=0.75pt}}
\begin{tikzpicture}[x=0.75pt,y=0.75pt,yscale=-1,xscale=1]
\draw    (59,60.25) -- (97,60.25) ;
\draw    (97,60.25) -- (117.75,71.25) ;
\draw    (38.25,49.25) -- (59,60.25) ;
\draw    (38.5,71) -- (59,60.25) ;
\draw    (97,60.25) -- (117.5,49.5) node[left,xshift=1.3cm,yshift=-0.18cm]{$\vdots~3+r$} node[right,xshift=-3.4cm,yshift=-0.18cm]{$3+r~\vdots$};
\end{tikzpicture}
}
\,.
\end{equation}
This diagram is infrared finite provided that $r\geq 0$, which explains the specific definition of $r$. A direct computation yields an expression much more complicated than \eqref{restreeamp}:
\begin{align}\label{scat}
 A_{3+r,3+r}&= \frac{\lambda^2 H^{2r}}{2} \sum_{l=0}^r b_l \frac{(k_{\text{in}}-E_{\text{in}})^{1+r-l}+(k_{\text{in}}+E_{\text{in}})^{1+r-l}}{2k_{\text{in}} (-E_{\text{in}}^2+k_{\text{in}}^2)^{1+r-l}} \partial^{r+l}\delta(E_{\text{in}}-E_{\text{out}}) \,,
\end{align}
where
\begin{subequations}
\begin{align}
 E_{\text{in}} &= \sum_{a=1}^{3+r} |\vec{k}_a |\,, & E_{\text{out}}&\equiv\sum_{a=4+r}^{6+2r} |\vec{k}_a |\,, \\
 \vec{k}_{\text{in}} &=\sum_{a=1}^{3+r} \vec{k}_a & \vec{k}_{\text{out}} & \equiv \sum_{a=4+r}^{6+2r} \vec{k}_a & b_l &\equiv \frac{r!}{l!} (-1)^{l} \,.
\end{align}
\end{subequations}
In particular, this involves a sum of terms each reminiscent of a Feynman propagator raised to some power with additional derivatives of a delta function. 
\end{mdexample}
The interesting claim now is that these de Sitter amplitudes obey the conventional optical theorem familiar from Minkowski QFT:
\begin{align}\label{optical}
 A_{i\to f}-A_{f\to i}^*=i\sum_X \int \d\Pi_X\, (2\pi)^4 \delta^{(3)}(\vec{k}_{\text{in}}-\vec{k}_X)A_{iX}A_{fX}^*\,,
\end{align}
where the summation is over all potential states, with the sole distinction from Minkowski space being the retention of energy-conserving delta functions, thus eliminating the need to explicitly add any on the right-hand side. A significant implication of this theorem is that the right-hand side is clearly positive for forward scattering thanks to our symmetric definition of in and out states, which non-perturbatively restricts the imaginary part of $A_{ii}$! The hope is to one day use this as a way to obtain positivity bounds for de Sitter cosmology. The optical theorem is satisfied in a somewhat non-trivial manner, yet this can be verified through explicit examples.
\begin{mdexample}
    \emph{(An explicit check)} We consider for example $4\to 4$ scattering ($r=1$). We have
    \begin{subequations}\label{eq:rhs1}
    \begin{align}
 \text{RHS of \eqref{optical}}&=i \int \frac{\d k_X^3}{(2\pi)^3} \frac{1}{2E_X}(2\pi)^4 \delta^{(3)}(\vec{k}_{\text{in}}-\vec{k}_X)|A_{4,1}|^2\\
 &=2\pi i\lambda^2 H^2 \delta'(E_{\text{out}}-E_{\text{in}})\frac{\delta'(E_{\text{in}}-\vec{k}_{\text{in}})}{2E_{\text{in}}}\,. \label{RHS}
\end{align}
\end{subequations}
and 
\begin{align}\label{eq:lhs1}
 \text{LHS of \eqref{optical}}&=i \frac{\lambda^2 H^2}{\vec{k}_{\text{in}}} \delta'(E_{\text{in}}-E_{\text{out}}) \Im \frac{1}{ (E_{\text{in}}-\vec{k}_{\text{in}}+i\e)^2}\,.
\end{align}
Admittedly, \eqref{eq:rhs1} and \eqref{eq:lhs1} appear quite different. However, integrating by parts and using the standard $i\varepsilon$ prescription to shift the energy pole results in a perfect match.
\end{mdexample}
\subsection{Outlook}
In the near future, it would be interesting to:
\begin{itemize}[label=$\diamond$]
    \item Use de Sitter amplitudes to derive positivity bounds. This requires understanding the analytic structure and asymptotics.
    \item  Find an IR regulator that preserves in-in = in-out.
    \item Find a dispersive representation of the time-ordered propagator for any mass in de Sitter space to then reduce the correlator integrals to Minkowski amplitude integrals.
    \item Understand de Sitter amplitudes factorization.
\end{itemize}


\section[Open Effective Field Theories]{\label{sec:open-effective-field-theories}Open Effective Field Theories\\
\normalfont{\textit{Enrico Pajer}}}\label{sec3}

The goal of primordial cosmology is to understand QFT and quantum gravity in (approximately, asymptotically) de Sitter spacetime.
On large scales ($\gg\mathrm{Mpc}$) cosmological surveys measure QFT correlators of metric fluctuations as (see \eqref{eq:important})
\begin{equation}
\left\langle\prod_{a=1}^n \delta\left(\vec{k}_a\right)\right\rangle \sim \left[\prod_{a=1}^n T^{(\delta)}\left(\vec{k}_a\right)\right]\left\langle\prod_{a=1}^n \zeta\left(\vec{k}_a\right)\right\rangle+\dots\, ,
\end{equation}
where the metric fluctuations $\delta\left(\vec{k}_a\right)$ are linked to observables such as CMB temperature fluctuations, density distribution of galaxies, dark matter, etc.\ and the curvature perturbations $\zeta\left(\vec{k}_a\right)$ are in de Sitter spacetime. 
In single-clock inflation models, the gravitational floor of non-Gau{\ss}ianity in the primordial fluctuations is $f_{N L}^{e q} \gtrsim 10^{-2}$ \cite{Cabass:2016cgp}, which provides a fundamental prediction to test against cosmological observations. 

We have several aspirations regarding what fundamental physics we can learn about from cosmology. For example, we would like to:
\begin{itemize}
    \item learn about new degrees of freedom and their interactions since inflation requires at least one degree of freedom and three energy scales beyond the standard model\footnote{These are: the energy scale of inflation, $H$, its first derivative, which is indirectly related to the duration of inflation and its second derivatives, which in the simplest models is measured in the tilt of the power spectrum},
    \item learn about the laws of gravity at short distances/high energies by probing GR and beyond at high energies,
    \item explore which QFTs are consistent in FLRW/de Sitter spacetime,
    \item and finally learn when QFT on curved space-time breaks down and quantum gravity is needed.
\end{itemize} 
Here we develop a non-standard point of view on the physics of the primordial universe: we describe gravity plus the inflaton sector as an open system, remaining agnostic about other constients of the universe. We begin with an invitation to the relevant formalism, namely the Schwinger--Keldysh. Then we apply this description to general open effective field theories during inflation. 


\subsection{Invitation to the Schwinger--Keldysh formalism}

Here, we present an invitation to open quantum systems for high-energy physicists.

Instead of a pure state, in a general open system, we have to deal with a density matrix. This might seem unfamiliar or unnecessarily complicated; however, we find it necessary to describe standard high-energy calculations, as we will show in this simply toy model.
Consider the following Lagrangian for two scalars\footnote{This Lagrangian appears often as a toy model of effective field theories. In this context, it was considered recently to discuss the field theoretic wavefunction \cite{Salcedo:2022aal}.}, one of which is heavy $\chi$ and the other light $\phi$
\begin{equation}
    \mathcal{L}_{\mathrm{UV}}=-(\partial \phi)^2 - (\partial \chi)^2 - m^2 \phi^2 - M^2 \chi^2-g M \phi^2 \chi\, .
\end{equation}
As a classic example of EFT, we have 
\begin{equation}
\mathcal{A}_{2 \rightarrow 2}\sim\begin{tikzpicture}[baseline=(current bounding box.center), >=stealth]
    \draw (-0.4,0) -- (0.4,0)
    node[above, midway]  {$\chi$}; 
    \draw (-0.4,0) -- (-1,0.68);
    \draw (0.4,0) -- (1,0.68);
    \draw (0.4,0)-- (1,-0.68);
    \draw (-0.4,0) -- (-1,-0.68);
\end{tikzpicture}\sim\frac{g^2 M^2}{s-M^2} \sim g^2 \sum_n \left(\frac{s}{M^2}\right)^n\, .
\end{equation}
Then we can expand in local unitary, interactions
\begin{equation}
    \mathcal{L}_\text{eff} = \left(\partial \phi\right)^2 - m^2 \phi^2 - g^2 M^2 \sum_n \phi^2 \left(\frac{\Box}{M^2}\right)^n \phi^2
\end{equation}
at tree level, which depends only on $M^2$. Hence, when computing amplitudes, there is no need to use an open quantum system description. 

What happens when we want to compute finite-time correlators? Let us consider the equivalent four-point correlator
\be
B_4 = \begin{tikzpicture}[baseline=(current bounding box.center)]
    \draw (0,0) -- (-0.5,1);
    \draw (0,0) -- (0.5,1);
    \draw (0,0) -- (2,0);
    \draw (2,0) -- (1.5,1);
    \draw (2,0) -- (2.5,1);
    \fill (1.5,1) circle (2pt);
    \fill (2.5,1) circle (2pt);
    \fill (0.5,1) circle (2pt);
    \fill (-0.5,1) circle (2pt);
\end{tikzpicture} = \langle \prod_{a=1}^{4} \phi(t,\vec{k}_a) \rangle = \frac{(gM)^2}{E_S \prod_{a=1}^4 E_a} \frac{2(E_T + E_S)}{E_L E_R E_T}\, ,
\ee
where we have
\begin{equation}
\begin{aligned}
& E_L=E_1 + E_2 + E_S\, , \\
& E_R=E_3 + E_4 + E_S\, ,\\
& E_S=\sqrt{\left|\vec{k}_1+\vec{k}_2\right|^2+M^2}\, ,\\
&E_T=\sum_{a=1}^4 E_a\, .
\end{aligned}
\end{equation}
Now we want to find the low-energy expansion of this as $M\to\infty$,
\be
\lim_{M \to \infty} B_4^{\text{UV}} = \frac{g^2}{\prod_{a=1}^4 E_a} \left[ \frac{M^2 - k_s^2 + E_{12} E_{34}}{E_T M^2} + \frac{E_{12} E_{34}}{M^3} + \mathcal{O}(1/M^4) \right]\,,
\ee
where $E_{ij} = E_i + E_j$. This object is not Lorentz invariant, but this should not surprise us because we have chosen to look at a non-Lorentz invariant observable. More interestingly, the above result cannot come from a unitary EFT. Why? We just derived the cutting rules for correlators in the previous section. One of the rules was that $B_4(E) + (-1)^4 B_4(-E) = 0$, see \eqref{notuni}, but the above formula is even in energy. 
This is related to the influence functional that we heard about in Ch. \cite{chapterHaehlRangamani}.

We can see by hand that the terms that are odd\footnote{The original Lagrangian only depended on $M^2$ and so should the final result. In the above formulae, we have written $\sqrt{M^2}$ simply as $M$, assuming a real and positive mass.} in $M$ do not come from a unitary EFT. The Lagrangian has to be a total derivative because otherwise it would contribute to the amplitude. A total derivative can also be written as a boundary term, and sees that the following would do the job,
\begin{equation}
\mathcal{S}_{\text {EFT}} \supset \lambda \int d^3 x  \dot{\phi}^2 \phi^2  \Rightarrow B_{\text{EFT}}^{(4)}=\begin{tikzpicture}[baseline=(current bounding box.center), scale=0.5]
    \draw (1,-1) -- (-1,1);
    \draw (-1,-1) -- (1,1);
    \fill (0,0) circle (2pt);
\end{tikzpicture}=\frac{\sum_{a, b} E_a E_b E_T}{E_T \prod_{a=1}^4 E_a} \operatorname{Re}(i \lambda)\,.
\end{equation}
However, to make this non-zero we need imaginary coupling $\lambda=i R $ which is non-unitary\footnote{More in detail one can show the Hamiltonian is not Hermitian}. Thus, we can construct an EFT but not a unitary EFT. While this toy model does display some interesting features of full-fledged cosmological open systems, it has also some shortcoming and subtleties. In particular, the non-unitary terms are contact terms\footnote{E.P. thanks Alberto Nicolis for pointing this out.} and so arguably they could well be missed by the EFT. We nevertheless find it a useful example of how open system physics can show up in very familiar settings.\footnote{E.P. is in debt to Cliff Burgess, Alberto Nicolis, Thomas Colas and Santiago Agui Salcedo for useful discussions on this example.}

The origin of the non-unitarity is that the quantum system is made of $\phi$ and $\chi$  but we are only observing the former. Since the two sectors are entangled in the interacting theory, tracing over one, say $\chi$ requires describing $\phi$ via the non-Hamiltonian evolution of a density matrix. Indeed, the correlator we are computing is schematically
\be
\langle \phi^n \rangle = \text{Tr}_{\phi,\chi}[\rho \phi^n] =\text{Tr}_{\phi,\chi}\sum_{\chi'}\left[\rho|\chi'\rangle\langle\chi'| \phi^n\right]=\text{Tr}_{\phi}[\rho_{\text{red}} \phi^n]\, .
\ee
We had just a pure state, but after taking the trace over $\chi$, the entanglement between $\phi$ and $\chi$ led to a mixed state. This situation in general requires to be studied with the Schwinger-Keldysh formalism, as we discuss below.


\begin{QA}
\question{ If we use the LSZ reduction formula, it will turn the correlator into the amplitude, for which we showed things worked well in a unitary local EFT. What's the deal?}
Let us write the expression in Fourier space:
\be
B_4^{\text{UV}}(p_1, p_2, p_3, p_4) = \prod_{a=1}^4 \frac{1}{p_a^2 + i\eps} \frac{1}{(s - M^2)}\,, 
\ee
In and EFT we would expand the off-sheel propagator as $ \frac{1}{s - M^2} \simeq \frac{1}{M^2} \sum_n\left(\frac{s}{M^2}\right)^n$. This expansion however has a finite radius of convergence and so is valid only when $s<M^2$. When we take the Fourier transform to compute the correlator in the time-wavenumber domain, namely
\be
B_4(t, \vec{p}_1, \vec{p}_2, \vec{p}_3, \vec{p}_4) = \int \d p_1^0 \d p_2^0 \d p_3^0 \d p_4^0 B_4^{\text{UV}}(p_1^0,p_2^0,p_3^0,p_4^0,\vec{p}_1,\vec{p}_2,\vec{p}_3,\vec{p}_4 )\,,
\ee
we necessarily integrate outside of the radius of convergence of the above Taylor expansion. Hence the result will not be correct if we expand before the Fourier transformation. Thus, working at fixed time forces us to integrate over energies outside of the radius of convergence of the EFT expansion, which here is $s=M^2$. 
\end{QA}

\subsection{Open Quantum Cosmology}

In general, more complicated systems arise from assembling simpler building blocks. In quantum mechanics, we build bigger Hilbert spaces $\mathscr{H}_{\text {tot }}$ from tensor products of smaller Hilbert spaces, $\mathscr{H}_{\text {tot }}=\otimes_i \mathscr{H}_i$. Open quantum systems arise when we observe only parts of $\mathscr{H}_{t o t}$. For the Universe, this is always the case. When studying the cosmology of the very early universe there are many parts of the system that we do not observe: momenta conjugate of fields, spectator fields, gravitons, massive particles, etc.. Hence, as a first step, we should always start with an open-system approach to cosmology, both at the quantum level and in the semi-classical limit. 

The defining characteristic of open systems is dissipation, which we use here as a catch-all word for a series of phenomena  including fluctuations, non-unitary evolution, dispersion, etc.. It is hard to give a general rule for when the presence of an environment is important and leads to sizable dissipative effects. A rule of thumb is the following: \emph{``Open systems are needed when there are many ``excited" degrees of freedom we do not observe or measure"}.
We can gives some examples of cosmological scenarios that always or only sometimes require an open system approach: 
\begin{itemize}[label=$\diamond$]
\item  \emph{Always:} particle production out of vacuum due to time-dependent background. In de Sitter, we expect that the "vacuum", as experienced by a comoving observer, is an environment with a thermal bath at the Hubble temperature\footnote{The finite temperature comes about very explicitly when one restricts the description to the static patch of a given observer.} $\e^{-E / T} \sim \e^{-2 \pi m / H}$. 
\item  \emph{Sometimes:} physical production of particles, warm inflation, non-adiabatic kicks (features), etc. The size of the effect of the environment is model-dependent and can be large.
\item  \emph{Sometimes:} long-short mode coupling, stochastic inflation. Again, the size of the effect is model dependent and can be large.
\end{itemize}

Let us look more closely at vacuum production in de Sitter that leads to curvature perturbations. The cosmological collider story considers the three-point function for the curvature perturbations $\zeta$ in the squeezed limit which in position space corresponds to two modes with a short separation distance $k_{\text{short}}$, and one mode at a long separation distance $k_{\text{long}}$, depicted as \cite{Arkani-Hamed:2015bza}:
\begin{equation}\adjustbox{valign=c}{
    \includegraphics[scale=0.8]{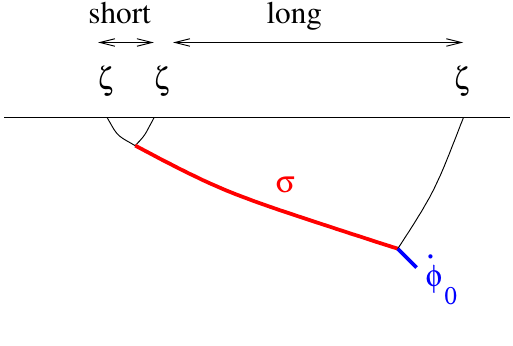}}
\end{equation}
Here $\sigma$ corresponds to a massive field that can decay into two inflatons, and is produced by the time-dependent background (de Sitter expansion) at a rate of $H \sim 2 \pi T_d S$. This corresponds to corrections to the inflaton 3-pt function. The leading-order effects are captured by unitary evolution where there is no large dissipation. 
Hence, low-energy EFT is unitary but non-local, leading to characteristic non-analytic signals, as, for example, in the already mentioned squeeze limit of the three-point function $B_3$.
This limit corresponds to small values of the ratio between the short and long separation distances. 
Considering, for example, the squeeze limit for a single particle with mass $m>\frac{3 H}{2}$ and spin $s$, we have \cite{Arkani-Hamed:2015bza, Noumi:2012vr}
\begin{equation}
\Delta_{ \pm}=\frac{3}{2} \pm i \mu, \quad \mu=\sqrt{\frac{m^2}{H^2}-\frac{9}{4}}\, ,
\end{equation}
\begin{equation}
\frac{\langle\zeta \zeta \zeta\rangle}{\langle\zeta \zeta\rangle_{\text {short }}\langle\zeta \zeta\rangle_{\text {long }}} \sim \epsilon \e^{-\pi \mu}|c(\mu)|\left[\e^{i \delta(\mu)}\left(\frac{k_{\text {long }}}{k_{\text {short }}}\right)^{\frac{3}{2}+i \mu}
\!\!{+}\,
\e^{-i \delta(\mu)}\left(\frac{k_{\text {long }}}{k_{\text {short }}}\right)^{\frac{3}{2}-i \mu}\right] \! P_s(\cos \theta),
\end{equation}
where $\Delta_i$ are conformal dimensions and $\epsilon$ is a slow roll parameter. This is a general picture for studying the spectrum of particles during inflation.

From here, we can consider warm inflation, wherein we can build a mechanism to continuously produce particles by draining energy from the background or inflaton. 
For example, a scalar (axion) coupled to $U(1)$ vector via $(\alpha / f) \phi F \tilde{F}$ leads to tachyonic production of $A_i$ particles, which back react on perturbations
\begin{equation}
\left(\frac{\partial^2}{\partial \tau^2}+k^2 \mp 2 a H k \xi\right) A_{ \pm}(\tau, k)=0, \quad \text { with } \quad \xi \equiv \frac{\alpha \dot{\phi}}{2 f H}.
\end{equation}
This induces, fluctuation, dissipation, and non-unitary evolution
\begin{equation}
\ddot{\phi}+3 H \dot{\phi}+\frac{d V}{d \phi} \simeq \frac{\alpha}{f}\langle\vec{E} \cdot \vec{B}\rangle.
\end{equation}
The low energy EFT is non-unitary. However, since most of the gauge fields are produced around horizon crossing, the low-energy dynamics of the inflaton around the Hubble scale is necessary non-local if we integrate out the gauge fields. This theory does not have a small parameter to organize an EFT expansion in derivative and hence it must be treated by assuming a concrete UV model.

There are many other models of particle production. An interesting one considered in \cite{Creminelli_2023} features the coupling between the inflaton $\phi$ and an additional degree of freedom $\chi$, both minimally coupled to gravity, 
\begin{equation}
\begin{split}
S = \int \mathrm{d}^4 x \sqrt{-g} & \left[ \frac{1}{2} M_{\mathrm{Pl}}^2 R - \frac{1}{2} (\partial \phi)^2 - V(\phi) - |\partial \chi|^2 + M^2|\chi|^2 \right. \\
& \left.- i \frac{\partial_\mu \phi}{f} \left(\chi \partial^\mu \chi^* - \chi^* \partial^\mu \chi\right) - \frac{1}{2} m^2 \left(\chi^2 + \chi^{*2}\right) \right]\, .
\end{split}
\end{equation}
In this case $\chi$ modes become tachyonic and are copiously produced. The big difference is that most modes are produced with a wavelength that is parametrically shorter than the Hubble radius.
The $\chi$ environment leads to fluctuations, dissipation, etc. in the dynamics of $\phi$, but crucially this can be described from the bottom up in terms of an open quantum system with a single degree of freedom and a \textit{local} action! The corresponding linearized equation of motion in Fourier space takes the typical form of a Langevin equation,
\begin{equation}
\ddot{\varphi}_{\vec{k}}+(3 H+\gamma) \dot{\varphi}_{\vec{k}}+\left(\frac{k^2}{a^2}+V^{\prime \prime}\right) \varphi_{\vec{k}}=-\frac{m^2}{f} \delta \mathcal{O}_S(\vec{k})\, ,
\end{equation}
where $\gamma$ is the friction coefficient that causes dissipation and $\delta \mathcal{O}_S$ are fluctuations.
\begin{figure}
    \centering
    \includegraphics[scale=0.25]{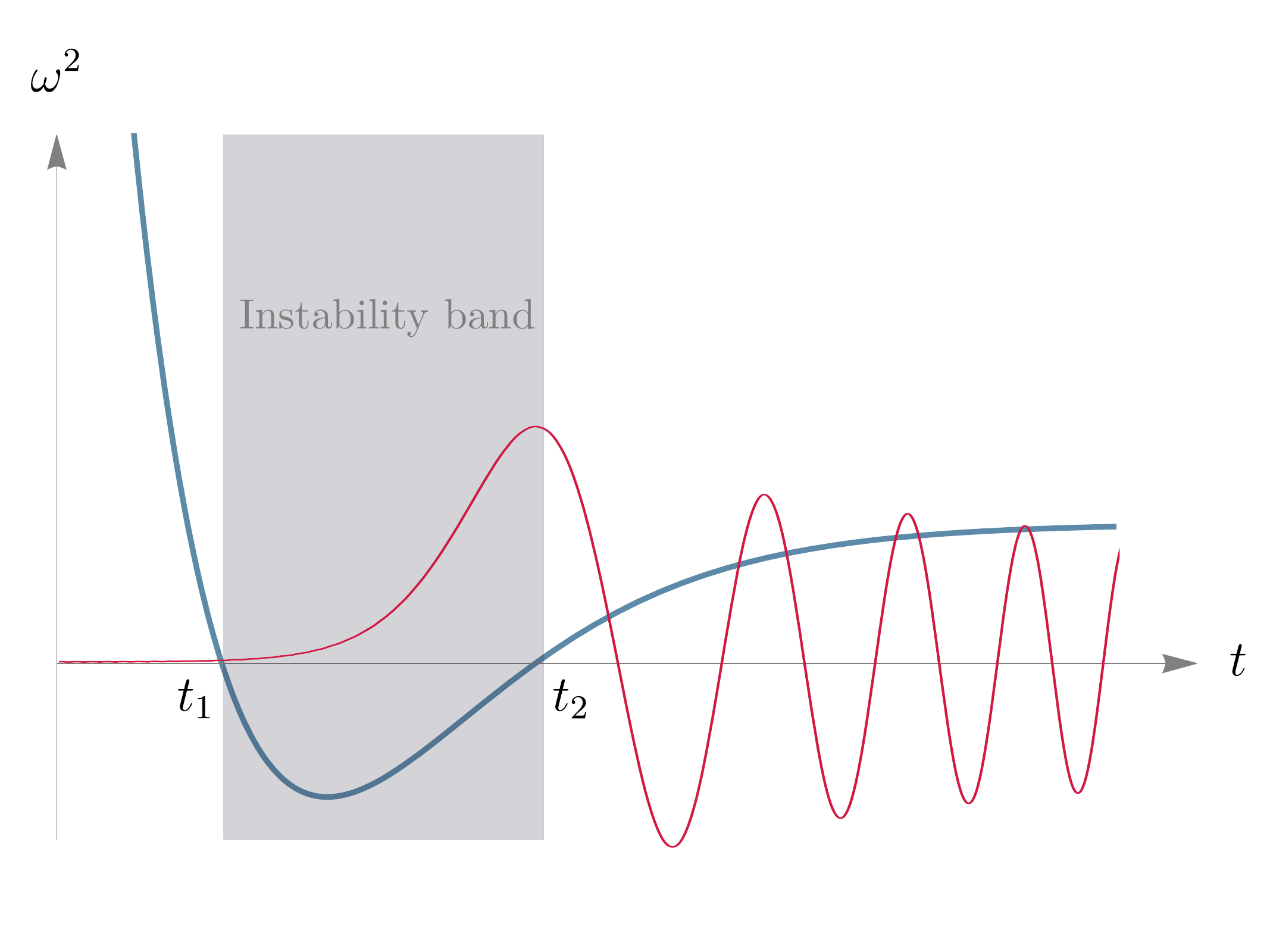}
    \caption{The blue line depicts WKB frequency over time while the red line depicts the mode function with an amplitude that grows exponentially during the instability band. Reproduced from \cite{Creminelli_2023}; for further details, see the original source.}
    \label{fig:cosmomodes}
\end{figure}

As a last example, we briefly discuss stochastic inflation \cite{Starobinsky1986,Starobinsky:1994bd}. Sometimes the environment is not a different field, as in $\mathscr{H}=\mathscr{H}_\phi \otimes \mathscr{H}_\chi$ but different Fourier modes of the same field as in $\mathscr{H}=\otimes_{\mathbf{k}} \mathscr{H}_{\mathbf{k}}$.
Then one ends up with a characteristic Langevin equation of the form
\begin{equation}
\frac{\mathrm{d} \bar{\varphi}}{\mathrm{d} t}=-\frac{V^{\prime}}{3 H}+\frac{
H^{3 / 2} }{2 \pi} \xi(t)\, .
\end{equation}
where the second time derivative term has been dropped and the last term gives the fluctuations that short modes induce on long modes. This is typically treated non-perturbatively but semi-classically. In this case, there is a local, semi-classical, low-energy EFT.

We close this section by comparing bottom-up versus top-down approaches:
\begin{itemize}
    \item Top-down pros: many explicit models exist which exhibit large open quantum system, a.k.a. "environmental", behavior.
    \item Top-down cons: calculations can be long and hard, it is difficult to find general patterns or phenomena, and models are delicate. By delicate we mean that small changes in the model can lead to intractable dynamics such as: non-locality in time, non-Markovian quantum dynamics, a breakdown of perturbation theory, instability, and large backreaction.
    \item Bottom-up pros: general features are visible, meaning that we can assume certain rules, such as stability, locality, perturbativity, etc. and we do not need to check the detailed implementation. 
    \item Bottom-up cons: it is not clear whether one misses some constraints on the low-energy effective couplings and, relatedly, whether the model can be UV-completed.
\end{itemize} 
Overall, our goal in what follows is to develop a systematic bottom-up approach to open quantum dynamics
during inflation.

\subsection{An Open Effective Field Theory of Inflation}
The main idea is to use the principles of Effective Field Theory (EFT) to build large classes of bottom-up open quantum systems.
The EFT rules are:
\begin{enumerate}
    \item Name the low-energy \textit{degrees of freedom}.
    \item Choose \textit{symmetries and principles} (e.g., locality) and write down the most generic symmetric action, which has infinitely many terms.
    \item Choose a radiatively stable \textit{power counting} and truncate the EFT to the desired precision with a \textit{finite} number of operators.
\end{enumerate}
We note that in the absence of locality in time and space, it is difficult to satisfy number 3 above\footnote{It is not impossible either: sometime one can Taylor expand in Fourier space around some fixed but non-zero momentum, as for the Fermi surface of fermions.}. The rest of this section follows closely \cite{Salcedo:2024smn} (see also \cite{Colas:2024lse} for a summary), which built upon previous results and especially \cite{LopezNacir:2011kk,Hongo:2018ant}.

We begin with number 1 above and consider the degrees of freedom of an open EFT. The simplest open EFT has a single degree of freedom, the Nambu--Goldstone boson of the spontaneous breaking of time translation by the inflaton background, as in the usual “unitarity” EFT of inflation.
We use a close-time path integral, a.k.a. in-in formalism, which prepares a density matrix, as depicted here:
\begin{equation}
    \tikzset{every picture/.style={line width=0.75pt}}
\begin{tikzpicture}[x=0.75pt,y=0.75pt,yscale=-1,xscale=1]
\draw  [draw opacity=0] (301.5,111) .. controls (307.3,111) and (312,115.7) .. (312,121.5) .. controls (312,127.3) and (307.3,132) .. (301.5,132) -- (301.5,121.5) -- cycle ; \draw   (301.5,111) .. controls (307.3,111) and (312,115.7) .. (312,121.5) .. controls (312,127.3) and (307.3,132) .. (301.5,132) node[right,midway,yshift=0.2cm]{$\mathscr{O}(t)$} node[pos=1,left,midway,yshift=0.5cm,xshift=-3.5cm]{$\rho(t_0)$};  
\draw    (181,111) -- (301.5,111) node[above,pos=0]{$t_0$} node[above,pos=1]{$t$};
\draw    (181,132) -- (301.5,132) ;
\draw[->]    (181,111) -- (241.25,111) node[above]{$+$};
\draw[-<]    (181,132) -- (241.25,132) node[below]{$-$};
\end{tikzpicture}
\end{equation}
This is equivalent to the master equation known in operator language and has already been used for 20 years in inflation. However, hardly anyone has allowed for
dissipation. It is convenient to perform the Keldysh rotation of the doubled fields, moving from the $+$ and $-$ contours of the path integral to the so-called retarded and advanced fields\footnote{These are also sometimes called ``classical'' and ``quantum'' fields. However the classical limit is subtle and we prefer to avoid this nomenclature.}
\begin{equation}
\begin{array}{lll}
\pi_{r}=\frac{\pi_{+}+\pi_{-}}{2} & \text { and } & \pi_{a}=\pi_{+}-\pi_{-} \\
\pi_{+}=\pi_{r}+\frac{\pi_{a}}{2} & \text { and } & \pi_{+}=\pi_{r}-\frac{\pi_{a}}{2} 
\end{array}
\end{equation}

We need to ensure our EFT satisfies several constraints, nicely summarise e.g., in \cite{Crossley:2015evo,Liu:2018kfw}. 
We want our open QFT to come from a unitary ``closed" UV theory with a Hermitian time evolution, so we have to satisfy 
\begin{equation}
\begin{aligned}
S_{\text {eff }}\left[\pi_{+}, \pi_{+}\right] & =0\,, & S_{\mathrm{eff}}\left[\pi_{r}, \pi_{a}=0\right] & =0\,, \\
S_{\mathrm{eff}}\left[\pi_{+}, \pi_{-}\right] & =-S_{\mathrm{eff}}^*\left[\pi_{-}, \pi_{+}\right]\,, & S_{\mathrm{eff}}\left[\pi_{r}, \pi_{a}\right] & =-S_{\mathrm{eff}}^*\left[\pi_{r},-\pi_{a}\right]\,, \\
\Im S_{\mathrm{eff}}\left[\pi_{+}, \pi_{-}\right] & \geq 0 \,,& \Im S_{\mathrm{eff}}\left[\pi_{r}, \pi_{a}\right] & \geq 0.
\end{aligned}
\end{equation}
These follow from conditions on a density matrix: normalisation requires $\operatorname{Tr}(\rho)=1$, hermiticity means $\rho=\rho^{\dagger}$, and positivity means $\rho \geq 0$. Further details on the derivation of these constraints can be found in \cite{Salcedo:2024smn, Colas:2024lse}. 

Then we aim to write a general in-in functional, including a ``unitary" part, $S_u\left(\pi_{ \pm}\right)$, and a general ``Feynman--Vernon" influence functional, $F\left(\pi_{+}, \pi_{-}\right)$, as
\begin{equation}
\begin{aligned}
Z\left[J_{ \pm}\right] & =\int D \pi_{+} D \pi_{-} \e^{i S\left(\pi_{+}, \pi_{-}\right)+i J_{ \pm} \pi_{ \pm}} \\
& =\int d\pi \int^\pi D \pi_{+} \int^\pi D \pi_{-} \, \e^{i\left[S_u\left(\pi_{+}\right)-S_u\left(\pi_{-}\right)+F\left(\pi_{+}, \pi_{-}\right)+J_{ \pm} \pi_{ \pm}\right]} \\
Z\left[J_{a},J_r\right]& =\int d\pi \int^\pi D \pi_{r} \int^0 D \pi_{a} \,  \e^{i S(\pi_r,\pi_a)+J_{ a} \pi_{ a}+J_{ r} \pi_{ r}} \,.
\end{aligned}
\end{equation}
In the Keldysh basis we see that $\pi_a$ is non-dynamical because both its initial and final conditions are fixed by the boundaries of the path integral. It is only $\pi_r$ that describes a dynamical field since its final condition is integrated over and hence arbitrary. 

Now we address number 2 of our EFT rules. We need to ensure that our open QFT is consistent with general physical principles such as unitarity, locality, and causality. 
Unitarity of the UV theory is encoded in the three conditions already mentioned in the density matrix: $\operatorname{Tr}(\rho)=1, \rho=\rho^{\dagger}$ and $\rho \geq 0$. 
We note that locality is not necessary because the environment can mediate interactions at a distance that are non-local in time
(non-Markovianity).
Nevertheless, we restrict ourselves to EFTs that are local in time and space, and hence we assume a separation of scales. These EFTs exist, as in the scalar warm inflation example. This is our strongest
assumption, which leads to major simplifications.
Finally, causality and analyticity are satisfied for the case we consider in the following, but additional constraints are expected to emerge from demanding causality of the UV completion as in amplitudes' positivity bounds \cite{Adams:2006sv}.

Furthermore, we study the symmetries of our EFT. 
As in the EFT considered when addressing number 1, a general breaking of $\mathrm{U}_t(1)$ time translations is incompatible with scale invariance. We also need an internal shift symmetry $\mathrm{U}_{\text{int}}(1)$ and we require a diagonal\footnote{Additionally one must assume that the degree of freedom enjoying the shift symmetry also evolves linearly, otherwise the low energy couplings depend on time, albeit in a fixed way \cite{Finelli:2018upr}.} is unbroken $\mathrm{U}_t(1) \times \mathrm{U}_{\text{int}}(1)\rightarrow \mathrm{U}_{\text {scale }}(1)$. This is depicted as
\begin{equation}
    \adjustbox{valign=c}{\tikzset{every picture/.style={line width=0.75pt}}
\begin{tikzpicture}[x=0.75pt,y=0.75pt,yscale=-1,xscale=1]
\draw  [draw opacity=0] (301.5,111) .. controls (307.3,111) and (312,115.7) .. (312,121.5) .. controls (312,127.3) and (307.3,132) .. (301.5,132) -- (301.5,121.5) -- cycle ; \draw   (301.5,111) .. controls (307.3,111) and (312,115.7) .. (312,121.5) .. controls (312,127.3) and (307.3,132) .. (301.5,132) node[right,midway,yshift=0.2cm]{$t_f$} node[pos=1,left,midway,yshift=0.5cm,xshift=-3.5cm]{$\bra{\text{in}}$} node[pos=1,left,midway,yshift=-0.2cm,xshift=-3.5cm]{$\ket{\text{in}}$};  
\draw    (181,111) -- (301.5,111);
\draw    (181,132) -- (301.5,132) ;
\draw[-<]    (181,111) -- (241.25,111);
\draw[->]    (181,132) -- (241.25,132);
\draw  [draw opacity=0][fill={rgb, 255:red, 0; green, 0; blue, 0 }  ,fill opacity=1 ] (309,120.75) .. controls (309,119.51) and (310.01,118.5) .. (311.25,118.5) .. controls (312.49,118.5) and (313.5,119.51) .. (313.5,120.75) .. controls (313.5,121.99) and (312.49,123) .. (311.25,123) .. controls (310.01,123) and (309,121.99) .. (309,120.75) -- cycle ;

\draw[dash pattern={on 0.84pt off 2.51pt}]    (271.5,110.5) -- (271.5,131.75) ;
\draw[dash pattern={on 0.84pt off 2.51pt}]   (249.5,110.5) -- (249.5,131.75) node[pos=1,below]{$\textcolor{RoyalBlue}{\epsilon_+}=\textcolor{Maroon}{\epsilon_-}=\epsilon_{\text{cl}}$};

\draw  [draw opacity=0][fill=RoyalBlue ,fill opacity=0.3 ] (247,131.75) .. controls (247,130.51) and (248.01,129.5) .. (249.25,129.5) .. controls (250.49,129.5) and (251.5,130.51) .. (251.5,131.75) .. controls (251.5,132.99) and (250.49,134) .. (249.25,134) .. controls (248.01,134) and (247,132.99) .. (247,131.75) -- cycle ;
\draw  [draw opacity=0][fill=RoyalBlue  ,fill opacity=1 ] (269,131.75) .. controls (269,130.51) and (270.01,129.5) .. (271.25,129.5) .. controls (272.49,129.5) and (273.5,130.51) .. (273.5,131.75) .. controls (273.5,132.99) and (272.49,134) .. (271.25,134) .. controls (270.01,134) and (269,132.99) .. (269,131.75) -- cycle ;

\draw  [draw opacity=0][fill=Maroon  ,fill opacity=0.3 ] (247.25,110.5) .. controls (247.25,109.26) and (248.26,108.25) .. (249.5,108.25) .. controls (250.74,108.25) and (251.75,109.26) .. (251.75,110.5) .. controls (251.75,111.74) and (250.74,112.75) .. (249.5,112.75) .. controls (248.26,112.75) and (247.25,111.74) .. (247.25,110.5) -- cycle ;
\draw  [draw opacity=0][fill=Maroon  ,fill opacity=1 ] (269.25,110.5) .. controls (269.25,109.26) and (270.26,108.25) .. (271.5,108.25) .. controls (272.74,108.25) and (273.75,109.26) .. (273.75,110.5) .. controls (273.75,111.74) and (272.74,112.75) .. (271.5,112.75) .. controls (270.26,112.75) and (269.25,111.74) .. (269.25,110.5) -- cycle ;

\end{tikzpicture}} \qquad
  \qquad  \adjustbox{valign=c}{\tikzset{every picture/.style={line width=0.75pt}}
\begin{tikzpicture}[x=0.75pt,y=0.75pt,yscale=-1,xscale=1]
\draw  [draw opacity=0] (301.5,111) .. controls (307.3,111) and (312,115.7) .. (312,121.5) .. controls (312,127.3) and (307.3,132) .. (301.5,132) -- (301.5,121.5) -- cycle ; \draw   (301.5,111) .. controls (307.3,111) and (312,115.7) .. (312,121.5) .. controls (312,127.3) and (307.3,132) .. (301.5,132) node[right,midway,yshift=0.2cm]{$t_f$} node[pos=1,left,midway,yshift=0.5cm,xshift=-3.5cm]{$\bra{\text{in}}$} node[pos=1,left,midway,yshift=-0.2cm,xshift=-3.5cm]{$\ket{\text{in}}$};  
\draw    (181,111) -- (301.5,111);
\draw    (181,132) -- (301.5,132) ;
\draw[-<]    (181,111) -- (241.25,111);
\draw[->]    (181,132) -- (241.25,132);
\draw  [draw opacity=0][fill={rgb, 255:red, 0; green, 0; blue, 0 }  ,fill opacity=1 ] (309,120.75) .. controls (309,119.51) and (310.01,118.5) .. (311.25,118.5) .. controls (312.49,118.5) and (313.5,119.51) .. (313.5,120.75) .. controls (313.5,121.99) and (312.49,123) .. (311.25,123) .. controls (310.01,123) and (309,121.99) .. (309,120.75) -- cycle ;

\draw[dash pattern={on 0.84pt off 2.51pt}]    (271.5,110.5) -- (271.5,131.75) ;
\draw[dash pattern={on 0.84pt off 2.51pt}]   (249.5,110.5) -- (249.5,131.75) node[pos=1,below]{$\textcolor{RoyalBlue}{\epsilon_+}=\frac{\epsilon_{\text{a}}}{2}$};
\draw[dash pattern={on 0.84pt off 2.51pt}]  (228.5,110.5) -- (228.25,131.75) node[pos=0,above]{$\textcolor{Maroon}{\epsilon_-}=-\frac{\epsilon_{\text{a}}}{2}$};

\draw  [draw opacity=0][fill=RoyalBlue ,fill opacity=0.3 ] (247,131.75) .. controls (247,130.51) and (248.01,129.5) .. (249.25,129.5) .. controls (250.49,129.5) and (251.5,130.51) .. (251.5,131.75) .. controls (251.5,132.99) and (250.49,134) .. (249.25,134) .. controls (248.01,134) and (247,132.99) .. (247,131.75) -- cycle ;
\draw  [draw opacity=0][fill=RoyalBlue  ,fill opacity=1 ] (269,131.75) .. controls (269,130.51) and (270.01,129.5) .. (271.25,129.5) .. controls (272.49,129.5) and (273.5,130.51) .. (273.5,131.75) .. controls (273.5,132.99) and (272.49,134) .. (271.25,134) .. controls (270.01,134) and (269,132.99) .. (269,131.75) -- cycle ;

\draw  [draw opacity=0][fill=Maroon  ,fill opacity=0.3 ] (247.25,110.5) .. controls (247.25,109.26) and (248.26,108.25) .. (249.5,108.25) .. controls (250.74,108.25) and (251.75,109.26) .. (251.75,110.5) .. controls (251.75,111.74) and (250.74,112.75) .. (249.5,112.75) .. controls (248.26,112.75) and (247.25,111.74) .. (247.25,110.5) -- cycle ;
\draw  [draw opacity=0][fill=Maroon  ,fill opacity=1 ] (226.25,110.5) .. controls (226.25,109.26) and (227.26,108.25) .. (228.5,108.25) .. controls (229.74,108.25) and (230.75,109.26) .. (230.75,110.5) .. controls (230.75,111.74) and (229.74,112.75) .. (228.5,112.75) .. controls (227.26,112.75) and (226.25,111.74) .. (226.25,110.5) -- cycle ;
\end{tikzpicture}}
\end{equation}
The coset-construction for in-in is still relatively under developed (however, see \cite{Hongo:2019qhi, Akyuz:2023lsm} for recent progress).
Under time translations the Goldstone boson $\pi_r$ transforms non-linearly, while the auxiliary advanced field $\pi_a$ transforms linearly,
\begin{equation}
\begin{aligned}
\pi_{r}(t, \boldsymbol{x}) \rightarrow &\pi_{r}(t, \boldsymbol{x})+\epsilon_{r}^0\left[1+\dot{\pi}_{r}(t, \boldsymbol{x})\right]+\mathcal{O}\left[\left(\epsilon_{r}^0\right)^2\right] \\
\pi_{a}(t, \boldsymbol{x}) \rightarrow &\pi_{a}(t, \boldsymbol{x})+\epsilon_{r}^0 \dot{\pi}_{a}(t, \boldsymbol{x})+\mathcal{O}\left[\left(\epsilon_{r}^0\right)^2\right]
\end{aligned} \quad \text{for} \quad \epsilon_{+}^0=\epsilon_{-}^0=\epsilon_{r}^0
\end{equation}
This tells us that $\pi_a$ should not be interpreted as a Goldstone boson. Conversely, the other linear combination of time translations, which would transform fields as
\begin{equation}
\begin{aligned}
\pi_{r}(t, \boldsymbol{x}) \rightarrow \pi_{r}^{\prime}(t, \boldsymbol{x}) & =\pi_{r}(t, \boldsymbol{x})+\frac{\epsilon_{a}^0}{2} \dot{\pi}_{a}(t, \boldsymbol{x})+\mathcal{O}\left[\left(\epsilon_{a}^0\right)^2\right] \\
\pi_{a}(t, \boldsymbol{x}) \rightarrow \pi_{a}^{\prime}(t, \boldsymbol{x}) & =\pi_{a}(t, \boldsymbol{x})+\epsilon_{a}^0\left[1+\dot{\pi}_{r}(t, \boldsymbol{x})\right]+\mathcal{O}\left[\left(\epsilon_{a}^0\right)^2\right]
\end{aligned} \quad \text{for} \quad \epsilon_{+}^0=-\epsilon_{-}^0=\frac{\overline{\epsilon_{a}^0}}{2}\, .
\end{equation}
is \textit{explicitly} broken by the presence of the environment and plays no role in the following discussion. We will discuss non-linear boosts later in this section.

We now turn to the free theory. At quadratic order in perturbations and up to one derivative per field we have
\begin{equation}
\begin{aligned} \mathcal{L}^{(2)}=&\left(\alpha_1-2 \alpha_2\right) \dot{\pi}_{r} \dot{\pi}_{a}-\alpha_1 \partial_i \pi_{r} \partial^i \pi_{a}-\alpha_0 \pi_{r} \pi_{a} \\
& \quad-2 \gamma_1 \dot{\pi}_{r} \pi_{a}+i\left[\beta_1 \pi_q^2-\left(\beta_2-\beta_4\right) \dot{\pi}_q^2+\beta_2\left(\partial_i \pi_{a}\right)^2\right]\, .
\end{aligned}
\end{equation}
The first line of $\mathcal{L}^{(2)}$ corresponds to unitarity dynamics\footnote{Unitary dynamics comes from $S(\pi_+)-S(\pi_-)$ and hence only generates terms that are odd in $\pi_a$.} and the second line contains terms with dissipation $\gamma_1$ and fluctuations $\beta_i$. (Notice the imaginary unit $i$ in front of the square brackets.)
Two point functions are given by the Keldysh propagator $G_K$ and the retarded/advanced propagators $G_{R, A}$ as
\begin{equation}
\left(\begin{array}{ll}
\left\langle\pi_{r}(x) \pi_{r}(y)\right\rangle & \left\langle\pi_{r}(x) \pi_{a}(y)\right\rangle \\
\left\langle\pi_{a}(x) \pi_{r}(y)\right\rangle & \left\langle\pi_{a}(x) \pi_{a}(y)\right\rangle
\end{array}\right)=\left(\begin{array}{cc}
i G_K(x, y) & -G_R(x, y) \\
-G_A(x, y) & 0
\end{array}\right)\, .
\end{equation}
Note $\pi_a$ does not propagate, as anticipated. Here $G_A=G_R^*$ are the advanced and retarded Green's functions of the dissipative equations of motion, while $G_K$ describes the perturbations in the system induced by fluctuations in the environment. 

Next, we look more closely at propagators for this free theory. We can look to flat-space examples as illuminating warm ups:
\begin{equation}
\begin{gathered}
G_{R / A}(k ; \omega)=-\frac{1}{\omega^2 \pm i \gamma \omega-c_s^2 k^2}=-\frac{1}{\left(\omega_{-}-\omega_{+}\right)}\left[\frac{1}{\omega-\omega_{-}}-\frac{1}{\omega-\omega_{+}}\right]\, , \\
\omega_{ \pm}=-i \frac{\gamma}{2} \pm E_k^\gamma \quad \text { and } \quad E_k^\gamma=\sqrt{c_s^2 k^2-\frac{\gamma^2}{4}}\, .
\end{gathered}
\end{equation}
Here retarded/advanced propagators feel dissipation, which shifts poles in the complex plane but are independent of fluctuations, since $G_{A,R}$ are state independent. Dissipation can be seen in the presence of a complex pole, leading to exponential suppression in the power spectrum. This represents erasure of memory from the distant past due to dissipation into the environment. 

The Keldysh propagator instead probes fluctuations
\begin{equation}
G_K(k ; \omega)=-G_R(k ; \omega) \widehat{D}_K(k ; \omega) G_A(k ; \omega)=-i \frac{\beta_1+\beta_2 \omega^2+\beta_3 k^2}{\left(\omega^2-c_s^2 k^2\right)^2+\gamma^2 \omega^2}
\end{equation}
for which the final spectrum is largely specified by fluctuations
\begin{equation}
P_k=\frac{2 \beta_1}{\gamma c_s^2 k^2}+\frac{2 \beta_2}{\gamma}+\frac{2 \beta_3}{c_s^2 \gamma}\, .
\end{equation}

Next, we consider de Sitter propagators. The mode functions are still Hankel functions, but now the index depends on dissipation. Indeed in de Sitter the Hubble expansion leads to a characteristic dissipative term proportional to the number of spatial dimensions. In the presence of dissipation due to an environment, $\gamma$  can be thought of as a change in the number of dimensions. Indeed, the asymptotic time dependence of fields is found to be
\begin{equation}
\lim_{\eta\to 0} \pi_k(\eta)=\widetilde{A} \eta^{\frac{3}{2}}+\frac{\gamma}{2 H} H_{\frac{3}{2}+\frac{\gamma}{2 H}}^{(1)}(\eta)\, ,
\end{equation}
where $\pi \sim \eta^{\Delta}$ and $\Delta=\left(0,3+\frac{\gamma}{H}\right)$. 
Here, massless scalars still freeze out $(\gamma \geq 0)$. 
The propagators are pretty messy, especially $G_K$ ($G_\gamma$ and $F_\gamma$ are combinations of ${ }_2 F_3$, see \cite{Salcedo:2024smn}). We find
\begin{equation}
G^R\left(k ; \eta_1, \eta_2\right)=-i \frac{\pi}{4} H^2\left(\eta_1 \eta_2\right)^{\frac{3}{2}}\left(\frac{\eta_1}{\eta_2}\right)^{\frac{\gamma}{2 H}} \Im \left[H_{\frac{3}{2}+\frac{\gamma}{2 H}}^{(1)}\left(-k \eta_1\right) H_{\frac{3}{2}+\frac{\gamma}{2 H}}^{(2)}\left(-k \eta_2\right)\right] \theta\left(\eta_1-\eta_2\right),
\end{equation}
\begin{equation}
\begin{aligned}
G_1^K\left(k ; \eta_1, \eta_2\right)=&-i \frac{\widetilde{\beta}_1 \pi^2}{8}\left(\eta_1 \eta_2\right)^{\frac{3}{2}+\frac{\gamma}{2 H}} \Re \left\{H_{\frac{3}{2}+\frac{\gamma}{2 H}}^{(1)}\left({-}k \eta_1\right) H_{\frac{3}{2}+\frac{\gamma}{2 H}}^{(1)}\left({-}k \eta_2\right)\left[F_\gamma\left(z_2\right){-}F_\gamma(\infty)\right]\right.\\
&\left.-H_{\frac{3}{2}+\frac{\gamma}{2 I I}}^{(1)}\left(-k \eta_1\right) H_{\frac{3}{2}+\frac{\gamma}{2 H I}}^{(2)}\left(-k \eta_2\right) G_\gamma\left(z_2\right)\right\}+\left(\eta_1 \leftrightarrow \eta_2\right) .
\end{aligned}
\end{equation}
Symmetries ensure $P \sim 1 / k^3$, so smoking-gun signals come from the bispectrum.

Now, we consider interactions for our open EFT. To cubic order in fluctuations, the in-in action is organized in powers of $\pi_q$. 
\begin{equation}
\begin{aligned}
\mathcal{L}_1^{(3)}= & \left(4 \alpha_3-3 \alpha_2\right) \dot{\pi}_{r}^2 \dot{\pi}_{a}+\alpha_2\left(\partial_i \pi_{r}\right)^2 \dot{\pi}_{a}+2 \alpha_2 \dot{\pi}_{r} \partial_i \pi_{r} \partial^i \pi_{a} \\
& +\left(4 \gamma_2-\gamma_1\right) \dot{\pi}_{r}^2 \pi_{a}+\gamma_1\left(\partial_i \pi_{r}\right)^2 \pi_{a}\, , \\
\mathcal{L}_2^{(3)}= & i\left[-\beta_3 \dot{\pi}_{r} \dot{\pi}_{a} \pi_q+\beta_3 \partial_i \pi_{r} \partial^i \pi_{a} \pi_{a}+2 \beta_4 \dot{\pi}_{r} \dot{\pi}_q^2-2 \beta_4 \partial_i \pi_{r} \partial^i \pi_{a} \dot{\pi}_{a}\right]\, , \\
\mathcal{L}_3^{(3)}= & \delta_1 \pi_{a}^3+\left(\delta_5-\delta_2\right) \dot{\pi}_{a}^2 \pi_{a}+\delta_2\left(\partial_i \pi_{a}\right)^2 \pi_{a}-\delta_4\left(\partial_i \pi_{a}\right)^2 \dot{\pi}_{a}+\left(\delta_4-\delta_6\right) \dot{\pi}_{a}^3\, .
\end{aligned}
\end{equation}
The real power of the EFT comes from relating operators at different orders because of the non-linearly realised boost, as noted early in the dissipative context in \cite{LopezNacir:2011kk}. The inflaton background breaks both t-translation and boosts, where the former are resurrected by a diagonal combination with shift symmetry, and the latter are non-linearly realized (``broken").

We consider non-linear symmetries in the EFT. In the flat-space decoupling, the most general symmetry transformation is
\begin{equation}
\begin{aligned}
\pi_{r}(t, \boldsymbol{x}) \rightarrow \pi_{r}^{\prime}(t, \boldsymbol{x}) & =\pi_{r}\left(\Lambda_{r \mu}^0 x^\mu, \Lambda_{r \mu}^i x^\mu\right)+\Lambda_{r \mu}^0 x^\mu-t\, , \\
\pi_{a}(t, \boldsymbol{x}) \rightarrow \pi_{a}^{\prime}(t, \boldsymbol{x}) & =\pi_{a}\left(\Lambda_{r}^0 x^\mu, \Lambda_{r \mu}^i x^\mu\right)\, .
\end{aligned}
\end{equation}
Clearly $\pi_{r}$ transforms non-linearly, hence relating $\mathscr{L}_2$ to $\mathscr{L}_3$, as already pointed out in \cite{LopezNacir:2011kk, Hongo:2018ant}. Once again notice that $\pi_a$ transforms as a spectator field and is hence not a Goldstone boson. To proceed from here, we can write down all operators and demand the above symmetry, or work with invariant combinations.
Using invariant combination we see the relation between quadratic and cubic orders in the Lagrangian
\begin{equation}
\begin{aligned} \mathcal{L}_1=\,&\mathcal{L}_1^{\mathrm{LO}}+\sum_{n=2}^{\infty} \gamma_n\left[-2 \dot{\pi}_{r}+\left(\partial_\mu \pi_{r}\right)^2\right]^n \pi_{a} \\
& -\sum_{n=2}^{\infty} \alpha_n\left[-2 \dot{\pi}_{r}+\left(\partial_\mu \pi_{r}\right)^2\right]^{n-1}\left(-\dot{\pi}_{a}+\partial^\mu \pi_{r} \partial_\mu \pi_{a}\right)\, ,
\end{aligned}
\end{equation}
where we neglected higher-order terms with two or more derivatives per field. 

We now turn to correlators.
Correlators can be computed in perturbation theory using the standard in-in rules, being careful about distinguishing between retarded and Keldysh propagators.
To build intuition, consider 3-point function in flat space which are mostly of the form \cite{Salcedo:2024smn}
\begin{equation}
B_3 \sim \frac{\operatorname{Poly}\left(E_1, E_2, E_3\right)}{\operatorname{Sing}_\gamma}\, ,
\end{equation}
\begin{equation}
\begin{aligned}
\operatorname{Sing}_\gamma= & \left|E_1^\gamma+E_2^\gamma+E_3^\gamma+\frac{3}{2} i \gamma\right|^2\left|-E_1^\gamma+E_2^\gamma+E_3^\gamma+\frac{3}{2} i \gamma\right|^2 \\
& \times\left|E_1^\gamma-E_2^\gamma+E_3^\gamma+\frac{3}{2} i \gamma\right|^2\left|E_1^\gamma+E_2^\gamma-E_3^\gamma+\frac{3}{2} i \gamma\right|^2\, ,
\end{aligned}
\end{equation}
\begin{equation}
E_k^\gamma=\sqrt{c_s^2 k^2-\frac{\gamma^2}{4}}\, .
\end{equation}
The 3-point function in flat space peaks when $k_i \pm k_j \pm k_l=0$, namely for folded triangles
\begin{figure}[tbp]
 \begin{minipage}{6in}
    \centering
    \raisebox{-0.5\height}{\includegraphics[width=.46\textwidth]{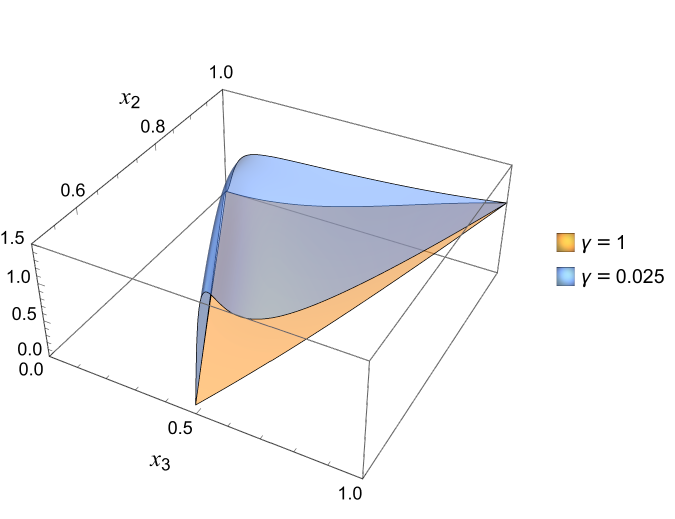}}
    \raisebox{-0.5\height}{\includegraphics[width=.46\textwidth]{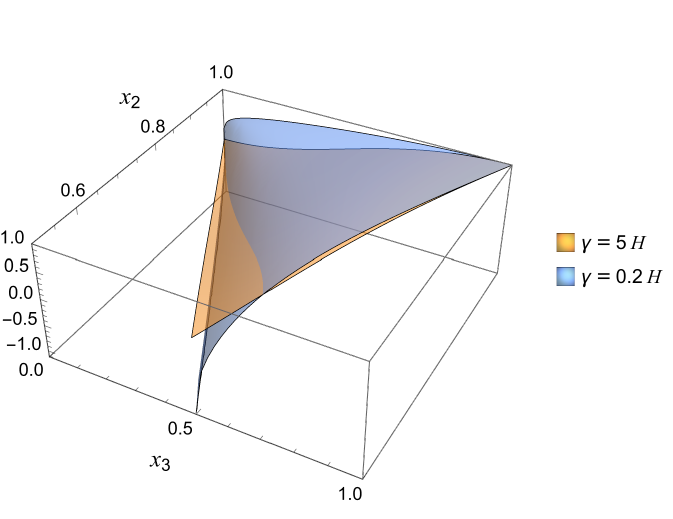}}
\end{minipage}
\caption{Shape function $S(x_2,x_3) \equiv (x_2 x_3)^2 [B(k_1, x_2 k_1, x_3 k_1)/B(k_1,k_1,k_1)]$ for the bispectrum corresponding to $\pi_a^3$ in Minkowski (\textit{left}) and in de Sitter (\textit{right}). Figure adapted from  \cite{Salcedo:2024smn}.\label{fig:foldedintro} }
\end{figure} 
The folded divergence is regulated by dissipation $\gamma \neq 0$, because interactions can build only over a finite amount of time before being erased by dissipation.

Cosmologists have used the in-in formalism for 20 years and yet the vast majority of papers assumed unitarity and non-dissipative dynamics. Here we have developed a systematic single-field open EFT for inflation, based on previous work by \cite{LopezNacir:2011kk, Hongo:2018ant}. Assuming locality in time and space, the resulting EFT is easy to write down and can be studied in perturbation theory. The smoking gun signal are peaks in the bispectrum near folded triangles. 
Our formalism is a starting point for studying aspects of quantum information of inflation such as entanglement growth, purity, decoherence, and more. 


\section[Bootstrapping the analytic wavefunction]{Bootstrapping the analytic wavefunction\\
\normalfont{\textit{Mang Hei Gordon Lee}}}\label{sec4}

The purpose of this section is to provide a quick overview on how to bootstrap the cosmological wavefunction. We will introduce the off-shell wavefunction coefficients and explain how they can be constrained by physical principles such as unitarity, locality, and scale invariance. We will also illustrate this by bootstrapping the three-point wavefunction of a single massless scalar. This discussion is based on \cite{Pajer:2020wxk,Goodhew:2020hob,Goodhew:2021oqg, Jazayeri:2021fvk,Melville:2021lst,Salcedo:2022aal}.

\subsection{Cosmological wavefunctions}
The story begins with the Bunch--Davies vacuum state, $|\text{BD} \rangle$, in the far past projected onto a field basis at time $\eta_0$ (usually $\eta_0 \to 0$, i.e., at the end of inflation)
\begin{equation}\label{eq:wftnDet}
\Psi[\phi_0,\eta_0]=\braket{\phi_0,\eta_0|\text{BD}}=\int_{\text{BD}}^{\phi(\eta_0)=\phi_0}\D\phi~\e^{i S[\phi;\eta_0]}\,,
\end{equation}
where $S$ denotes the action of the theory. From this we can determine the correlator of $n$ (scalar) fields:
\begin{equation}
    \braket{\phi(\Vec{k}_1)\ldots\phi(\Vec{k}_n)}=\int \D\phi~|\Psi(\phi)|^2~\phi(\Vec{k}_1)\ldots\phi(\Vec{k}_n)\,.
\end{equation}
To make the definition of the cosmological wavefunction in \eqref{eq:wftnDet} more tractable, we usually expand it as follows \cite{Goodhew:2020hob, Baumann2023}
\begin{equation}
    \Psi[\phi;\eta_0]=\exp\left[-\sum_{n=2}^\infty \int \prod_{i=1}^{n}\left(\frac{d^D\vec{k}_i}{(2\pi)^D}\right)~\psi_n(\vec{k}_1,...,\vec{k}_n;\eta_0)\prod_{i=1}^n\phi(\vec{k}_i)\right]\,,
\end{equation}
where $n$ denotes the number of vertices with external legs attached. The \emph{on-shell} wavefunction coefficients $\psi_j$'s, which are functions of external energies ($\omega_k = \sqrt{\vec{k}^2 + m^2}$) and spatial momenta ($\vec{k}$), are what we wish to bootstrap. What is meant by ``bootstrap'' will be clarified soon. Moreover, $\omega_k$ is the variable we will eventually continue in to go off-shell.

Let us mention that at tree level, correlators contain the same information as the wavefunctions, e.g., \cite{Goodhew:2020hob,Baumann2023}
\begin{subequations}
    \begin{align}
        \braket{\phi(\vec{k}_1)\phi(\vec{k}_2)\phi(\vec{k}_3)}&\sim \frac{1}{\prod_{a=1}^3 2 \text{Re}(\psi_2(\vec{k}_a))}[2 \text{Re}(\psi_3)]\,,
        \\
        \braket{\phi(\vec{k}_1)\phi(\vec{k}_2)\phi(\vec{k}_3)\phi(\vec{k}_4)}&\sim \frac{1}{\prod_{a=1}^4 2 \text{Re}(\psi_2(\vec{k}_a))}\Big[2 \text{Re}(\psi_4)+\frac{\text{Re}(\psi_3)\text{Re}(\psi_3)}{\text{Re}(\psi_2)}\Big]\,.
    \end{align}
\end{subequations}
Here $\text{Re}(\psi_n)$ refers to the real part of the wavefunction coefficient. This is no longer true at loop level, for instance, at one loop we have:
\begin{equation}
    \begin{split}
\braket{\phi(\vec{k}_1)\phi(\vec{k}_2)\phi(\vec{k}_3)\phi(\vec{k}_4)}\sim &\frac{1}{\prod_{a=1}^4 2 \text{Re}(\psi_2(\vec{k}_a))}\Big[2 \text{Re}(\psi_4^{1L})\\&+\int \frac{\d^\D p}{(2\pi)^\D}\frac{1}{2\text{Re}(\psi_2(\vec{p}))}\psi_6^{\text{tree}}+\dots\Big]\,.
    \end{split}
\end{equation}
The wavefunction is a more fundamental object than the usual in-in correlators since it contains more information than correlators of fields. It also knows about correlators between fields and their conjugate momentum. If we want to compute the correlator between the fields and its conjugate momentum $\pi(\vec{k})$, we can obtain this from the imaginary part of the wavefunction. For example \cite{Cespedes:2020xqq}:
\begin{equation}
    \braket{\pi(\vec{k}_1)\phi(\vec{k}_2)\phi(\vec{k}_3)}\sim \frac{1}{\prod_{a=1}^3 2 \text{Re}(\psi_2(\vec{k}_a))}[2 \text{Im}(\psi_3)]\,.
\end{equation}
Before we continue, let us review the Feynman rules in the context of cosmological correlators.

\subsection{Feynman rules}
Perturbatively, given a Feynman diagram, we have the following (schematic) rules:
\begin{itemize}[label=$\diamond$]
    \item To each \textcolor{RoyalBlue}{external line} is associated a bulk-to-boundary propagator $\textcolor{RoyalBlue}{K}$. The form of $K$ depends on the spacetime considered and whether the particle propagating is massive. For example:
    \begin{equation}
        K_\omega=\begin{sqcases}
            \e^{i\omega \eta} & \text{(flat space)}\\
            (1-i\omega \eta)\e^{i\omega \eta} & \text{(de Sitter massless)}\\
             (-\eta)^{3/2} H_{iv}^{(2)}(-\omega \eta)  & \text{(de Sitter massive)} \\
             -i\partial_\omega \e^{i\omega \eta} & \text{(conformally coupled)}
        \end{sqcases}\,,
    \end{equation}
where $H^{(2)}_w$ denotes the Hankel function of the second kind of order $w=iv(\D)$, where $v(\D)=\sqrt{\frac{\D^2}{4}-m^2}$.
    \item To each \textcolor{Maroon}{internal line} is associated a bulk-to-bulk propagator $\textcolor{Maroon}{G}$. This propagator can generally be expressed in terms of $K$'s as follows
    \begin{equation}\label{eq:GasF}
        G_\omega(\eta,\eta')=\frac{K_\omega^*(\eta_2)K_\omega(\eta_1)\theta(\eta_1-\eta_2){+}(\eta_1\leftrightarrow \eta_2){-}K_\omega(\eta_1)K_\omega(\eta_2)}{2\text{Re}(\psi_2)}\,.
    \end{equation}
    \item To each vertex are associated a factor $F$ and a time integral from $-\infty < \eta < \eta_0$. In particular, the associated wavefunction will depend only on events happening in its past light cone, which is a signature of causality.
\end{itemize}
For a more detailed discussion of these rules, see \cite{Goodhew:2020hob,Baumann2023,Goodhew:2021oqg,Salcedo:2022aal}. 

\begin{mdexample}
For example, applying the above rules to the following graph (time is going up) gives  
\begin{equation}
\begin{split}
          \psi_4(\{\omega_k\},\{\vec{k}\})&=\adjustbox{valign=c}{\begin{tikzpicture}[baseline={([yshift=2ex]current bounding box.center)},scale=1.0001]
    \coordinate (a) at (0,0);
    \coordinate (b) at (0.5,0);
    \coordinate (c) at (1.5,0);
    \coordinate (d) at (2.5,0);
    \coordinate (e) at (3,0);
    \coordinate (la) at (0.5,-1);
    \coordinate (lb) at (1.5,-1);
    \coordinate (lc) at (2.5,-1);
    \draw[thick] (a) -- (e) (a) --++ (180:0.2) (e) --++ (0:0.2);
    \draw[thick, color=RoyalBlue] (a) -- (la) node[left,midway]{$1$}; 
    \draw[thick, color=RoyalBlue] (b) -- (la) node[right,midway]{$2$};
    \draw[thick, color=RoyalBlue] (d) -- (lc) node[left,midway]{$3$};
    \draw[thick, color=RoyalBlue] (e) -- (lc) node[right,midway]{$4$}; 
    \draw[thick, color=Maroon] (la) -- (lc) node[below,midway]{$s$};
    \draw[thick,fill=black] (a) circle (1pt)  (b) circle (1pt)  (d) circle (1pt)  (e) circle (1pt);
    \draw[thick,fill=black] (la) circle (1pt) (lc) circle (1pt) ;
    \end{tikzpicture}}\\&=\int_{-\infty}^{\eta_0}\d \eta_{\text{L}}\int_{-\infty}^{\eta_0}\d \eta_{\text{R}}F_{\text{L}}F_{\text{R}}\textcolor{RoyalBlue}{K_{\omega_1}}\textcolor{RoyalBlue}{K_{\omega_2}}\textcolor{Maroon}{G_{\omega_s}}\textcolor{RoyalBlue}{K_{\omega_3}}\textcolor{RoyalBlue}{K_{\omega_4}}\,.
\end{split}\label{eq:eq:exLLg0}
\end{equation}
Here, ``L'' and ``R'' label the left and right vertices, respectively. Note that if the exponentials in the $K$'s converge for negative $\eta$, then we have analyticity in the lower half $\omega$ planes for the associated wavefunction. We therefore treat $\{\omega_k\}$ as separate variables in which we analytically continue wavefunctions off-shell.
\end{mdexample}

We can say more about the analytic properties of the wavefunction than what was said above. In particular, perturbatively, we can say that generally singularities are localized at (partial) energy poles \cite{Salcedo:2022aal}. It is relatively easy to see that this is at least true for flat-space wavefunctions. We also expect this to hold for massless/conformally coupled scalars in de Sitter: their mode functions can be written as derivatives of a plane wave, which means the wavefunction coefficients can be written as the derivative of the flat-space wavefunction.

\begin{mdexample}
When energy going into a subdiagram vanishes, there is a singularity. Thus, given a Feynman diagram, we circle all subdiagrams and equate the energy going in and out. For example, the following are singular loci of the associated diagram
\begin{subequations}
\begin{align}
    \adjustbox{valign=c}{\begin{tikzpicture}[baseline={([yshift=2ex]current bounding box.center)},scale=1.0001]
    \coordinate (a) at (0,0);
    \coordinate (b) at (0.5,0);
    \coordinate (c) at (1.5,0);
    \coordinate (d) at (2.5,0);
    \coordinate (e) at (3,0);
    \coordinate (la) at (0.5,-1);
    \coordinate (lb) at (1.5,-1);
    \coordinate (lc) at (2.5,-1);
    \draw[thick] (a) -- (e) (a) --++ (180:0.2) (e) --++ (0:0.2);
    \draw[thick, color=RoyalBlue] (a) -- (la) node[left,midway]{$1$}; 
    \draw[thick, color=RoyalBlue] (b) -- (la) node[right,midway]{$2$};
    \draw[thick, color=RoyalBlue] (d) -- (lc) node[left,midway]{$3$};
    \draw[thick, color=RoyalBlue] (e) -- (lc) node[right,midway]{$4$}; 
    \draw[thick, color=Maroon] (la) -- (lc) node[below,midway]{$s$};
    \draw[thick,fill=black] (a) circle (1pt)  (b) circle (1pt)  (d) circle (1pt)  (e) circle (1pt);
    \draw[thick,fill=black] (la) circle (1pt) (lc) circle (1pt) ;
    \draw[ForestGreen,thick] (la) circle (4pt);
    \begin{scope}[xshift=0.75cm,yshift=-1cm]
        \draw[Orange, thick] (0.75, 0) ellipse (1.35cm and 0.3cm);
    \end{scope}
    \end{tikzpicture}}&\leadsto \begin{sqcases}
        \textcolor{ForestGreen}{\omega_1+\omega_2+\omega_s=0}\,,\\
        \textcolor{orange}{\omega_1+\omega_2+\omega_3+\omega_4=0}\,,\label{eq:exTLg}
    \end{sqcases}
    \\
       \adjustbox{valign=c}{\begin{tikzpicture}[baseline={([yshift=2ex]current bounding box.center)},scale=1.0001]
    \coordinate (a) at (0,0);
    \coordinate (b) at (0.5,0);
    \coordinate (c) at (1.5,0);
    \coordinate (d) at (2.5,0);
    \coordinate (e) at (3,0);
    \coordinate (la) at (0.5,-1);
    \coordinate (lb) at (1.5,-1);
    \coordinate (lc) at (2.5,-1);
    \draw[thick] (a) -- (e) (a) --++ (180:0.2) (e) --++ (0:0.2);
    \draw[thick, color=RoyalBlue] (a) -- (la) node[left,midway]{$1$}; 
    \draw[thick, color=RoyalBlue] (b) -- (la) node[right,midway]{$2$};
    \draw[thick, color=RoyalBlue] (d) -- (lc) node[left,midway]{$3$};
    \draw[thick, color=RoyalBlue] (e) -- (lc) node[right,midway]{$4$}; 
      \begin{scope}[xshift=0.75cm,yshift=-1cm]
        \draw[Maroon, thick] (0.75, 0) ellipse (1cm and 0.3cm);
    \end{scope}
    \draw[thick,fill=black] (a) circle (1pt)  (b) circle (1pt)  (d) circle (1pt)  (e) circle (1pt);
    \draw[thick,fill=black] (la) circle (1pt) (lc) circle (1pt) ;
    \draw[ForestGreen,thick] (la) circle (4pt);
    \begin{scope}[xshift=-1,yshift=-57,scale=0.033]
    \draw[Orange, thick]    (47.5,32.5) .. controls (21.5,32.5) and (18.5,18.5) .. (11.02,26.87) .. controls (3.54,35.25) and (26.5,42.5) .. (46,43.5) .. controls (65.5,44.5) and (89.47,34.03) .. (80.98,26.76) .. controls (72.5,19.5) and (73.5,32.5) .. (47.5,32.5) -- cycle ;
    \end{scope}
    \end{tikzpicture}}&\leadsto \begin{sqcases}
        \textcolor{ForestGreen}{\omega_1+\omega_2+\omega_s+\omega_{s'}=0}\,,\\
        \textcolor{orange}{\omega_1+\omega_2+\omega_3+\omega_4+2\omega_{s'}=0}\,.\label{eq:exLLg}
    \end{sqcases}
    \end{align}
\end{subequations}
   Note that all of the above equations have support outside the physical region where $\omega_j > 0$ for all $j$. This is because there are no physical out states in this formalism, but only in states with positive energy (by convention). 

   For loop diagrams, the internal energies also depend on the loop momentum $\textbf{p}$, so we also need to minimize the energy with respect to $\textbf{p}$. For example, in \ref{eq:exLLg}, after minimization we obtain (for the flat-space wavefunction of a scalar with mass $m$):
   \begin{align}
       \textcolor{ForestGreen}{\omega_1+\omega_2+\sqrt{s^2+4m^2}}&\textcolor{ForestGreen}{=0}\,,\\
       \textcolor{Orange}{\omega_1+\omega_2+\omega_3+\omega_4+2m}&\textcolor{Orange}{=0}\,.
   \end{align}
   Here $s$ is the magnitude of the momentum $\textbf{k}_1+\textbf{k}_2$. Notice how the first singularity can be rewritten as $(\omega_1+\omega_2)^2-s^2=4m^2$.
   
   Moreover, as we saw in \cite{chapterBenicasaVazao}, the amplitude $A$ is the total energy pole of the wavefunction, i.e., 
    \begin{equation}
        \psi\sim \frac{A}{\sum_i \omega_i}\,.
    \end{equation}
At the location of the partial energy poles associated to a subdiagram $\gamma$ of the original diagram $\Gamma$, the wavefunction instead factorizes on lower point amplitudes, i.e., 
\begin{equation}
        \psi_\Gamma\sim \frac{A_\gamma\times \Tilde{\psi}_{\Gamma\setminus \gamma}}{\big[\sum_{i\in \gamma} \omega_i\big]^{\#_\gamma}}\,.
    \end{equation}
\end{mdexample}

\subsection{Bootstrap constraint 1: unitarity}

Quite like what we have for flat-space scattering amplitudes, we can write the evolution operator~\cite{Goodhew:2021oqg} as
\begin{equation}
    U(\eta_0)=\exp\Big[-i\int_{-\infty}^{\eta_0}\d \eta H_{\text{int}}(\eta)\Big]\,,
\end{equation}
where $H_{\text{int}}$ is the time-dependent interaction Hamiltonian of the system. As a consequence of perturbative unitarity and of being in the Bunch--Davies vacuum, the time evolution operator $U$ is unitary:
\begin{equation}
    U^\dagger U =\mathbbm{1} \,.
\end{equation}
Decomposing $U = \mathbbm{1} + \delta U$, unitarity implies that
\begin{equation}
    \delta U + \delta U^{\dagger} = - \delta U \delta U^\dagger\,.
\end{equation}
So far, the discussion has been pretty much the same as the standard flat-space one discussed in textbooks. Things differ when we insert bras and kets:
\begin{equation}
    \underbracket[0.4pt]{\langle n | \delta U | 0 \rangle}_{\text{(I)}} + \underbracket[0.4pt]{\langle n | \delta U^\dag | 0 \rangle}_{\text{(II)}} = -\sumint_X\underbracket[0.4pt]{\bra{n} \delta U | X \rangle}_{\text{(III)}} \langle X | \delta U^\dag\ket{0}\,.
\end{equation}
In perturbation theory (when $\delta U$ is small), it is clear that term (I) corresponds to the wavefunction $\psi_n(\{\omega\})$. What (II) corresponds to is \emph{not} $\psi_n^*(\{\omega\})$, but rather $\psi_n^*(\{-\omega^*\})$. This observation makes it rather manifest that we \emph{need} analyticity in the energies to define (II) in the first place! Finally, (III) is clearly not $\psi_n({\omega})$ since $\ket{X}$ generally differs from the Bunch--Davies vacuum ($\ket{X} \neq \ket{0}$).

At this stage, let us introduce the \defQ{discontinuity operator} in the energy $\omega$: 
\begin{equation}
\begin{split}
    \Disc_{\omega_s} \psi_n(\omega_1, \ldots, \omega_s, \ldots, \omega_n; \{\vec{k}\}) &= \psi_n(\omega_1, \ldots, \omega_s, \ldots, \omega_n; \{\vec{k}\}) \\& \quad - \psi_n^*(-\omega_1^*, \ldots, \omega_s, \ldots, -\omega_n^*; \{\vec{-k}\})\,.
\end{split}
\end{equation}
Based on this definition, we ask what can we learn about the wavefunction coefficients $\phi_j$ from this operator?
\begin{mdexample}
    Let us consider once more the four-point tree-level diagram in \eqref{eq:eq:exLLg0}. The discontinuity in $\omega_s$ is given by 
    \begin{equation}
        \hspace{-0.3cm}\text{Disc}_{\omega_s} \psi_4(\omega_1,\ldots, \omega_4,\omega_s; \{\vec{k}\})=\text{Disc}_{\omega_s}\int_{-\infty}^{\eta_0}\d \eta_{\text{L}}\d \eta_{\text{R}}F_{\text{L}}F_{\text{R}}\textcolor{RoyalBlue}{K_{\omega_1}}\textcolor{RoyalBlue}{K_{\omega_2}}\textcolor{Maroon}{G_{\omega_s}}\textcolor{RoyalBlue}{K_{\omega_3}}\textcolor{RoyalBlue}{K_{\omega_4}}\,.
    \end{equation}
    Unitarity ensures that the couplings in $H_{\text{int}}$ are real, and so the associated vertex factors satisfy $F(\{\vec{k}\})=F^*(\{-\vec{k}\})$, while the fact that we start with the Bunch--Davies vacuum ensures $K(\{\omega\})=K^*(\{-\omega^*\})$. To see this, remember that the Bunch--Davies vacuum implies $K(\omega)\sim \e^{i\omega \eta}$ in the far past, and it is clear that both $K(\omega)$ and $K^\ast(-\omega^\ast)$ are the same. It turns out that under some mild assumptions about the equations of motion, this property is preserved under time evolution (see \cite{Goodhew:2021oqg} for more details). As an example, for a massless scalar in de Sitter, we have:
    \begin{equation}
        K^\ast(-\omega^\ast)=[(1+i\omega^\ast \eta)\e^{-i\omega^\ast\eta}]^\ast=(1-i\omega\eta)\e^{i\omega\eta}=K(\omega)\,.
    \end{equation}
    Using this together with \eqref{eq:GasF} gives
    \begin{equation}
        \text{Im} G_{\omega_s}(\eta_{\text{L}},\eta_{\text{R}})=\frac{1}{2\text{Re}(\psi_2)}\text{Im}(K_{\omega_s}(\eta_{\text{L}}))\text{Im}(K_{\omega_s}(\eta_{\text{R}}))\,,
    \end{equation}
    which gives
        \begin{equation}
        \begin{split}
        \hspace{-0.3cm}\text{Disc}_{\omega_s} \psi_4(\omega_1,\ldots, \omega_4,\omega_s; \{\vec{k}\})&=\frac{1}{2\text{Re}(\psi_2)}\int_{-\infty}^{\eta_0}\d \eta_{\text{L}}F_{\text{L}}\textcolor{RoyalBlue}{K_{\omega_1}}\textcolor{RoyalBlue}{K_{\omega_2}}\textcolor{Maroon}{\text{Im}(K_{\omega_s}(\eta_{\text{L}}))}
        \\&\quad
        \times
        \int_{-\infty}^{\eta_0}\d \eta_{\text{R}}F_{\text{R}}\textcolor{RoyalBlue}{K_{\omega_3}}\textcolor{RoyalBlue}{K_{\omega_4}}\textcolor{Maroon}{\text{Im}(K_{\omega_s}(\eta_{\text{R}}))}\,.
        \end{split}
    \end{equation}
    Diagrammatically, we have
    \begin{equation}
            \text{Disc}_{\omega_s}\adjustbox{valign=c}{\begin{tikzpicture}[baseline={([yshift=2ex]current bounding box.center)},scale=1.0001]
    \coordinate (a) at (0,0);
    \coordinate (b) at (0.5,0);
    \coordinate (c) at (1.5,0);
    \coordinate (d) at (2.5,0);
    \coordinate (e) at (3,0);
    \coordinate (la) at (0.5,-1);
    \coordinate (lb) at (1.5,-1);
    \coordinate (lc) at (2.5,-1);
    \draw[thick] (a) -- (e) (a) --++ (180:0.2) (e) --++ (0:0.2);
    \draw[thick, color=RoyalBlue] (a) -- (la) node[left,midway]{$1$}; 
    \draw[thick, color=RoyalBlue] (b) -- (la) node[right,midway]{$2$};
    \draw[thick, color=RoyalBlue] (d) -- (lc) node[left,midway]{$3$};
    \draw[thick, color=RoyalBlue] (e) -- (lc) node[right,midway]{$4$}; 
    \draw[thick, color=Maroon] (la) -- (lc) node[above,midway]{$s$};
    \draw[thick,fill=black] (a) circle (1pt)  (b) circle (1pt)  (d) circle (1pt)  (e) circle (1pt);
    \draw[thick,fill=black] (la) circle (1pt) (lc) circle (1pt) ;
    \end{tikzpicture}}=
    \text{Disc}_{\omega_s}
    \adjustbox{valign=c}{\begin{tikzpicture}[baseline={([yshift=2ex]current bounding box.center)},scale=1.0001]
    \coordinate (app) at (-0.2,0);
    \coordinate (a) at (0,0);
    \coordinate (b) at (0.5,0);
    \coordinate (bp) at (1,0);
    \coordinate (bpp) at (1.2,0);
    \coordinate (la) at (0.5,-1);
    \draw[thick] (app) -- (bpp);
    \draw[thick, color=RoyalBlue] (a) -- (la) node[left,midway]{$1$}; 
    \draw[thick, color=RoyalBlue] (b) -- (la) node[left,midway,xshift=0.1cm,yshift=0.2cm]{$2$};
    \draw[thick, color=Maroon] (bp) -- (la) node[right,midway]{$s$};
    \draw[thick,fill=black] (a) circle (1pt)  (b) circle (1pt)  (bp) circle (1pt);
    \draw[thick,fill=black] (la) circle (1pt);
    \end{tikzpicture}}
    \text{Disc}_{\omega_s}
     \adjustbox{valign=c}{\begin{tikzpicture}[baseline={([yshift=2ex]current bounding box.center)},scale=1.0001]
    \coordinate (app) at (-0.2,0);
    \coordinate (a) at (0,0);
    \coordinate (b) at (0.5,0);
    \coordinate (bp) at (1,0);
    \coordinate (bpp) at (1.2,0);
    \coordinate (la) at (0.5,-1);
    \draw[thick] (app) -- (bpp);
    \draw[thick, color=RoyalBlue] (a) -- (la) node[left,midway]{$3$}; 
    \draw[thick, color=RoyalBlue] (b) -- (la) node[left,midway,xshift=0.1cm,yshift=0.2cm]{$4$};
    \draw[thick, color=Maroon] (bp) -- (la) node[right,midway]{$s$};
    \draw[thick,fill=black] (a) circle (1pt)  (b) circle (1pt)  (bp) circle (1pt);
    \draw[thick,fill=black] (la) circle (1pt);
    \end{tikzpicture}}\,.\label{eq:DISC1}
    \end{equation}

    Similar pictures also hold at loop-level. For example, if we take the discontinuity in the loop of the diagram in \eqref{eq:exLLg}, we find \cite{Melville:2021lst}
    \begin{equation}
    \begin{split}
    \text{Disc}\adjustbox{valign=c}{\begin{tikzpicture}[baseline={([yshift=2ex]current bounding box.center)},scale=1.0001]
    \coordinate (a) at (0,0);
    \coordinate (b) at (0.5,0);
    \coordinate (c) at (1.5,0);
    \coordinate (d) at (2.5,0);
    \coordinate (e) at (3,0);
    \coordinate (la) at (0.5,-1);
    \coordinate (lb) at (1.5,-1);
    \coordinate (lc) at (2.5,-1);
    \draw[thick] (a) -- (e) (a) --++ (180:0.2) (e) --++ (0:0.2);
    \draw[thick, color=RoyalBlue] (a) -- (la) node[left,midway]{$1$}; 
    \draw[thick, color=RoyalBlue] (b) -- (la) node[right,midway]{$2$};
    \draw[thick, color=RoyalBlue] (d) -- (lc) node[left,midway]{$3$};
    \draw[thick, color=RoyalBlue] (e) -- (lc) node[right,midway]{$4$}; 
      \begin{scope}[xshift=0.75cm,yshift=-1cm]
        \draw[Maroon, thick] (0.75, 0) ellipse (1cm and 0.3cm);
    \end{scope}
    \draw[thick,fill=black] (a) circle (1pt)  (b) circle (1pt)  (d) circle (1pt)  (e) circle (1pt);
    \draw[thick,fill=black] (la) circle (1pt) (lc) circle (1pt) ;
    \end{tikzpicture}}&=
    \text{Disc}\adjustbox{valign=c}{\begin{tikzpicture}[baseline={([yshift=2ex]current bounding box.center)},scale=1.0001]
    \coordinate (app) at (-0.4,0);
    \coordinate (a) at (-0.15,0);
    \coordinate (b) at (0.15,0);
    \coordinate (c) at (0.75,0);
    \coordinate (bp) at (1,0);
    \coordinate (bpp) at (1.2,0);
    \coordinate (la) at (0.5,-1);
    \draw[thick] (app) -- (bpp);
    \draw[thick, color=RoyalBlue] (a) -- (la) node[left,midway]{$1$}; 
    \draw[thick, color=RoyalBlue] (b) -- (la) node[right,midway,xshift=-0.1cm,yshift=0.2cm]{$2$};
    \draw[thick, color=Maroon] (c) -- (la) node[right,midway]{};
    \draw[thick, color=Maroon] (bp) -- (la) node[right,midway]{};
    \draw[thick,fill=black] (a) circle (1pt)  (b) circle (1pt) (c) circle (1pt)  (bp) circle (1pt);
    \draw[thick,fill=black] (la) circle (1pt);
    \end{tikzpicture}}
    \text{Disc}\adjustbox{valign=c}{\begin{tikzpicture}[baseline={([yshift=2ex]current bounding box.center)},scale=1.0001]
    \coordinate (app) at (-0.4,0);
    \coordinate (a) at (-0.15,0);
    \coordinate (b) at (0.15,0);
    \coordinate (c) at (0.75,0);
    \coordinate (bp) at (1,0);
    \coordinate (bpp) at (1.2,0);
    \coordinate (la) at (0.5,-1);
    \draw[thick] (app) -- (bpp);
    \draw[thick, color=RoyalBlue] (a) -- (la) node[left,midway]{$3$}; 
    \draw[thick, color=RoyalBlue] (b) -- (la) node[right,midway,xshift=-0.1cm,yshift=0.2cm]{$4$};
    \draw[thick, color=Maroon] (c) -- (la) node[right,midway]{};
    \draw[thick, color=Maroon] (bp) -- (la) node[right,midway]{};
    \draw[thick,fill=black] (a) circle (1pt)  (b) circle (1pt) (c) circle (1pt)  (bp) circle (1pt);
    \draw[thick,fill=black] (la) circle (1pt);
    \end{tikzpicture}}
    \\&
    \quad +\text{Disc}\adjustbox{valign=c}{\begin{tikzpicture}[baseline={([yshift=2ex]current bounding box.center)},scale=1.0001]
    \coordinate (a) at (0,0);
    \coordinate (b) at (0.5,0);
    \coordinate (bp) at (1,0);
    \coordinate (c) at (1.5,0);
    \coordinate (dp) at (2,0);
    \coordinate (d) at (2.5,0);
    \coordinate (e) at (3,0);
    \coordinate (la) at (0.5,-1);
    \coordinate (lb) at (1.5,-1);
    \coordinate (lc) at (2.5,-1);
    \draw[thick] (a) -- (e) (a) --++ (180:0.2) (e) --++ (0:0.2);
    \draw[thick, color=RoyalBlue] (a) -- (la) node[left,midway]{}; 
    \draw[thick, color=RoyalBlue] (b) -- (la) node[right,midway]{};
    \draw[thick, color=Maroon] (bp) -- (la) node[right,midway]{};
    \draw[thick, color=RoyalBlue] (d) -- (lc) node[left,midway]{};
    \draw[thick, color=Maroon] (dp) -- (lc) node[left,midway]{};
    \draw[thick, color=RoyalBlue] (e) -- (lc) node[right,midway]{}; 
    \draw[thick, color=Maroon] (la) -- (lc) node[above,midway]{$s$};
    \draw[thick,fill=black] (a) circle (1pt)  (b) circle (1pt)  (bp) circle (1pt) (dp) circle (1pt)  (d) circle (1pt)  (e) circle (1pt);
    \draw[thick,fill=black] (la) circle (1pt) (lc) circle (1pt) ;
    \end{tikzpicture}}
    \\&
    \quad +\text{Disc}\adjustbox{valign=c}{\begin{tikzpicture}[baseline={([yshift=2ex]current bounding box.center)},scale=1.0001]
    \coordinate (a) at (0,0);
    \coordinate (b) at (0.5,0);
    \coordinate (bp) at (1,0);
    \coordinate (c) at (1.5,0);
    \coordinate (dp) at (2,0);
    \coordinate (d) at (2.5,0);
    \coordinate (e) at (3,0);
    \coordinate (la) at (0.5,-1);
    \coordinate (lb) at (1.5,-1);
    \coordinate (lc) at (2.5,-1);
    \draw[thick] (a) -- (e) (a) --++ (180:0.2) (e) --++ (0:0.2);
    \draw[thick, color=RoyalBlue] (a) -- (la) node[left,midway]{}; 
    \draw[thick, color=RoyalBlue] (b) -- (la) node[right,midway]{};
    \draw[thick, color=Maroon] (bp) -- (la) node[right,midway]{};
    \draw[thick, color=RoyalBlue] (d) -- (lc) node[left,midway]{};
    \draw[thick, color=Maroon] (dp) -- (lc) node[left,midway]{};
    \draw[thick, color=RoyalBlue] (e) -- (lc) node[right,midway]{}; 
    \draw[thick, color=Maroon] (la) -- (lc) node[above,midway]{$s'$};
    \draw[thick,fill=black] (a) circle (1pt)  (b) circle (1pt)  (bp) circle (1pt) (dp) circle (1pt)  (d) circle (1pt)  (e) circle (1pt);
    \draw[thick,fill=black] (la) circle (1pt) (lc) circle (1pt) ;
    \end{tikzpicture}}\,.
    \end{split}
    \end{equation}
    The pattern of discontinuities becomes manifest from these two examples: the discontinuity is given by the sum over all ways of cutting the master diagram. 
\end{mdexample}
Unitarity already places significant constraints on the bootstrap of the wavefunction coefficients. However, there are additional factors to consider. In the following, we discuss two other constraining properties.

\subsection{Bootstrap constraint 2: manifest locality}
For the rest of this section we will focus mainly on a massless scalar or spin-$2$ tensor in de Sitter. Since these correspond to scalar curvature perturbations (which seed the fluctuations we observe in the CMB) as well as primordial gravitational waves, these are the most relevant observables in inflationary cosmology.

The bulk-to-boundary propagator for massless scalar in de Sitter looks like \cite{Jazayeri:2021fvk}
\begin{equation}
    K_\omega=(1-i\omega \eta)\e^{i\omega \eta}\,.
\end{equation}
Clearly, $K_\omega$ satisfies $\partial_\omega K_\omega|_{\omega=0}=0$. Therefore, if we assume that all the interactions in $H_{\text{int}}$ are built out of fields (and their derivatives) at the same spacetime point, we have obtained a constraint on the wavefunction coefficients which we refer to as the manifest locality test (MLT):
    \begin{equation}\label{eq:LOCALITY}
    \partial_{\omega_e}\psi_n(\{\omega\})|_{\omega_e=0}=0 \quad \text{for any external energy $\omega_e$}\,.
\end{equation}
Since a massless spin-$2$ tensor also shares the same bulk-to-boundary propagator, MLT also holds for a massless spin-$2$ tensor.

\subsection{Bootstrap constraint 3: scale invariance}

Massless scalars and spin-$2$ tensors in dS$_{1+3}$ scale in the following way:
\be
\psi_3 \sim \omega^3\,.
\ee

The reason is the following: in de Sitter, mass and scaling dimension are related by
\be
m^2=\Delta(3-\Delta)\,.
\ee
Clearly, if $m^2=0$ this implies $\Delta=0$ or $\Delta=3$. Consequently a scalar near the future conformal boundary behaves as:
\begin{equation}
    \psi_3(\textbf{x})\sim \mathcal{O}_{\Delta=0}+\eta^3\mathcal{O}_{\Delta=3}\,.
\end{equation}
As $\eta\rightarrow 0$ only the first term survives, and it is easy to see that this implies $\psi_3\sim \omega^3$ after a Fourier transform. 

Since we are in de Sitter, one may wonder if we should impose the full de Sitter isometry, the SO(4,1) group, as a constraint in our bootstrap. However, it can be shown that correlators (and subsequently wavefunction coefficients) for the scalar curvature perturbation are suppressed by the slow roll parameters if they respect the full de Sitter isometry \cite{Green:2020ebl}\footnote{One could also the full de Sitter isometry in the cosmological bootstrap. Initially, the approach is to write down Ward identities from the symmetry, which give rise to differential equations for the wavefunction coefficients \cite{Arkani-Hamed:2018kmz, Baumann:2019oyu,Baumann:2020dch, Aoki:2024uyi}. However differential equations have also been found for more general cosmological spacetime \cite{Arkani-Hamed:2023kig,Arkani-Hamed:2023bsv,Grimm:2024mbw, He:2024olr, Benincasa:2024ptf, Grimm:2024tbg}, which leads to the concept of "kinematic flow". }. This means that correlators that are relevant for observations in the near future generally break some of the de Sitter isometry. However, since we do observe a power spectrum which scales almost as $k^{-3}$ (as mentioned in Sec.~\ref{sec:five-things-cosmology}) we still impose scale invariance as a constraint.

Unitarity, locality, and scale invariance are sufficient to strongly constrain wavefunction coefficients. In the next section, we consider a simple example.

\subsection{Bootstrapping \texorpdfstring{$\psi_3$}{psi3} for a single scalar}

Let us start by summarizing what we want $\psi_3$ to satisfy:
\begin{itemize}[label=$\diamond$]
    \item Scale invariance: $\psi_3 \sim \omega^3$.
    \item Locality: $\partial_\omega \psi_3|_{\omega=0} = 0$.
    \item Symmetric under permutations of $\omega_{1,2,3}$.
\end{itemize}
Note that while we have not included unitarity in this list, it will be used in the following section to construct more interesting wavefunction coefficients by ``gluing'' simpler ones.

The simplest ansatz satisfying these properties is 
\be\label{eq:ansatz}
\psi_3^{(p)} = \frac{1}{\omega_T^p} \sum_{m,n\ge 0}c_{mn}\omega_{T}^{3+p-2m-3n}e_2^me_3^n ~ \text{where} ~ \begin{sqcases}
   p\in\mathbb{N}\\
    (3+p-2m-3n)\ge 0\\
    \omega_T = \omega_1 + \omega_2 + \omega_3\\
    e_2 = \omega_1 \omega_2 + \omega_1 \omega_3 + \omega_2 \omega_3\\
    e_3 = \omega_1 \omega_2 \omega_3
\end{sqcases}\,.
\ee
This ansatz corresponds to the wavefunction coefficients for tree level diagrams. It is written in $\omega_T,e_2, e_3$ which is symmetric in external energies, and its overall scaling is $\omega^3$. Note that $p$ is generally related to the number of derivatives in an interaction.

The only tree-level diagram for $\psi_3$ is the contact diagram, and so we expect singularities only at total energy poles (as this is the only subdiagram for a contact diagram). Therefore, in our ansatz we only consider the case where $m,n\geq 0$ and only allow for total energy poles.

Let us consider some explicit examples.
\begin{mdexample}
    In de Sitter, let us try to fix \eqref{eq:ansatz} for $p=0$. We start with the ansatz 
\be
\psi_3^{(0)} = c_1 \omega_T^3 + c_2 \omega_T e_2 + c_3 e_3\,.
\ee
Imposing locality gives 
\begin{equation}
    \begin{split}
        \partial_{\omega_1}\psi_3^{(0)}|_{\omega_1=0}=0 \implies c_2=-3 c_1~\text{and}~c_3=3c_1\,,
    \end{split}
\end{equation}
such that 
\be
\psi_3^{(0)} = c_1 (\omega_T^3-3\omega_T e_2+3e_3)\,.
\ee
This is the result expected from the free theory ($H_{\text{int}}=0$) after field redefinition $\phi\mapsto \phi+\phi^2$.

We can also modify the ansatz as follows:
\be
\psi_3^{(0)} = c_1 \omega_T^3 + c_2 \omega_T e_2 + c_3 e_3 + \log(- \omega_T \eta_0) (\tilde c_1 \omega_T^3 + \tilde c_2 \omega_T e_2 + \tilde c_3 e_3)\,.
\ee
Repeating the above exercise, we obtain the correct $\phi^3$ interaction tree-level wavefunction coefficient.

If we consider $p=3$, we find only two polynomials which satisfy all the bootstrap requirements:
\begin{subequations}
        \begin{align}
	\psi_3^{\text{EFT1}}&=\frac{e_3^2}{\omega_T^3}\,,\\
	\psi_3^{\text{EFT2}}&=\frac{1}{\omega_T^3}(\omega_T^6-3\omega_T^4 e_2+11\omega_T^3e_3-4\omega_T^2e_2^2-4\omega_Te_2e_3+12e_3^2)\,.
    \end{align}
\end{subequations}
Interestingly they correspond to the wavefunction coefficients computed from $\dot{\phi}^3$ and $\dot{\phi}(\nabla_i\phi)^2$ interactions respectively. Both interactions have exactly three derivatives. This is a general feature: $p$ tell us the number of derivatives on the scalar fields \cite{Jazayeri:2021fvk}.
\end{mdexample}

From this example, we observe that all that we have obtained are tree-level objects. Why is this the case? This is because the ansatz proposed was constructed from simple polynomials. If we aim to derive objects that might emerge from a loop diagram, the ansatz needs to be modified. One loop wavefunctions generally include polylogarithmic or even elliptic functions \cite{Salcedo:2022aal}, so we need to add those into the ansatz while taking into account the different possible singularities. Similarly to Feynman integrals \cite{Bourjaily:2022bwx}, in very general cases, it is not even clear what class of functions is needed for the ansatz.

\subsection{Gluing procedure}
The purpose of this section is to explain how to ``glue" tree-level wavefunction coefficients together to create more interesting ones. Below, we illustrate this procedure by gluing $\psi_3$ with $\psi_4$. The general approach is as follows. 

Starting with \eqref{eq:DISC1}, we have (schematically)
\begin{equation}\label{eq:disc2}
\begin{split}
    \psi_4(\omega_1, \ldots,\omega_4, \omega_s) -& \psi_4^*(-\omega_1, \ldots,-\omega_4, \omega_s) \\&= [\psi_{3,\text{L}}(\omega_1,\omega_2, \omega_s) - \psi_{3,\text{L}}^*(-\omega_1,-\omega_2, \omega_s)]\\& \qquad \times [\psi_{3,\text{R}}(\omega_3,\omega_4, \omega_s) - \psi_{3,\text{R}}^*(-\omega_3,-\omega_4, \omega_s)]\,.
\end{split}
\end{equation}
Denoting
\begin{equation}
    \omega_1+\omega_2+\omega_s=E_{\text{L}}
    \quad \text{and} \quad 
    \omega_3+\omega_4+\omega_s=E_{\text{R}}\,,\label{eq:defElEr}
\end{equation}
as the energy flowing in the left and right vertices respectively and taking the residue at $E_{\text{L}}=0$ on both sides of \eqref{eq:disc2} gives
\begin{equation}\label{eq:disc3}
    \text{Res}_{E_{\text{L}}=0}\psi_4(\omega_1, \ldots,\omega_4, \omega_s)=\psi_{3,\text{R}}(\omega_3,\omega_4, \omega_s) - \psi_{3,\text{R}}^*(-\omega_3,-\omega_4, \omega_s)\,.
\end{equation}
Of course, a similar expression holds for the residue at $E_{\text{R}}=0$. Writing the wavefunction coefficients as functions of $E_{\text{L}}$ and $E_{\text{R}}$, we can perform the shift $(E_{\text{L}},E_{\text{R}}) \mapsto (E_{\text{L}}+z,E_{\text{R}}-z)$ to write
\begin{equation}
    B=\oint_\gamma \frac{\d z}{z} \psi_4=\psi_4+\text{Res}_{z=-E_{\text{L}}}\psi_4+\text{Res}_{z=E_{\text{R}}}\psi_4\,.
\end{equation}

The contour integral simply picks up the partial energy poles, which are obtained in \eqref{eq:disc3}. By demanding $\psi_4$ to satisfy the locality test, $B$ can be fixed, and eventually this fixes $\psi_4$ completely \cite{Jazayeri:2021fvk}.

The takeaway point of this discussion is that, at tree level, unitarity and an ansatz of the form \eqref{eq:ansatz} are sufficient to completely bootstrap the wavefunction coefficients.

\begin{mdexample}
As an example of how the gluing procedure works, let us glue two contact $\psi_3$ from $\phi^3$ interaction to obtain an exchange $\psi_4$. Remember the contact $\psi_3$ is given by: 
\begin{equation}
    \psi_3(\omega_1,\omega_2,\omega_3)=\frac{1}{\omega_1+\omega_2+\omega_3}.
\end{equation}
First we compute the Disc, which is:
\begin{equation}
    \text{Disc}_{\omega_s}\psi_3(\omega_1,\omega_2,\omega_s)=\frac{1}{E_L}-\frac{1}{E_L-2\omega_s},
\end{equation}
where $E_L, E_R$ are defined in \eqref{eq:defElEr}. Now we look at the following:
\begin{equation}
    \Xi=\frac{1}{2\omega_s}\text{Disc}_{\omega_s}\psi_3(\omega_1,\omega_2,\omega_s)\text{Disc}_{\omega_s}\psi_3(\omega_3,\omega_4,\omega_s),
\end{equation}
and shift the energies by $(E_{\text{L}},E_{\text{R}}) \mapsto (E_{\text{L}}+z,E_{\text{R}}-z)$. As a result we obtain:
\begin{equation}
    \Xi=\frac{1}{2\omega_s}\left[\frac{1}{E_L+z}-\frac{1}{E_L+z-2\omega_s}\right]\left[\frac{1}{E_R-z}-\frac{1}{E_R-z-2\omega_s}\right].
\end{equation}
Clearly the residue at $z=-E_L$ is given by:
\begin{equation}
    \frac{1}{(E_R+E_L)(E_R+E_L-2\omega_s)}.
\end{equation}
The residue at $z=E_R$ is obtained similarly, and so we have:
\begin{align}
    \psi_4&=\frac{1}{E_L}\frac{1}{(E_R+E_L)(E_R+E_L-2\omega_s)}+\frac{1}{E_R}\frac{1}{(E_R+E_L)(E_R+E_L-2\omega_s)}+B\nonumber\\
    &=\frac{1}{\omega_T E_L E_R}+B,
\end{align}
and we notice that $B=0$ gives us the correct answer for the exchange $\psi_4$ in flat space.

Generically in de Sitter, $\Xi$ has higher order poles in $E_L$ and $E_R$, and $B$ would not be zero. See section 6 of \cite{Jazayeri:2021fvk} for more examples. 

\end{mdexample}

\subsection{Cancellation of singularities}
There is an interesting story about the singularities of in-in correlators versus the singularities of the wavefunction. As an example, consider the one-loop flat-space wavefunction with a single vertex. 

\begin{equation}
    \includegraphics[scale=1.0]{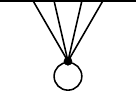}
\end{equation}

The wavefunction is given by:
\begin{equation}
    \psi_4\sim\log(\omega_T)\,.
\end{equation}

If we try to compute the correlator for the same graph, we find that it is given by:
\begin{equation}
    \langle \phi(k_1)\dots\phi(k_4)\rangle = \int\frac{\d^3 p}{(2\pi)^3}\frac{1}{2p}.
\end{equation}

Crucially, no matter how we choose to regularize this integral, we can never obtain a $\log(\omega_T)$ term. If we write down the one-loop in-in correlator in terms of wavefunction coefficients we obtain:
\begin{equation}
    \langle \phi(k_1)\dots\phi(k_4)\rangle_{\text{1-Loop}}\sim \psi_4^{\text{1-Loop}}+\int\frac{\d^3p}{(2\pi)^3}\frac{1}{2\text{Re}\psi_2(p)}\psi_6^{\text{tree}}.
\end{equation}
The integral over the tree-level $\psi_6$ cancels the logarithmic term from the one-loop $\psi_4$. 

This is an important lesson on the analytic structure of the wavefunction versus the analytic structure of in-in correlators. When we go to loop level, an $n$-point correlator is no longer only given by $\psi_n$, but also integrals of higher-point wavefunctions (with lower loop order). This can result in cancellation of singularities: for instance, we find that any branch point in $\omega_T$ from a wavefunction coefficient is never present in an in-in correlator \cite{AguiSalcedo:2023nds}. Interestingly, in flat space and at one loop order, the remaining singularities can be mapped to singularities from an amplitude with the same Feynman graph \cite{Lee:2023kno}. It would be interesting to better understand this story, particularly in the context of de Sitter spacetime\footnote{The flat space result was obtained by considering the spectral representation of the bulk-to-bulk propagator, which give rise to an integral representation of the wavefunction coefficients. Naturally one could try developing similar integral representations \cite{Chowdhury:2023arc}, spectral representations or dispersive relations \cite{Meltzer:2021zin,Sleight:2020obc, Melville:2024ove, Werth:2024mjg, Liu:2024xyi} in de Sitter.}.

\newpage
\bibliographystyle{jhep}
\bibliography{references}

\end{document}